\documentclass[12pt]{article}

  \usepackage{graphicx}
  \usepackage{epsfig}
  \usepackage{amssymb}
  \usepackage{subfigure}
  \usepackage{amsmath} 

\evensidemargin - .5truecm
\allowdisplaybreaks[1]

\def \lsim
{\raisebox{-3pt}{$\>\stackrel{<}{\scriptstyle\sim}\>$}}
\def \gsim
{\raisebox{-3pt}{$\>\stackrel{>}{\scriptstyle\sim}\>$}}

\begin{document}

\newcommand{\CC}{{\mathbb C}}
\newcommand{\RR}{{\mathbb R}}
\newcommand{\ZZ}{{\mathbb Z}}
\newcommand{\QQ}{{\mathbb Q}}
\newcommand{\NN}{{\mathbb N}}
\newcommand{\beq}{\begin{equation}}
\newcommand{\eeq}{\end{equation}}
\newcommand{\beal}{\begin{align}}
\newcommand{\eeal}{\end{align}}
\newcommand{\nn}{\nonumber}
\newcommand{\bea}{\begin{eqnarray}}
\newcommand{\eea}{\end{eqnarray}}
\newcommand{\ba}{\begin{array}}
\newcommand{\ea}{\end{array}}
\newcommand{\bfig}{\begin{figure}}
\newcommand{\efig}{\end{figure}}
\newcommand{\bc}{\begin{center}}
\newcommand{\ec}{\end{center}}

\newenvironment{appendletterA}
{
  \typeout{ Starting Appendix \thesection }
  \setcounter{section}{0}
  \setcounter{equation}{0}
  \renewcommand{\theequation}{A\arabic{equation}}
 }{
  \typeout{Appendix done}
 }
\newenvironment{appendletterB}
 {
  \typeout{ Starting Appendix \thesection }
  \setcounter{equation}{0}
  \renewcommand{\theequation}{B\arabic{equation}}
 }{
  \typeout{Appendix done}
 }

%
%
%
%

\begin{titlepage}
\nopagebreak

\renewcommand{\thefootnote}{\fnsymbol{footnote}}
\vskip 2cm
\begin{center}
\boldmath

{\Large\bf Winter (or $\delta$-shell) Model at Small and Intermediate Volumes}

\unboldmath
\vskip 1.cm
{\large  U.G.~Aglietti and A.~Cubeddu}
\vskip .3cm
{\it Dipartimento di Fisica, Universit\`a di Roma ``La Sapienza''}
\end{center}
\vskip 0.7cm

\begin{abstract}

We consider Winter (or $\delta$-shell) model at finite volume,
describing a small resonating cavity weakly coupled to a large one, for small and intermediate volumes (lengths).
By defining $N$ as the ratio of the length of
the large cavity over the small one,
we study the symmetric case $N=1$,
in which the two cavities actually have the
same length, as well as the cases $N=2,3,4$.

By increasing $N$ in the above range, the transition from
a simple quantum oscillating system
to a system having a resonance spectrum
is investigated.
We find that each resonant state is represented,
at finite volume, by a cluster of states,
each one resonating in a specific coupling region,
centered around a state resonating at very small
couplings.

We derive high-energy expansions for the 
particle momenta in the above $N$ cases,
which (approximately) resum their perturbative series
to all orders in the coupling among the cavities.
These new expansions converge rather quickly 
with the order, provide, surprisingly, a uniform 
approximation in the coupling
and also work, again surprisingly, at low energies.

We construct a first resummation scheme 
having a clear physical picture, which is based
on a function-series expansion, as well as 
a second scheme based on a recursion equation.
The two schemes coincide at leading order,
while they differ from next-to-leading order on.
In particular, the recursive scheme realizes an 
approximate resummation of the function-series
expansion generated within the first scheme.

\vskip .4cm

\end{abstract}
\vfill
\end{titlepage}    

\setcounter{footnote}{0}

\newpage

\tableofcontents

\newpage


\section{Introduction}

%
%

Resonances, also called unstable or metastable states, occur in any branch of quantum physics
as basic dynamical mechanisms
--- let's say from solid-state to particle physics ---, so their relevance is hard to over-estimate. 
As well known, resonances occur in quantum mechanics
and in quantum field theory,
when a discrete state is immersed and
weakly-coupled to a continuum of states
\cite{gamow}-\cite{Torrontegui};
A bound state is instead a discrete state 
lying below the continuous spectrum,
which cannot decay for kinematic reasons. 
In general, if the coupling of the resonance
to the continuum is sent to zero,
the lifetime of the resonance diverges,
so the latter turns from an approximate to
an exact eigenfunction of the Hamiltonian. 

The quantum theory of resonances can be generalized
to finite volume (in physical space), by coupling a discrete state to a quasi-continuum of states, i.e. 
to a set of closely-spaced levels 
(in energy space), rather than to a continuum
\cite{fano}-\cite{gaveau}. 
In physical language, we may say that we
consider "resonances in a box".
As reasonably expected, the quantum theory of resonances
at finite volume is considerably more
elaborated that the standard theory at infinite volume,
as typically an additional parameter enters the dynamics,
namely the level spacing of the final decay states.
As is often the case in physics,
one can study either general properties of
resonances in a box, or 
peculiar properties in specific resonance
models, which one may eventually try
to abstract later. In this paper we follow
the second route, by studying the generalization 
of Winter model to finite volume \cite{FVWMme},
describing a small resonating
cavity weakly coupled to a large one.

%
%

In general, there are various motivations to study resonances 
in a box, which are of different nature. 
There is certainly a mathematical-physics
interest in that, as finite-volume models
can be considered some sort of 
"infrared regularization" of the 
original infinite-volume models.
In general, quantum models
describing resonances in a box exhibit a new 
and rich mathematical structure.
We will study such structure in the case
of Winter model at finite volume.

Resonances in a box also present peculiar 
quantum-mechanical interference phenomena,
which do not occur in the usual, infinite-volume
limit. 
Since the energy spectrum of a finite-volume
model is discrete, quantum
recurrence phenomena do occur: There exist
large, but finite, times $t_R$'s, at which the wavefunction $\psi(t)$ of the system
comes arbitrarily close, in $L^2$ norm, to the initial
wavefunction $\psi_{ini}\equiv\psi(t_{ini})$,
$\left\| \psi(t_R)-\psi_{ini} \right\| < \epsilon \ll 1$.
In physical terms, after very long, specific times, 
the unstable state approximately goes back to its initial form.
In this context, one may study, for example, the distribution of the 
recurrence times $t_R$'s as functions of the parameters of the model,
as well as of the properties of the initial state.

In the standard, infinite-volume case,
if the initial wavefunction of the resonant state
is orthogonal to the discrete spectrum of the model%
\footnote{
A trivial example is a model with no discrete spectrum, such as standard Winter model
for $g>0$, i.e. for a repulsive coupling.
},
time evolution completely empties the initial
state asymptotically.
On the contrary, at finite volume 
(even in the case where the model, in the infinite volume limit, for example, has no discrete spectrum),
limited-decay phenomena may occur,
i.e. times where the amplitude in the 
initially occupied region
(typically a cavity or a potential well) 
approximately vanishes, do not exist.
In physical language, we may say that times at which
(almost) all the particles of the system have decayed,
do not exist.

As far as the phenomenological relevance is concerned,
let us observe that, in many physical situations, 
the initial
unstable state and/or the final decay products
are subjected to space restrictions.
Consider for example the $\beta$-decay of 
a nucleon (a proton or a neutron) inside a 
nucleus. The fact that the decaying nucleon
is not isolated in space, but is 
contained in a bound state (and
is surrounded by the other nucleons composing the nucleus), 
has substantial effects on its decay.
These effects are often modeled by the so-called
"Fermi motion", i.e. by assigning a simple
momentum distribution of the nucleon inside 
the nucleus; 
A common choice is a Gaussian distribution with a tunable width (the nucleus is assumed to be at rest).
That is equivalent to put the nucleon
in a harmonic potential well (a harmonic trap).
The general idea is that a particle in a bound state
can be roughly described as a particle
inside a box of proper size.
This treatment of the initial bound state 
is obviously rather rough and phenomenological 
in nature;
A detailed  treatment of initial bound-state 
effects on nucleon decay is clearly much more complicated.
In a shell model, for example, protons and neutrons 
are placed into one-particle states in
potential wells, representing average
nucleon interactions.
The initial decaying nucleon $N$ is usually 
contained in the 
lowest-energy available state, while
the produced nucleon $N'$ goes into one
of the available low-energy states.

Let us also briefly discuss an example of decays in restricted volumes of space, coming from particle physics.
The main decay mechanism of hadrons containing a heavy quark, i.e. in practice a charm or a beauty quark, is the fragmentation of the latter.
Consider for example a $B^-$ meson, 
i.e. a meson composed of a beauty ($b$)
quark and a light up anti-quark $(\bar{u})$.
The main semileptonic decays of the $B^-$'s originate
from the fragmentation of its
constituent $b$ quark into a charm $(c)$ quark
and a lepton pair, such as for example 
\beq
b \, \to \, c \, + \, e^- \, + \, \bar{\nu}_e,
\eeq
where $e^-$ is an electron and $\bar{\nu}_e$
is an electron anti-neutrino.
The fragmenting $b$ quark is immersed
in the intense color field created by the $\bar{u}$ 
antiquark, which acts as a spectator of the
fragmentation, 
and is contained in a small region $\approx$ 1 Fermi
(1 Fermi = $10^{-13}$ cm).
In the above decay, 
in order to form the real final hadronic states,
the final $c$ quark  must at the end combine 
with the spectator $\bar{u}$, or with eventual quarks
or anti-quarks created from the vacuum
by the strong interaction.
Because of color confinement,
all the quarks and antiquarks involved in the above 
decay do not move freely at asymptotic times
$(t \to \pm \infty)$, but come from initial-state
hadrons (color-singlet bound states of quarks) 
and go into final-state hadrons.
Taking into account initial and final bound-state 
effects is therefore crucial in order to understand
the decays of heavy hadrons.
Actually, there exists a popular model of hadrons,
the so-called "bag model", in which the 
hadrons are modeled as almost-free sets of
quarks contained in "bags", i.e. boxes 
of size $\approx$ 1 Fm \cite{TDLee}. 
As already noted, a particle in a potential
well can be described, to a first approximation, 
as a particle in a box.
In the bag model, the dominant decay of a $B$ meson
is described as the fragmentation of 
its constituent $b$ quark, treated  as a
free particle inside a box.

Further motivation to study resonances in a box
comes from the possibility of creating them
and observing their time evolution
in nanostructures. This possibility has been
discussed in detail in ref.\cite{FVWMme}, so we do not
duplicate the discussion.

%
%

%
%

The paper is organized as follows.
In section \ref{Winter_Intro} we introduce the
finite-volume generalization of Winter model. 
Standard Winter model 
\cite{flugge}-\cite{segre}
is a one-dimensional quantum-mechanical model,
describing a non-relativistic particle 
contained in a resonant cavity
having an impenetrable wall and a 
slightly penetrable one.
The coupling of the model, 
let's call it $g$, describes
the penetrability of the latter wall,
i.e. the coupling of the
cavity to a half-line outside --- the continuum.
In the free limit, $g \to 0$,
also the penetrable wall becomes
impenetrable, so that the system decomposes
into two non-interacting subsystems:
a particle in an isolated cavity, i.e. a box
(having a discrete spectrum only), 
and a particle in a half-line
(having a continuous spectrum only).
By going from the free limit $g=0$ to the
small-coupling domain, $0 < |g|\ll 1$,
the box eigenstates disappear from
the spectrum and, because of adiabatic continuity, 
turn into long-lived resonant states.
The conclusion is that Winter model contains, in
the weak-coupling domain, an
infinite non-degenerate resonance spectrum. 
In the opposite situation of infinite coupling
of the model, $g \to \infty$, the
penetrable barrier completely disappears 
and the system reduces to a particle on 
a half line.
Even though Winter model is not so realistic,
it is the basic Hamiltonian model 
for analytic studies of resonance decays.

In ref.\cite{FVWMme}, Winter model has been
generalized to a finite, large volume (length).
We may say that the generalized model
describes a small resonating cavity coupled
to a large resonating cavity, the latter replacing the
half line (having a continuum of states).
If we denote by $N$ the ratio of the length
of the large cavity over the small one,
standard Winter model is recovered by 
sending to infinity the length of the large cavity,
i.e. by taking the limit $N \to \infty$.
In the infinite-coupling limit
of the model, $g \to \infty$, the
penetrable barrier disappears and the system
reduces to a particle in a box of length
equal to the sum of the cavities lengths.
In the complementary free limit $g \to 0$,
the two cavities completely decouple from 
each other, so that the system reduces to the union
of two non-interacting cavities.
A combined analysis of the limiting cases $g \to 0$ and $g \to \infty$ of finite-volume Winter model
will offer us the possibility
of understanding in qualitative way the interacting
model, $0 < |g| < \infty$, which is our primary 
concern.
In agreement with the general observation above,
we may note that, while usual Winter model is a one-parameter model, namely the coupling $g$,
its finite-volume generalization 
gives rise to a two-parameter model,
the second parameter, $N$, being related to the density
of the quasi-continuous spectrum.


In section \ref{Sect_Spect} we study the momentum spectrum of finite-volume Winter model.
Because of the reflecting walls, the momenta $k$ of the particle are defined up to a sign, so we can assume
for example, without any generality loss, $k>0$.
Since the spectrum of the model is not degenerate, 
studying the momentum spectrum of the model
is completely equivalent to studying the energy 
spectrum ($E=k^2$), but is technically more
convenient.
It turns out that the momentum spectrum 
of the particle has, in
general, three different components.
There is an exceptional component,
with eigenvalues and eigenstates which
do not depend on the coupling $g$.
Such eigenfunctions also exist 
in the continuum limit (in momentum space), 
but in the latter case they have zero (Lebesgue) measure, so 
can be omitted. At finite-volume,
the measure is discrete and such eigenfunctions
cannot be neglected.
Second, there is a resonant component of the spectrum,
containing eigenfunctions having
a pronounced resonant behavior for very small couplings.
Finally, there is a non-resonant component,
containing momentum eigenfunctions which
do not exhibit any resonance behavior
at very small couplings. 
In some cases, the non-resonant eigenfunctions
show a moderate resonance behavior for larger couplings
(still smaller than one).
We present a perturbative expansion,
i.e. an expansion in powers of the coupling $g$,
of the resonant and non-resonant momenta
up to fifth order.
These expansions will be useful later,
when we will use them to check the correctness 
of new kinds of expansions.


In section \ref{Winter_Phys} we review the physics
of finite-volume Winter model in the 
large-volume limit $N \gg 1$ (in physical space)
or, equivalently in the quasi-continuum
limit in momentum (or energy) space,
where the level spacing is $ \approx 1/N \ll 1$.
We will find properties of resonances 
at very large volumes, which we will try later
to recognize also in the intermediate-volume 
cases.
In other words, being at large $N$,
we will easily identify finite-volume 
resonance properties, which later we will try
to find also in the intermediate volume
cases.
In the usual (infinite volume) case,
a given resonance $n$ such as, let's say,
the fundamental one ($n=1$),
is related to a {\it single} generalized eigenfunction
in the resonance spectrum. 
In the complex plane of the particle momentum
(or of the energy),
a resonance is represented by a simple pole 
$k_n=k_n(g) \in \CC$, 
which is an analytic function of the coupling $g$.
On the contrary, in the quasi-continuum case ($N \gg 1$),
each resonance is related to a set of many
different contiguous momentum eigenstates,
containing a resonant levels and close
non-resonant levels. 
In general, for each value of the coupling $g$
in the weak-coupling domain,
the resonant behavior is exhibited by a single state
of this set.
By increasing $g$
from zero up to values reasonably smaller than one
$(|g|\ll 1)$,
the resonant behavior is transferred from one state
of this set to the other.


Sections \ref{sec_symm_case}$-$\ref{sec_quadr_case} 
are devoted to the investigation of finite-volume Winter 
model for the specific values of $N=1,2,3,4$.
These sections are the central ones of the paper,
as they contain most of the original material.
In ref.\cite{FVWMme} finite-volume Winter model
has been investigated mostly in the quasi-continuum case $N \gg 1$.
In this paper we aim at studying 
this model in the complementary case, 
in which $N$ is fixed and is not much larger than one.
With the techniques available to us, we will be able
to investigate analytically the smallest-$N$ case $N=1$,
the symmetric one, as well as the moderately large-$N$ cases $N=2,3,4$.
By increasing $N$ in the above range, we will see 
how resonance dynamics progressively emerges
from a simple quantum oscillating behavior 
of the $N=1$ model, in which the two 
resonant cavities have the same length.


In section \ref{sec_symm_case} we consider  
finite-volume Winter model 
in the symmetric case $N=1$. 
This is by far the simplest case,
as it does not involve non-resonant levels,
but only resonant and exceptional levels.
We derive a high-energy expansion for the momenta of the particle which is, as far as we know, new.
This expansion realizes an approximate resummation of the 
perturbative series for the particle momenta $k=k(g)$, to all orders in the coupling $g$.
This perturbative, high-energy expansion turns out to be {\it a posteriori} a 
very good one, i.e. much better than expected.
Surprisingly, the new expansion converges rather quickly with the order of approximation to the exact momenta,
{\it uniformly} in the coupling 
$g\in\RR$, even though it was derived 
by looking at the ordinary perturbative expansion,
i.e. upon the assumption of weak coupling, $|g|\ll 1$.
Furthermore, our expansion also works for low-energy states, even though it was constructed by
assuming to be at high energy. 
The latter are two "bonus" of our expansion, 
which  allow us to say that we have come close to the 
exact analytic solution of finite-volume Winter model 
for the above values of $N$. 
We construct a first resummation scheme based on a
function series expansion, which has a clear
physical limit, as well as a second scheme
based on a recursion equation.
A comparison between the two schemes is made.
It turns out that the schemes exactly coincide
at first order, i.e. at the lowest-order non-trivial
approximation, while they slightly differ at higher 
orders. In the large-coupling region $(|g|\gg 1)$,
the recursive scheme is perhaps better 
than the function-series one.
That is because the recursive scheme realizes
an approximate resummation of the function series;
In a suggestive language, it "resums the resummation" 
realized by the function-series expansion.

It is clear that finite-volume Winter model for $N=1$ cannot
be considered, in any sense, a large-$N$
approximation of standard Winter model ($N=\infty$).
In particular, if we consider the time evolution
of an initial box eigenfunction contained in
one of the two cavities, we do not expect
the existence a temporal region where an approximate exponential decay with time occurs.
The $N=1$ model simply describes an oscillating system,
as the amplitude is constantly transferred with time,
back and forth, from one cavity to the other.  


In section$\,$\ref{sec_double_case} we consider 
finite-volume Winter model in the
double case $N=2$, in which the large cavity is
two times larger than the small one.
By going from $N=1$ to $N=2$, we encounter,
for the first time, the non-resonant levels, 
which did not exist at $N=1$.
At $N=2$, the level density of the large cavity
(in momentum space) 
is two times larger than the density of the small cavity,
so that resonant and non-resonant levels 
alternate, i.e. separate each other.
The non-resonant levels do not exhibit
a resonant behavior in any coupling region.
Even though $N=2$ cannot be reasonably 
considered a large number,
this case can be thought of as the lowest-$N$ 
case where some mild resonance 
dynamics may be expected.
The high-energy expansion constructed
for the resonant momenta
in the previous $N=1$ model,
is extended to the $N=2$ case,
both for resonant as well as non-resonant levels.
Both resummation schemes are used
and simple analytic formulae are obtained.

%
%

In section$\,$\ref{sect_gen_meth} we discuss, in
general and abstract form, the method
for resumming the perturbative series 
of the particle momenta, 
based on a high-energy expansion.
We decided to insert this section before the
sections devoted to the cases $N=3$ and $N=4$,
because the latter (especially the case $N=4$) 
are much more complex and ramified 
than the cases $N=1$ and $N=2$. 
Very cumbersome and not-intuitive formulae 
are indeed obtained in the $N=3,4$ cases.
For the derivation of the high-energy
expansion in the $N=3,4$ cases,
it is therefore convenient to refer
to a general abstract, previously constructed, 
scheme. 

%
%

In section$\,$\ref{sec_triple_case} we consider 
finite-volume Winter model in the
triple case $N=3$, in which the large cavity is
three times larger than the small one.
By going from $N=2\mapsto 3$,
the main novelty is the "differentiation"
of the non-resonant levels.
A resonance of standard Winter model
is naturally associated, in the $N=3$ model, 
to triplets of states,
consisting of a resonant level
and two contiguous non-resonant levels.
The latter exhibit indeed a mild resonant behavior.
In general, the transition $N=2\mapsto 3$ is a significant one and a much closer behavior to the infinite-volume limit 
($N=\infty$) is expected at $N=3$ than at $N=2$.

%
%

In section$\,$\ref{sec_quadr_case} we consider 
finite-volume Winter model in the 
quadruple case $N=4$. This is the largest-$N$
case which we can treat with our method by means
of elementary functions. 
The extension of our method to the cases $N>4$
would indeed require the introduction of special
functions.
The qualitative changes of the momentum spectrum
by going from $N=3$ to $N=4$ are noticeable,
but are not as great as in the transitions 
$N=1 \mapsto 2$ or $N=2 \mapsto 3$.
At $N=4$, there are three non-resonant levels
for each resonant level and a further
differentiation of the non-resonant levels
occurs. 
A resonance of the continuum ($N=\infty$)
is naturally associated, at $N=4$,
to a triplet of states, as in the case $N=3$. 
In going from $N=3$ to $N=4$, the main qualitative change 
is the occurrence of new non-resonant levels, 
which do not show a resonant behavior
in any coupling region, and which have the role
of separating different resonance triplets.

%
%

Finally, section$\,$\ref{Sect_Concl} contains the
conclusion of our analysis, together
with a discussion about possible future developments.
The extension of our high-energy expansions
to the cases $N>4$ is in principle feasible,
but it requires the introduction of special
functions, as the zeros of general polynomials
of degree $N$ have to be computed. 
A natural evolution of our analysis
also involves the computation
of the temporal evolution of initial
box eigenfunctions contained in the small cavity.
The occurrence of the typical signal of resonance decay,
namely an exponential decay with time, can
be explicitly investigated.
One can also study recurrence and limited-decay
phenomena, as well as small-time
decay effects (the so-called Zeno effect).

%
%

The paper also contains three appendices, presenting
material which, we believe, can be useful to many
(potential) readers.
In appendix \ref{app_tang_mult_ang}, 
we derive a general formula for the
reduction of the tangent of a multiple angle,
namely $\tan(N \pi k)$ with $N$ integer,
to a rational function in
$\tan(\pi k)$ (of degree $N$).
This formula allows the transformation of the transcendental
equation determining the momentum spectrum of the model, to
an algebraic equation in $\tan(\pi k)$ of degree $N$.
This reduction is a crucial step in our method
for analytically evaluating the model spectrum.

In appendix \ref{app_Card_ord3} 
we derive the formula for solving
a general third-order algebraic equation,
i.e. with general complex coefficients
--- the so called third-order Cardano's formula.
This formula, containing nested squared and
cubic roots, is needed to determine the momentum
spectrum of the particle in the $N=3$ model.
It is much more complicated than 
the standard formula for solving second-order 
equations.
In particular, even if the coefficients
and the zeroes of the equation are all real
(the so-called irreducible case),
the roots can be obtained only
by passing through the complex numbers.

Finally, in appendix \ref{app_Card_ord4} we present a sketchy
derivation of the formula for the zeroes of a general fourth-order
algebraic equation, given by the fourth-order Cardano's 
formula.
Within our expansion method, this formula is essential 
to determine the momentum spectrum
of the $N=4$ model.
We decided to devote two separate appendices to the derivation of the third-order and the fourth-order Cardano's formulae, because the latter are not so often used in physics.


\section{Winter model at finite volume}
\label{Winter_Intro}

In this paper we study Winter (or $\delta$-shell)
model, generalized at finite volume,
whose Hamiltonian operator,
after proper rescaling, may be written:
\beq
\label{H_Winter}
\hat{H} \, = \, 
- \, \frac{\partial^2}{\partial x^2} \, + \, 
\frac{1}{\pi g } \, \delta(x-\pi);
\qquad
0 \, \le \, x \, \le \, L;
\eeq
where $L> \pi$ is the total length of the
system and $g \ne 0$ is a real coupling constant.
Vanishing boundary conditions are imposed 
to the wavefunction of the particle
$\psi(x; t)$ at the boundary points:
\beq
\psi(x=0; \, t) \, \equiv \,  0;
\qquad
\psi(x=L; \, t) \, \equiv \,  0;
\qquad
- \, \infty \, < \, t < \, + \, \infty.
\eeq
Note that the Hamiltonian operator above is written in
such a way that it has a simple pole
in the free limit $g\to 0$. 

To simplify the computations, let us assume that
the total length of the system $L$ is an exact
integer multiple of $\pi$, the small cavity
length:
\beq
L \, = \, M \pi; \qquad M = 2,3,4,\cdots.
\eeq
This choice is sufficiently general to allow
the study of the infinite-volume limit
by taking the limit $M \to \infty$, as well as
the study of the quasi-continuum case $M \gg 1$.

The system drastically simplifies in two limiting cases
given by: 1) the strong-coupling limit $g \to \infty$
and 2) the free limit $g \to 0$, which are treated
separately in the next two sections.


\subsection{The strong-coupling limit $g\to \infty$}

In the strong (infinite) coupling limit 
$g\to \infty$, the Hamiltonian operator
of Winter model simplifies into the Hamiltonian
operator of a free particle in the segment $[0,L]$,
with vanishing boundary conditions:
\beq
\label{H_Winter_g_infty}
\hat{H}_{g=\infty} \, = \, 
- \, \frac{\partial^2}{\partial x^2} ;
\qquad
0 \, \le \, x \, \le \, L;
\eeq
with:
\beq
\psi(x=0; \, t) \, \equiv \,  0;
\qquad
\psi(x=L; \, t) \, \equiv \,  0;
\qquad
- \, \infty \, < \, t \, < \, + \, \infty.
\eeq
This system has a non-degenerate discrete
spectrum with quantized momenta
\beq
k \, = \, k_s \, = \, \frac{s}{M}; 
\eeq
energies
\beq
\epsilon_s \, = \, k_s^2 \, = \, \frac{s^2}{M^2};
\eeq
and normalized eigenfunctions
\beq
\psi_s(x,t) \, = \, \sqrt{\frac{2}{L}} \, 
\sin\left(\frac{s \, x}{M} \right) 
\, 
\exp\left( - \, i \frac{s^2}{M^2} \, t \right);
\eeq
where the index $s$ is a (strictly) positive integer,
\beq
s \, = \, 1,\,2,\,3,\,\cdots.
\eeq
The spacing of momentum levels is
constant and is given by:
\beq
\delta k_s \, \equiv \,  k_{s+1} \, - \, k_s
\, = \, \frac{1}{M}. 
\eeq
Note that, when the index $s$ is a multiple of 
the total system length $M$ (in units of $\pi$),
\beq
s \, = \, n \, M; 
\qquad n \, = \, 1,\, 2,\, 3,\, \cdots, 
\eeq
the particle momentum $k$ is integer:
\beq
k_{s=nM} \, = \, n.
\eeq
The corresponding eigenfunction,
\beq
\psi_{s=nM}(x,t) \, = \, \sqrt{\frac{2}{L}} 
\, \sin\left( n x \right) \, 
\exp\left( - \, i n^2 t \right),
\eeq
exactly vanishes at the potential wall
at $x=\pi$ of the general model $(|g|<\infty)$.
The above eigenfunctions, which do not "see"
the Dirac $\delta$-potential with support at $x=\pi$, are also eigenfunctions
of the general finite-volume model, as we are going to see
later.


\subsection{The free limit $g\to 0$}

In the free limit $g\to 0$, the potential wall,
located at $x=\pi$, becomes impenetrable,
implying the vanishing of wavefunctions at
this point:
\beq
g \, \to \, 0 
\quad
\Rightarrow
\quad 
\psi(x=\pi; \, t) \, \equiv \,  0;
\qquad
- \, \infty \, < \, t \, < \, + \, \infty.
\eeq
Therefore,
in this limit,  the system
decomposes into the following two non-interacting
subsystems: 
\begin{enumerate}
\item
A particle in the box
$[0,\pi]$, having a non-degenerate discrete spectrum only,
with particle momenta
\beq
k \, = \, k_n \, = \, n,
\eeq
and corresponding (normalized to one) eigenfunctions
\beq
\psi_n(x,t) \, = \, \sqrt{\frac{2}{\pi}} \, 
\sin(n x) \, \exp\left(- \, i \, n^2 t \right); 
\qquad 0 \, \le \, x \, \le \, \pi;
\eeq
where the index $n$ is a positive integer,
\beq
n \, = \, 1, \, 2, \, 3, \, \cdots.
\eeq
The momentum spacing is constant and equal to one:
\beq
\Delta k_n \, \equiv \, k_{n+1} \, - \, k_n 
\, = \, 1;
\eeq
\item
A particle in the box $[\pi,L]$, 
with $L=M\pi=(N+1)\pi$,
having a 
non-degenerate discrete momentum spectrum 
\beq
p \, = \, p_h \, = \, \frac{h}{N};
\eeq
and normalized (to one) eigenfunctions
\beq
\phi_h(x,t) \, = \,  
\sqrt{\frac{2}{N\pi}} \, 
\sin\left[ \frac{h \, ( x- \pi )}{N} \right]
\, \exp\left( - \, i \, \frac{h^2}{N^2} \, t \right) ;
\qquad \pi \, \le \, x \, \le \, L;
\eeq
where the index $h$ is a positive integer,
\beq
h \, = \, 1, \, 2, \, 3, \, \cdots,
\eeq
and
\beq
N \, \equiv \, M \, - \, 1 \, = \, 1,\, 2,\, 3, \,\cdots
\eeq
is the length of the large cavity $[\pi,M\pi]$
in units of (i.e. divided by) $\pi$.

Note that, if the index $h$ is a multiple of $N$,
\beq
h \, = \, n \, N; 
\qquad n \, = \, 1, \, 2, \, 3, \, \cdots;
\eeq
the momentum $p_h$ of the particle in the large cavity 
is exactly integer,
\beq
p_{\,nN} \, = \, n,
\eeq
being therefore equal to an allowed momentum 
of the small cavity.
Since
\beq
\sin[n(x-\pi)] \, = \, (-1)^n \, \sin(nx),
\eeq
we obtain, in this case, smooth eigenfunctions
at the point $x=\pi$, which coincide with
the eigenfunctions of the infinite-coupling limit
with $s=nM$.

Also in this case the momentum spacing is constant and is given by:
\beq
\Delta p_h \, \equiv \, 
p_{h+1} \, - \, p_h \, = \, \frac{1}{N}.
\eeq
As expected, $\Delta p \equiv \Delta p_h$ tends to zero
in the infinite-volume limit $N \to \infty$.
Note also that $\Delta p$ is larger than the 
momentum spacing $\delta k$ of the infinite-coupling limit
which, as we have seen, is given by 
$\delta k = 1/M = 1/(N+1)$.
This fact, combined with the degeneracy of the
zero-coupling limit at integer momenta,
implies that the average level density in momentum space,
\beq
\frac{\Delta n}{\Delta k},
\eeq
is the same in both limits \cite{FVWMme}.
\end{enumerate}
%
The norm of a wavefunction $\psi(x)$ of the total system
(small cavity + large cavity) 
is naturally defined, in the free limit $g\to 0$, as:
\beq
\| \psi \|^2  \, \equiv \,
\int\limits_0^\pi \left| \psi(x) \right|^2 dx
\, + \,
\int\limits_\pi^L \left| \psi(x) \right|^2 dx.
\eeq
Note that one can consider, in particular, wavefunctions
identically vanishing in the large cavity, for which
\beq
\| \psi \|^2  \, = \,
\int\limits_0^\pi | \psi(x) |^2 \, dx,
\qquad 
\psi(x) \, \equiv \, 0 
\,\,\, \mathrm{for} \,\,\, 
\pi \, < \, x \, < \, L,
\eeq
as well as wavefunctions identically vanishing
in the small cavity, for which
\beq
\| \psi \|^2  \, = \,
\int\limits_\pi^L | \psi(x) |^2 \, dx,
\qquad 
\psi(x) \, \equiv \, 0 
\,\,\, \mathrm{for} \,\,\, 
0 \, < \, x \, < \, \pi.
\eeq
Of course, the wavefunction cannot identically vanish
on both cavities (the particle has to be somewhere in 
the segment $[0,L]$). 

If we consider time-dependent wavefunctions,
i.e. $\psi=\psi(x,t)$, 
in the free limit $g \to 0$,
as already noted,
the potential barrier at $x=\pi$ becomes impenetrable,
so that there are no transitions among the cavities
with time, implying that:
\beq
\frac{d}{dt} \int\limits_0^\pi | \psi(x,t) |^2 \, dx 
\, = \, 0;
\qquad
\frac{d}{dt} \int\limits_\pi^L | \psi(x,t) |^2 \, dx 
\, = \, 0.
\eeq
The momentum (or energy) spectrum of the
free limit ($g \to 0$)  of finite-volume
Winter model is degenerate. 
Indeed, the momentum (or energy) eigenfunction 
\beq
\psi_S(x) \, \equiv \, \sqrt{\frac{2}{\pi}} \, \theta(\pi-x) \, \sin(nx),
\qquad n \, = \, 1, \, 2, \, 3, \, \cdots,
\eeq
identically vanishing in the large cavity,
has exactly the same momentum $k=n$ 
(or energy $E=n^2$) of the eigenfunction
\beq
\psi_L(x) \, \equiv \, \sqrt{\frac{2}{\pi \, N}} \, \theta(x-\pi) \, \sin(nx),
\eeq
identically vanishing in the small cavity.
The function (formally, distribution) 
$\theta(y) \equiv 1$ 
for $y>0$ and zero otherwise is the standard 
Heaviside step function.


\subsection{Generic values of the coupling $g$}

Having discussed in some depth the two limiting cases
$g\to\infty$  and $g \to 0$ of finite-volume
Winter model, we can understand
some qualitative properties of the spectrum
of the model for generic values of $g$.
Because of continuity, for a small but non-zero coupling,
\beq
0 \, < \, g \, \ll \, 1,
\eeq 
the eigenfunctions $\psi(x)$ of the system
are small, but do not vanish 
exactly anymore, at the potential
support,
\beq
0 \, < \, |\psi(x=\pi)| \, \ll \,  1.
\eeq
As a consequence, the small-cavity eigenstates weakly interact with the large-cavity ones
and the probabilities for the particle of being
inside the small cavity or inside the large one
are no more constant with time.

In the large-coupling region, $g \gg 1$,
the potential barrier is low and is not capable,
for example, of keeping the particle inside
the small cavity for some time.
Therefore Winter model at finite volume does
describe resonances in a box
only in the small-coupling region,
as it happens at infinite volume,
in complete agreement with physical intuition.


\section{Spectrum}
\label{Sect_Spect}

Finite-volume Winter Model has a discrete 
(or point) spectrum only.
The eigenfunctions of the system are naturally classified
according to whether:
\begin{enumerate}
\item[1)]
They do (exactly) vanish at the point $x=\pi$,  where the 
$\delta$-potential is supported, i.e. it is concentrated,
\beq
\psi(x=\pi) \, \equiv \, 0.
\eeq	
In this case  we have the "exceptional" part of the spectrum;
\item[2)]
They do not (exactly) vanish at this point,
\beq
\psi(x=\pi) \, \ne \, 0.
\eeq
In this case we have instead the "normal" part of the spectrum.
\end{enumerate}
Let us consider these two components in the
next sections.


\subsection{Exceptional spectrum}

It is immediately checked that the (normalized) wavefunctions
\beq
\label{eigen_k_integer}
\varphi_n(x) \, = \,
\sqrt{\frac{2}{L}} \, \sin\left(p_n \, x\right);
\qquad
0 \, \le \, x \, \le \, L;
\eeq
having the exactly-integer momenta,
\beq
\label{exc_mom}
p_n \, \equiv \, n \, = \, 1, 2, 3, \cdots,
\eeq
form an infinite sequence of
eigenfunctions of finite-volume Winter model.
Note that the eigenvalues $p_n$'s,
as well as the related eigenfunctions $\varphi_n(x)$'s, 
do not depend
on the coupling $g$, because the $\varphi_n(x)$'s
exactly vanish at the point $x=\pi$. 


\subsection{Normal spectrum}

The normal eigenfunctions,
as functions of the (still unrestricted) momentum $k$,
are given by:
\beq
\label{old_ef}
\psi(k; x) \, = \,
\mathcal{N}_N(k) \Big\{
\theta(\pi-x) \sin(\pi N k) \sin(kx)
\, + \,
\theta(x-\pi) \sin(\pi k) \sin\big[ k (L - x ) \big]
\Big\} ;
\eeq
As already defined, $N$ is the length, divided by $\pi$,
of the segment $[\pi, L]$,
i.e. the length of the large cavity $[\pi,L]$
(to which the small cavity $[0,\pi]$ is coupled 
for $g \ne 0$).
By normalizing to one the above eigenfunction,
the normalization constant $\mathcal{N}_N(k)$
turns out: 
\beq
\label{eq_normalization}
\mathcal{N}_N(k) 
\, \equiv \, 
\Bigg\{
\sin^2(\pi N k)
\left[
\frac{\pi}{2}
- \frac{\sin(2\pi k)}{4k}
\right]
\, + \,
\sin^2(\pi k)
\left[
\frac{\pi N }{2}
- \frac{\sin(2\pi N k)}{4k}
\right]
\Bigg\}^{-1/2}.
\eeq
It is trivial to check that the
eigenfunction in eq.(\ref{old_ef}) verifies the boundary conditions,
i.e. that it vanishes both at $x=0$ and at $x=L=(N+1)\pi$,
for any value of $k$ and $N$.
It is also immediate to check that 
$\psi(k;x)$, as a function of $x$,
is continuous across the point $x=\pi$.

The equation for the quantization
of the particle momentum $k$ reads:
\beq
\label{eq_basic_prim}
\sin[\pi(N+1)k] \, + \, 
\frac{\sin(\pi k) \sin(\pi N k)}{g \pi k }
\, = \, 0
\qquad (g \ne 0).
\eeq
By assuming that $k$ is not a fraction
of the form $s/N$, where $s$ is an integer,
we can divide eq.(\ref{eq_basic_prim}) 
by the numerator of the second term
on its left-hand-side (l.h.s.),
obtaining the simpler equation
\beq
\label{eq_basic}
\cot(\pi k) \, + \, \cot(\pi N k) 
\, + \, \frac{1}{\pi g k} \, = \, 0
\qquad (g \ne 0).
\eeq
The above equation
can be explicitly solved with respect to 
the variable (coupling) $z \equiv -g$ as:
\beq
\label{from_k_to_g}
z \, = \, h_N(w);
\eeq
where:
\beq
h_N(w) \, \equiv \, \frac{1}{w \, 
\left[ \cot(w) \, + \, \cot(N w)\right]};
\eeq
with:
\beq
w \, \equiv \, \pi \, k.
\eeq
Because of the presence of cotangent functions 
in $h_N(w)$, namely of the "block"
\beq
\cot(w) \, + \, \cot(N w),
\eeq
which is periodic in $w$ of period, 
\beq
T \, = \, \pi,
\eeq
the function $z = h_N(w)$ is not one-to-one.
That implies that the formal inverse,
\beq
w \, = \, h_N^{-1}(z) \, \equiv \, f_N(z) 
\eeq
is a multi-valued function of $z$, having an 
infinite number of branches.

By means of a simple trigonometric transformation,
the above function can also be written in the
alternative form:
\beq
h_N(w) \, = \, 
\frac{ \sin(w) \sin(N w) }{
w \sin[(N+1) w]}.
\eeq
According to the above representation,
$h_N(w)$ is a meromorphic function of $w$
with simple poles at the momenta
of the form
\beq
\qquad\qquad
w \, = \, \pi \, \frac{h}{N+1},
\qquad\qquad
h \, \notin \, (N+1) \ZZ;
\eeq
namely $h$ an integer not multiple of $N+1$,
so that $w$ is not an integer times $\pi$.
As usual,
\beq
\ZZ \, \equiv \, 
\left\{ \cdots,\,-\,2,\,-\,1,\,0,\,+\,1,\,+\,2,\,+\,3,\cdots\right\}
\eeq
is the ring of the integers, so that $(N+1) \ZZ$
is the set of all the multiples of $N+1$.
The above values of the momentum correspond
to the infinite-coupling limit $z \to \infty$
(no potential barrier at all).

The allowed momenta $k$ (or, equivalently, $w$) 
of the particle are the real solutions
of eq.(\ref{eq_basic}), determining the
coupling-dependent, non-integer momenta of the model.
Since, according to this equation,
the momentum $k$ of a given eigenfunction
is a continuous (actually differentiable) 
function of the coupling $g$,
let us consider this momentum as a function
of the coupling: $k=k(g)$.
In the free limit, $g\to 0$, the function
\beq
\frac{1}{\pi g k}
\eeq
diverges, implying that at least one of the
cotangent functions in eq.(\ref{eq_basic}) diverges.
Equivalently, the function $z \, = \, h_N(w)$ has simple zeroes at 
the momenta 
\beq
\qquad\qquad
w \, = \, \pi \, \frac{s}{N},
\qquad\qquad
s \, \notin \, \ZZ^* \, \equiv \, \ZZ \setminus \{ 0 \},
\eeq
namely $w$ not a non-zero integer times $\pi$.
The above values of the particle momentum are the free-theory
$(z \to 0)$ momenta, where the potential barrier separating 
the cavities becomes impenetrable.


\subsubsection{Resonant case}

Since $N$ is assumed to be an integer
if, in the limit $g \to 0$, 
the particle momentum tends to an integer,
\beq	
k \, \mapsto \, n \, = \, 1,\,2,\,3,\,\cdots,
\eeq
then both cotangent functions in eq.(\ref{eq_basic})
diverge; We say that, in this case, both cavities resonate.
Because of continuity,
for $|g|\ll 1$, we expect the momentum $k$
to be close to an integer, i.e. to be of the form:
\beq
k \, = \, n \, \left( 1 \, + \, \Delta_n \right);
\eeq
where $\Delta_n$ is a very small momentum correction:
\beq
\left| \Delta_n \right| \, \ll \, 1.
\eeq
The eigenvalue equation (\ref{eq_basic})
simplifies in this case to the equation:
\beq
\label{eq_basic_ris}
\cot(\pi n \, \Delta_n) \, + \, \cot(\pi n N \Delta_n) 
\, + \, \frac{1}{\pi g n (1 \, + \, \Delta_n)} \, = \, 0;
\eeq
where we have taken into account the $\pi$-periodicity
of both the cotangent functions.
One then assumes a power-series expansion in $g$ for 
$\Delta_n$:
\beq
\Delta_n \, = \, \Delta_n(g;N) \, = \,
\sum_{l=1}^\infty c_n^{(l)}(N) \, g^l,
\eeq
and replaces the latter on the l.h.s. of eq.(\ref{eq_basic_ris}).
Both cotangent functions are then Laurent expanded
around the origin according to:
\beq
\label{eq_expand_cot}
\cot(x) \, = \, \frac{1}{x} \, - \, \frac{x}{3} \, - \, \frac{x^3}{45}
\, - \, \frac{2}{945} \, x^5 \, + \,
\cdots; \qquad |x| \, < \, \pi.
\eeq
By recursively imposing that the coefficients
of the different powers of $g$ in the momentum
equation are zero, one obtains for the lowest-order
coefficients:
\bea
\label{coef_ris_straight}	
c^{(1)}_n(N) &=& - \left( 1 \, + \, \frac{1}{N} \right);
\\
c^{(2)}_n(N) &=& + \left(1 \, + \, \frac{1}{N}\right)^2 ;
\nonumber\\
c^{(3)}_n(N) &=& + \,
\left( 1 + \frac{1}{N} \right)^3
\left( \frac{\,\nu^2}{3} \, N \, - \, 1 \right) =
\nonumber\\
&=&
+ \, \frac{1}{3} \, \nu^2 \, N 
\, + \, \nu^2 \, - \, 1 
\, + \, \frac{\nu^2-3}{N}
\, + \, \frac{\nu^2-9}{3 N^2}
\, - \, \frac{1}{N^3} ;
\nonumber\\
c^{(4)}_n(N) &=& - \, \left( 1 + \frac{1}{N} \right)^4
\left(
\frac{4}{3} \, \nu^2 N \,  - \, 1
\right); 
\nonumber\\
c^{(5)}_n(N) &=& + \, \frac{1}{45} \left( 1 + \frac{1}{N} \right)^5
\Big[
\nu^4 N^3 \,  - \, 11 \nu^4 N^2 
\, + \, \nu^2 \left(150 \, + \, \nu^2 \right) N \, - \, 45
\Big];
\nonumber
\eea
where we have defined 
\beq
\nu \, \equiv \, \pi \, n.
\eeq
Let us briefly comment on the above formulae.
The first three orders have been computed in \cite{FVWMme}, while the fourth and the fifth orders are new.
Odd powers of $\nu$ are absent from the above formulae.
By looking at the form of the above coefficients,
we find it convenient to define the new coupling
\beq
\zeta \, \equiv \, - \, \left( 1 + \frac{1}{N} \right) g. 
\eeq
The two lowest-order coefficients do not depend
on $n$, the resonance index, and $N$ only
appear as a positive power of $1/N$.
Things dramatically change from third order included on:
In this case the coefficients also depend 
explicitly on $n$
and terms proportional to $N$, as well
as to $1/N$, do appear.
To be specific, $c^{(3)}_n(N)$ and $c^{(4)}_n(N)$
contain terms proportional to $N$ of the
form, respectively,
\beq
\frac{1}{3} \, \nu^2 \, N;
\qquad
- \, \frac{4}{3} \, \nu^2 N.
\eeq
The fifth-order coefficient, $c^{(5)}_n(N)$,
once removed the overall factor
$(1+1/N)^5$, is a third-order polynomial in $N$,
with the leading term in $N$
\beq
\frac{1}{45} \, \nu^4 N^3.
\eeq
Such contributions to the coefficients
of the perturbative series, 
power enhanced for $N \gg 1$,
occur at any order in $g$.
They tend to spoil the convergence
of the (necessarily truncated) 
ordinary perturbative series.
However, it is possible to resum such
enhanced terms to all orders in $g$,
obtaining an improved perturbative
expansion, which allows to 
consistently describe 
the large-$N$ cases \cite{FVWMme}.

Putting pieces together,
the ordinary perturbative expansion for a
resonant level finally reads:
\beq
k \, = \, k_n(g;N) \, = \, 
n \big[ 
1 \, + \, \Delta_n(g;N) 
\big]
\, = \, n
\left[
1 \, + \, \sum_{i=1}^\infty
g^i \, c^{\,(i)}_n(N)
\right].
\eeq
As it stems from the above formula:
\beq
\lim_{g \to 0} k_n(g;N) \, = \, n;
\eeq
i.e. the correct limit is obtained in the
free limit.
In the weak-coupling regime, it is natural
to label a given momentum level, thought as a function of
the coupling $g$, by means of the value assumed for 
$g \to 0$.


\subsubsection{Non-resonant case} 

If, for $g\to 0$, the particle momentum $k$ tends instead 
to a non-apparent fraction of the form $s/N$,
\beq	
g \, \mapsto \, 0:
\qquad 
k \,\, \mapsto \,\, \frac{s}{N} \,\, = \,\, 
\frac{1}{N}, \,\, \frac{2}{N}, \,\, \frac{3}{N}, \,\, \cdots,
\,\, \frac{N-1}{N}, \,\, \frac{N+1}{N}, \,\, \cdots,
\eeq
where $s$ is a positive integer which is not a multiple of $N$,
\beq
s \,\, \ne \,\, 0, \,\, N , \,\, 2N, \,\, 3N, \,\, \cdots,
\eeq
then only the function $\cot(\pi N k)$ in eq.(\ref{eq_basic}) diverges;
In this case, only the large cavity resonates.
By considering the small cavity,
we call the above $k$'s non-resonant momenta.
By means of the euclidean division of $s$ by $N$,
one can write
\beq
s \, = \, n \, N \, + \, l ;
\eeq
where $n$ is the quotient and $l$ in the (non-zero)
remainder.
One can take for these indices, for example,
the ranges:%
\footnote{
An equivalent choice for these indices 
(which is preferable in resonance studies,
as we are going to see in the next section)
involves a (quasi-)symmetric remainder:
\beq
n \, = \, 1,\, 2,\, 3,\, \cdots; 
\qquad 
- \, \frac{N}{2} \, < \, l \, \le \, + \, \frac{N}{2};
\quad l \, \ne \, 0.
\eeq
}
\beq
n \, = \, 0,\, 1,\, 2,\, \cdots; 
\qquad l \, = \, 1, \, 2, \, \cdots, \, N-1.
\eeq
In the weak-coupling regime, $|g|\ll 1$, we assume for 
the non-resonant momenta $k$ an expression of the form
\beq
k \, = \, k_{s/N}(g;N)
\, = \, \frac{s}{N} 
\left[ 1 \, + \, \Delta_{s/N}(g;N) \right];
\qquad
\left| \Delta_{s/N}(g;N) \right| \, \ll \, 1.
\eeq
By replacing the above expression in eq.(\ref{eq_basic}),
one obtains the (still exact) non-resonant momentum equation
\beq
\label{eq_basic_non_ris}
\cot\left[
\pi \, \frac{s}{N} \, + \, \pi \, \frac{s}{N} \, 
\Delta_{s/N}(g;N) \right] 
\, + \, 
\cot\left[ \pi s \, \Delta_{s/N}(g;N) \right] 
\, + \, \frac{N}{
\pi g \, s \left[ 1 \, + \, \Delta_{s/N}(g;N) \right] } \, = \, 0.
\eeq
As remarked above, 
\beq
\lim_{g \to 0} \cot\left[ \pi \, k_{s/N}(g;N) \right] 
\, = \, 
\cot\left(\pi \, \frac{s}{N} \right)
\, = \, 
\cot\left(\pi \, \frac{l}{N} \right);
\eeq
where the r.h.s. of the above equation is a (finite) constant.
As in the resonant case,
we assume for the momentum correction
an ordinary power series expansion in $g$:
\beq
\Delta_{s/N}(g;N)
\, = \, 
\sum_{h=1}^\infty
g^h \, c^{\,(h)}_{n + l/N}(N).
\eeq
The lowest-order coefficients of the perturbative expansion 
have the following explicit expressions:
\bea
\label{eq_Neq2_nonris}
c_{n + l/N}^{\,(1)}(N) &=& - \, \frac{1}{N};
\\ \nonumber \\
c_{n + l/N}^{\,(2)}(N) &=& 
+ \, \frac{\pi}{N} 
\left(n + \frac{l}{N}\right) \cot\left(\frac{\pi l}{N}\right)
+
\frac{1}{N^2};
\nonumber \\ \nonumber \\
c_{n + l/N}^{\,(3)}(N) &=& 
- \, \frac{\pi^2}{N} \left( 1 - \frac{1}{N} \right) 
\left(n + \frac{l}{N}\right)^2 \cot^2\left(\frac{\pi l}{N}\right) \, +
\nonumber \\
&& - \, \frac{3 \pi}{N^2} \left(n + \frac{l}{N}\right) 
\cot\left(\frac{\pi l}{N}\right)
+ \frac{\pi^2}{3N} \left(1+\frac{3}{N}\right) 
\left(n + \frac{l}{N}\right)^2
- \frac{1}{N^3};
\nonumber
\eea
%
%
\bea
c_{n + l/N}^{\,(4)}(N) &=&
\frac{\pi^3}{N} \left( 1-\frac{3}{N} + \frac{1}{N^2}\right)
\left(n + \frac{l}{N}\right)^3 \cot^3\left(\frac{\pi l}{N}\right)
+
\nonumber\\
&+&\frac{2\pi^2}{N^2} \left(3-\frac{2}{N}\right) 
\left(n + \frac{l}{N}\right)^2 \cot^2\left(\frac{\pi l}{N}\right)
+
\nonumber\\
&-& \frac{\pi}{N} 
\left[
\pi^2 \left( 1 + \frac{3}{N} - \frac{1}{N^2}\right)
\left(n + \frac{l}{N}\right)^2
-
\frac{6}{N^2}
\right]
\left(n + \frac{l}{N}\right)
\cot\left(\frac{\pi l}{N}\right) +
\nonumber\\
&-& \frac{4\pi^2}{3N^2} \left( 1 + \frac{3}{N} \right)
\left(n + \frac{l}{N}\right)^2 
+ \frac{1}{N^4};
\nonumber
\eea
%
%
\bea
c_{n + l/N}^{\,(5)}(N) &=&
- \frac{\pi^4}{N} \left( 1 - \frac{6}{N} + \frac{6}{N^2}
- \frac{1}{N^3} \right)
\left(n + \frac{l}{N}\right)^4 
\cot^4\left(\frac{\pi l}{N}\right) +
\nonumber\\
&-& \frac{5\pi^3}{N^2} \left( 2 - \frac{4}{N} + \frac{1}{N^2} \right) \left(n + \frac{l}{N}\right)^3 
\cot^3\left(\frac{\pi l}{N}\right) +
\nonumber\\
&+& \frac{2\pi^2}{N} 
\left[
\frac{\pi^2}{3}
\left(
3 + \frac{7}{N} - \frac{12}{N^2} + \frac{2}{N^3}
\right) 
\left(n + \frac{l}{N}\right)^2
- \frac{5}{N^2} \left( 2 - \frac{1}{N}\right)
\right] \times
\nonumber\\
&& \qquad\qquad\qquad\qquad\qquad\qquad\qquad\qquad
\times \left(n + \frac{l}{N}\right)^2 \cot^2\left(\frac{\pi l}{N}\right) +
\nonumber\\
&+& \frac{5\pi}{N^2} 
\left[
\frac{\pi^2}{3} \left( 4 + \frac{12}{N} - \frac{3}{N^2}
\right) 
\left(n + \frac{l}{N}\right)^2
- \frac{2}{N^2}
\right] 
\left(n + \frac{l}{N}\right)
\cot\left(\frac{\pi l}{N}\right) +
\nonumber\\
&-& \frac{\pi^4}{15 N}\left( 3 + \frac{20}{N} + \frac{30}{N^2}   - \frac{5}{N^3}\right) \left(n + \frac{l}{N}\right)^4 +
\nonumber\\
&+& \frac{10\pi^2}{3N^3} \left( 1+ \frac{3}{N} \right)
\left(n + \frac{l}{N}\right)^2 - \frac{1}{N^5}.
\nonumber
\eea
%
%
%
The first three orders have been computed in \cite{FVWMme}, while the fourth and the fifth
orders are new.
The cotangent function always appears in the "block"
\beq
\left(n + \frac{l}{N}\right)
\cot\left(\frac{\pi l}{N}\right)
\, = \,
\left(n + \frac{l}{N}\right)
\cot\left[\pi \left(n + \frac{l}{N}\right)\right]
.
\eeq
By increasing the order of the expansion,
the expression of the coefficients becomes quickly
very cumbersome.
The coefficient of order $n$ is a polynomial
of order $n-1$ in  $\cot\left(\pi l/N\right)$.
Note that the first-order coefficient 
$c_{n + l/N}^{\,(1)}(N)=-1/N$
is suppressed by a factor
$1/(N+1)$ with respect to
the first-order resonant coefficient
$c_{n}^{\,(1)}(N)=-1-1/N$.
In the latter coefficient,
we may think that the contributions 
$-1$ and $-1/N$ 
originate from the small and
the large cavity respectively.

Note that the function $\cot\left(\pi l/N\right)$
becomes very large in the quasi-continuum
limit $N \gg 1$ for the non-resonant levels
close to the resonant ones, i.e. for:
\beq
0 \, < \, |l| \, \ll \, N.
\eeq
One has indeed:
\beq
\cot\left(\frac{\pi l}{N}\right) \approx \frac{N}{\pi \, l} 
\, \gg \, 1.
\eeq
Therefore there are $N$-enhanced contributions
to the coefficients of the perturbative series
also of the non-resonant levels, which are close 
to the resonant ones.
Such enhanced terms can be resummed to all
orders in $g$ also in this case \cite{FVWMme}.

The perturbative expansion for
non-resonant momenta finally reads:
\beq
\qquad
\qquad
k \, = \, k_{n+l/N}(g;N) \, = \, 
\left( n \, + \, \frac{l}{N} \right)
\left[
1 \, + \, \sum_{i=1}^\infty
g^i \, c^{\,(i)}_{n + l/N}(N)
\right]
\qquad
\qquad
(l \ne 0).
\eeq	
As in the resonant case, it is natural to
label the curve of a given momentum level
$k=k(g)$ in the $g$-$k$ plane by means 
of the value assumed
in the free limit, i.e.
\beq
k \, = \, k_{s/N}(g;N);
\eeq
with:
\beq
\lim_{g \to 0} k_{s/N}(g;N) \, = \, \frac{s}{N}.
\eeq

 
\section{Physics of the model in the quasi-continuum limit $(N \gg 1)$}
\label{Winter_Phys}

In this section we try to explain
resonance dynamics of finite-volume Winter model
in the quasi-continuum limit $(N \gg 1)$
in a simple way.
As we are going to show, while the analysis
is necessarily rather long and detailed
(at our present understanding of the matter), 
the physical conclusions are very simple.

As well known from resonant scattering theory
\cite{newton}, there are two typical resonant behaviors:
\begin{enumerate}
\item
The production cross section of the resonance,
$\sigma_{res} = \sigma_{res}(E)$,
having a high and thin peak as the function
of the initial (real) energy $E$, as we cross
the resonance mass (or energy) $M$:
\beq
\sigma_{res}(E) \, \approx \, 
\frac{c^2}{(E-M)^2 \, + \, \Gamma^2/4},
\qquad
|E\, - \,M| \, \lsim \, \Gamma,
\eeq
where $\Gamma>0$ is the decay width
and $c$ is a real constant
specifying the coupling of the initial
state to the resonance. 
\item
The phase of the resonant state,
$\delta = \delta(E)$,
quickly crossing the value
$\pi/2$ (modulo $\pi$), again as a
function of the initial energy $E$,
when we cross the resonance mass $M$:
\beq
\delta(E) \, \approx \, 
\arctan\left[ \frac{\Gamma}{2(M-E)} \right].
\eeq
Indeed:
\beq
\lim_{E \to M} \delta(E) \, = \, \frac{\pi}{2}
\qquad (\mathrm{mod}\,\,\pi)
\eeq
and
\beq
\frac{d\delta}{dE} \, = \, \frac{\Gamma}{2}
\, \frac{1}{(E-M)^2 \, + \, \Gamma^2/4}
\,\,\, \mapsto \,\,\,
\frac{2}{\Gamma} \, \gg \, 1
\qquad \mathrm{for}\,\, E \, \mapsto \, M
\eeq
when $\Gamma \ll 1$.
\end{enumerate}
%
%
\begin{figure}[ht]
\begin{center}
\includegraphics[width=0.5\textwidth]{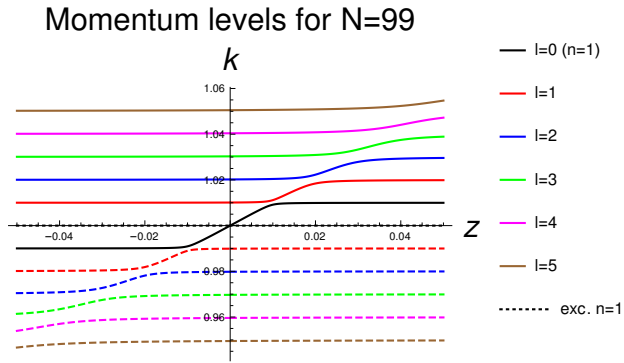}
\footnotesize
\caption{
\label{fig_PkTOT}
\it
Plot of the momentum $k$ of the first resonance $(n=1,\,l=0)$ and of the ten closest
non-resonant levels $(n=1, \,l=\pm 1, \pm 2, \pm 3, \pm 4, \pm5)$,
as functions of the coupling $z\equiv -g$.
The first exceptional level $k_{exc}\equiv 1$ is also
plotted (the black dotted line, overlapping with the $x$-axis).
The levels with $l<0$ are dashed and have the same
color of the corresponding $l>0$ levels;
For example, the red dashed line is the plot
of the $l=-1$ level.
}
\end{center}
\end{figure}
%
%
Both above behaviors are obtained
with a resonant amplitude $A_{res}$ having,
as a function of the complex energy $E$,
a simple pole at $M-i\Gamma/2$
with residue equal to $c$:
\beq
A_{res} \, \approx \, \frac{c}{E \, - \, M \, + \, i \, \Gamma/2}.
\eeq
We have indeed:
\beq
\sigma_{res} \, \approx \, |A_{res}|^2;
\qquad 
\tan\left(\delta\right) \, \approx \, 
\frac{\mathrm{Im}(A_{res})}{\mathrm{Re}(A_{res})} .
\eeq
Our aim is to identify the resonant behavior
in finite-volume Winter model.
The above effects cannot be used in
a straightforward manner because,
in our case, the
energy $E$ cannot be varied continuously,
as the energy spectrum is 
discrete.
The idea is to consider a whole set of 
momentum levels
$k=k_{n+l/N}(z)$ for a given value of $n$
(specifying the resonance order)
and small values of $|l|$ around
zero, $l=0,\pm 1,\pm 2,\pm 3, \cdots$, $|l|\ll N$,
as functions of the coupling $z$.
The latter replaces the variable energy $E\in \RR$ 
and is our "tuning parameter".

By looking at the eigenfunctions $\psi_k(x)$'s
of the Hamiltonian of our model, we find that:
\begin{enumerate}
\item
The first condition above naturally
translates into looking at the values
of $l$ and $z$ for which
the inside amplitude
($0 \le x \le \pi$) is
much larger than the outside 
one ($\pi \le x \le L$);
\item
The second condition
translates into finding the values
of $l$ and $z$ for which
the difference between the phase of the outside
amplitude and the inside one 
--- hereafter the phase shift --- 
quickly crosses the value $\pi/2$ (modulo $\pi$),
as a function of $z$. 
\end{enumerate}
Let us consider the above effects in turn.
In order to be close to the standard
resonances in the continuum (in momentum space),
let us assume to be in the quasi-continuum case,
\beq
N \, \gg \, 1.
\eeq 
We will often identify $N+1$ with $N$,
when the difference between these
quantities can be considered as a correction
of order $1/N$.

%
\begin{figure}[ht]
\begin{center}
\includegraphics[width=0.5\textwidth]{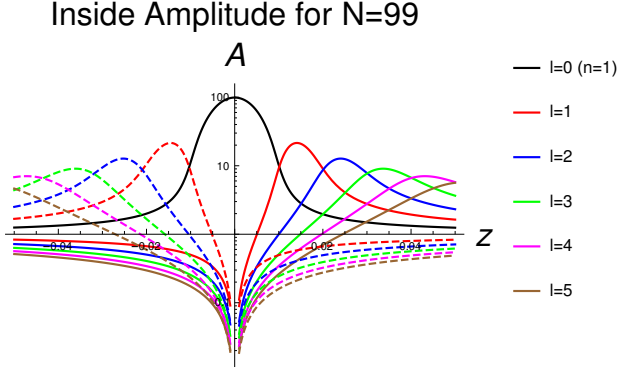}
\footnotesize
\caption{
\label{fig_PLogIA}
\it
Ratio $A$ of the inside amplitude over the outside one,
as a function of the coupling $z$, for $N=99$,
for the first resonance $(n=1)$ and the ten closest
non-resonant levels.
The vertical scale is logarithmic, so that the
downwards vertical asymptote for $z \to 0$ correspond
to a vanishing amplitude ratio $A$ in that limit.
The levels with $l<0$ are dashed and have the same
color of the corresponding $l>0$ levels.
}
\end{center}
\end{figure}
%


\subsection{Amplitude Ratio}

The ratio of the inside amplitude over the
outside one, as a function
of the particle momentum $k$
(thought as an independent variable), reads: 
\beq
A_N(k) \, = \, 
\left| \frac{\sin(\pi N k)}{\sin(\pi k)} \right|.
\eeq
The function $A_N(k)$ is periodic of unitary period,
\beq
A_N(k+1) \, = \, A_N(k),
\eeq
and is identically equal to one at $N=1$,
\beq
A_{N=1}(k) \, \equiv \, 1.
\eeq
That implies that the model with $N=1$
does not exhibit a resonant or an anti-resonant
behavior.
The function $A_N(k)$ has the following
critical points:
\begin{enumerate}
\item
The amplitude ratio has {\it absolute minima} when the sine function
at the numerator, namely $\sin(\pi N k)$,
vanishes, while the sine function at
the denominator, $\sin(\pi k)$, does not, 
i.e. when:
\beq
\label{eq_minima_0}
k \, = \, \frac{s}{N},
\quad
k \, \ne \, \mathrm{integer}.
\eeq
That is to say that the index $s$ is any integer which is not multiple of 
$N$, 
\beq
s \,\, \ne \,\, \cdots, \,\, - \, N, \,\, 0, \,\, + \, N, \,\, + \, 2 N, \,\, \cdots,
\eeq
or, more explicitly:
\beq
s \,\, = \,\, \cdots, \,\,-N-1, \,\,-N+1, \,\, \cdots, 
\,\,-2,\,\,-1,\,\,+1,\,\,+2,\,\,\cdots,\,\, N-1, \,\, N+1, \,\, \cdots.
\eeq
It is often more convenient to rewrite
the above equation in the form:
\beq
\label{eq_minima}
\qquad\qquad
k \, = \, n \, + \, \frac{l}{N},
\quad l \ne 0, \quad 
\qquad (\mathrm{amplitude\,\,absolute\,\,minima}),
\eeq
where $n$ is the quotient and $l$ is the
(non-zero) remainder of the euclidean division of $s$
by $N$, namely $s=nN+l$.
Let us assume the remainder to be in the
quasi-symmetric range 
\beq
- \, \frac{N}{2} \, < \, l \, \le \, + \, \frac{N}{2},
\qquad l \, \ne \, 0.
\eeq
\item
The amplitude ratio has {\it absolute maxima} 
when $k$ is an integer, let's say $n$:
\beq
\label{eq_abs_max}
\qquad
k \, = \, n \, = \, 1,2,3,\cdots 
\qquad (\mathrm{amplitude\,\,absolute\,\,maxima}),
\eeq
because:
\beq
\lim_{k \to n} A_N(k) \, = \, N.
\eeq
\item
The amplitude ratio has {\it local maxima} when the sine function
at the numerator is close to $\pm 1$,
namely when:
\beq
\label{eq_loc_max}
k \, \simeq \, n \, + \, \frac{u \, + \, 1/2}{N},
\qquad u \, \ne \, 0, \, - \, 1
\qquad (\mathrm{amplitude\,\,local\,\,maxima}).
\eeq
The index $u$ is any (signed) integer different from $0$
and $-1$:
\beq
u \,\, = \,\,
\cdots, \,\, - \, 3,\,\, - \, 2,\,\, + \, 1,\,\, + \, 2,\,\, 
+ \, 3, \,\, \cdots
\eeq
The momenta $k \simeq n \pm 1/(2N)$
are not local maxima because, at these points,
the effect of the variation of the sine function at the denominator of $A_N(k)$ cannot be neglected.
\end{enumerate}


\subsubsection{Resonant levels}

The so-called resonant levels, 
whose perturbative expansion
in the coupling $z$ up to first order reads 
\beq
\label{eq_mom_reson_case}
k \, = \, k_n(z) \, = \,
n \, + \, n \left(1 \, + \, \frac{1}{N} \right) z 
\, + \, \mathcal{O}\left(z^2\right),
\qquad
n \, = \, 1, \, 2, \, 3, \, \cdots,
\eeq
posses basically three different regions,
as far as the behavior with respect to the
coupling $z$ is concerned (see fig.$\,$\ref{fig_PkTOT}):
\begin{enumerate}
\item
The negative coupling region
below the critical coupling
of order minus one,
\beq
\label{coupl_neg}
- \, \infty \,\, < \,\, z \,\, \lsim \,\, z_c^{(-1)} 
\, \equiv \, - \, \frac{1}{n \, N},
\eeq
where we have defined the $j$-th critical coupling
\beq
z_c^{(j)} \, \equiv \, \frac{j}{n \, N};
\qquad j \,\, = \,\, \pm \, 1, \, \pm \, 2, \, \pm \, 3, \, \cdots.
\eeq
In this region, the resonant momentum is roughly
constant:
\beq
k_n(z) \, \simeq \, n \, - \, \frac{1}{N}
\, \simeq \, n \, - \, \frac{1}{N \, + \, 1}
\, = \, k_n(z=-\infty) .
\eeq
Therefore, according to eq.(\ref{eq_minima}) for $l=-1$, 
the resonant level $k_n(z)$ does not exhibit any resonant
behavior in the coupling region (\ref{coupl_neg}),
as far as inside-amplitude enhancements are concerned.
\item
The {\it small-coupling region}, symmetric
with respect to (and including) the origin, 
\beq
\label{coupl_res_res_lev}
z_c^{(-1)} \,\,\, \lsim \,\,\, z \,\,\, \lsim \,\,\, z_c^{(+1)}.
\eeq
A resonant behavior occurs inside this region because
the resonant levels approach an integer value
in the free limit,
\beq
\lim_{z \to 0} k_n(z) \, = \, n,
\eeq
where, because of eq.(\ref{eq_abs_max}), the amplitude ratio takes an absolute maximum,
\beq
\lim_{z \to 0} A_N\left[ k_n(z) \right] \, = \, N.
\eeq
Instead, at the first critical points 
$z_c^{(\mp 1)}$,
the resonant momentum $k_n(z)$,
according to eq.(\ref{eq_mom_reson_case}),
takes the values:
\beq
k_n\left(z=z_c^{(\mp 1)}\right) 
\, \simeq \, 
n \, \mp \, \frac{1}{N}.
\eeq
Again according to eq.(\ref{eq_mom_reson_case}),
the resonant momentum $k_n(z)$
rises roughly linearly with the coupling $z$ in the region
(\ref{coupl_res_res_lev}),
from the lower value $k \simeq n - 1/N$ up to the upper value
$k \simeq n + 1/N$.
The above momentum values, 
according to eq.(\ref{eq_minima})
for $l = \mp 1$ respectively,
are the (absolute) minima of the amplitude ratio 
closest to the principal maximum:
\beq
A_N\left[ k_n\left(z_c^{(\mp 1)}\right) \right] 
\, \simeq \, 0. 
\eeq
Therefore, by increasing the coupling $z$
from the minus-one critical coupling 
$z_c^{(-1)}$ up to the plus-one critical point 
$z_c^{(+1)}$, we hit the point $z=0$, where
a strong resonance behavior manifests itself.

To summarize, a strong resonant behavior of the
amplitude ratio is found inside the
small-coupling region (\ref{coupl_res_res_lev}).

Note that the critical couplings $z_c^{(\pm 1)}$
are inversely proportional to the resonance
order (quantum number)
$n$ because, as already remarked \cite{FVWMme}, 
the effective coupling of the $n$-th
resonance to the quasi-continuum is $nz$, rather
than $z$.
That, in turn, originates from the fact that the
$\delta$-function potential produces a finite
discontinuity in the first derivative of the
wavefunction, which is proportional to the
particle momentum $k$ (see eq.(\ref{eq_basic_prim})).
The second resonance ($n=2$), for example,
roughly exhibits at the coupling value $z$  
the same phenomena which the
fundamental resonance ($n=1$) exhibits at $2z$
\cite{FVWMme}. 

Note also that the resonant level 
$k_n(z)=n \, + \, \mathcal{O}(z)$
intersects the exceptional (constant)
level $k_n^{exc} \equiv n$ at the point $z=0$,
which is therefore a singular point.
\item
For larger couplings than the first critical
point,
\beq
\label{lev_ris_large_coupl}
z_c^{(+1)} \,\, \lsim \,\, z \,\, < \,\, + \, \infty,
\eeq
the resonating momentum is roughly
constant:
\beq
k_n(z) \, \simeq \, n \, + \, \frac{1}{N}
\, \simeq \, n \, + \, \frac{1}{N+1}
\, = \, k_n(z=+\infty) .
\eeq
Therefore, again
according to eq.(\ref{eq_minima}) for $l=+1$, 
the resonant level $k_n(z)$ does not exhibit any resonant
behavior in the region (\ref{lev_ris_large_coupl}),
as far as amplitudes are concerned.

Note that the range of $k_n(z)$ 
(a monotonically-increasing function of $z\in \RR$)
is given by:
\beq
\label{eq_range_res}
\mathrm{Range}\left[k_n\right]
\, \equiv \,
k_n(z=+\infty) \, - \, k_n(z=-\infty)
\, = \, \frac{2}{N+1} \, \simeq \, \frac{2}{N}.
\eeq
\end{enumerate}
A few comments are in order.
Roughly speaking, we may summarize the 
above discussion by saying that
the resonant level $k_n(z)$,
by going from $z=-\infty$ up to $z=+\infty$,
has:
\begin{enumerate}
\item
A plateau at $k \simeq n-1/N$ for $z \lsim -1/(nN)$,
with no resonant behavior;
\item
A linearly-rising behavior with $z$ from $k\simeq n-1/N$ 
up to $k\simeq n+1/N$,
for $z$ going from $-1/(nN)$ up to $+1/(nN)$,
with a strong resonant behavior.
These $z$ values form a small-coupling region, 
symmetric with respect to the origin;
\item
A plateau at $k\simeq n+1/N$ for $z \gsim 1/(nN)$,
with no resonant behavior.
\end{enumerate}
As we have just shown,
the so-called "resonant levels" $k_n(z)$, $n=1,2,3,\cdots$,
actually do have a resonant behavior
for very small couplings only.
As we are going to show in the next section, 
for larger (still perturbative) couplings, the resonant
behavior is "transferred" to different
levels, namely the non-resonant levels 
$k_{n+l/N}(z)$,
$l \ne 0$.
Therefore,
the "resonant levels" $k_n(z)$ should be more
properly called "resonant levels
at very small couplings" or
"resonant levels around zero coupling".
However, for consistency reasons, we will
continue to call the levels $k_n(z)$
simply "resonant levels", without further
specification, as in previous work \cite{FVWMme}.


\subsubsection{Non-resonant momentum levels}

Let us now consider the so-called non-resonant levels
as functions of the coupling $z$,
whose perturbative expansion up to first order in $z$
reads:
\beq
\label{eq_lev_non_res_pert_1}
k \, = \, k_{n+l/N}(z)
\, = \, \left( n + \frac{l}{N} \right)
\left( 1 + \frac{z}{N} \right) 
\, + \, \mathcal{O}\left(z^2\right) ,  
\qquad  l \, = \, \pm \, 1, \, \pm \, 2, \, \pm \, 3, \, \cdots.
\eeq
In order to investigate the resonant behavior,
let us assume that the momentum level
sub-index $l$ is much smaller in size than the (large) 
cavity size $N$:
\beq
0 \, < \, |l| \, \ll \, N,
\eeq
so that:
\beq
0 \, < \, \frac{|l|}{N} \, \ll \, n
\, = \, \mathcal{O}(1).
\eeq
As in the resonant case,
there are basically three different coupling 
regions.
For clarity's sake, it is convenient
to treat separately the cases $l>0$
and $l<0$, which are discussed in turn
in the next sections.


\subsubsection{Non-resonant levels with $l>0$}

For $0 < l \ll N$, i.e. for the momentum levels 
$k_{n+l/N}(z)$ right
above the resonant one $k_n(z)$ 
(to investigate resonance behavior, unlike $l$,
$n$ is fixed),
one finds the following coupling regions
(see fig.$\,$\ref{fig_PkTOT}):
\begin{enumerate}
\item
For smaller couplings that the $l$-th
critical coupling,
\beq
\label{eq_non_res_small_coup}
- \, \infty \,\,\, < \,\,\, z \,\,\, \lsim \,\,\, z_c^{(l)} 
\, \equiv \, \frac{l}{n \, N}
\qquad (l>0),
\eeq
the momentum level has roughly a flat region
with a value equal to the free-theory limit ($z\to 0$)
or the $z\to-\infty$ limit:
\beq
k_{n+l/N}(z) \, \simeq \, n \, + \, \frac{l}{N}
\, = \, k_{n+l/N}(z=0)
\, \simeq \, n \, + \, \frac{l}{N+1}
\, = \, k_{n+l/N}(z=-\infty).
\eeq
Unlike resonant levels,
non-resonant levels are quite flat
around $z=0$, because the coefficient
of their first-order correction in $z$,
according to eq.(\ref{eq_lev_non_res_pert_1}),
is $1/N$, i.e. it is very small
(in the resonant case, as already remarked, the corresponding
coefficient is instead $1+1/N$). 
Because of that,
the $\mathcal{O}(z)$ correction to the
free-theory momentum evaluated at the 
$l$-th critical coupling $z=z_c^{(l)}$, 
\beq
\frac{z_c^{(l)}}{N} \, = \,
\frac{l}{n \, N^2} \, = \,  
\mathcal{O}\left(\frac{1}{N^2}\right),
\eeq
is very small, so that $k_{n+l/N}(z_c^{(l)}) \approx k_{n+l/N}(0) = n+l/N$.
According to eq.(\ref{eq_minima}),
in this coupling region, the amplitude ratio
is much smaller than one, so there is not any 
resonant behavior.
\item
In the coupling window
\beq
\label{coupl_interv_reson}
\qquad\qquad
z_c^{(l)} \,\,\, \lsim \,\,\, z 
\,\,\, \lsim \,\,\, z_c^{(l+1)}
\qquad\qquad (l>0),
\eeq
the momentum level $k_{n+l/N}(z)$ rises
roughly linearly with the coupling $z$, from
the value $k=n+l/N$ up to the next value $k=n+(l+1)/N$,
\beq
\label{k_non_res_range}
z: z_c^{(l)} \mapsto z_c^{(l+1)}
\qquad \Rightarrow \qquad
k_{n+l/N}(z): 
\,\, n \, + \, \frac{l}{N} 
\,\,\, \mapsto \,\,\,
n \, + \, \frac{l \, + \, 1}{N}
\qquad
(l \mapsto l+1).
\eeq
To a first approximation,
we may write in this region
(where the approximation in eq.(\ref{eq_lev_non_res_pert_1}) does not hold anymore \cite{FVWMme}):
\beq
k_{n+l/N}(z) \, \simeq \, 
n \, + \, \frac{l}{N} 
\, + \, n \left( z-z_c^{(l)} \right)
\qquad\quad (l>0).
\eeq
Note that the above equation can also
be equivalently rewritten in terms of the higher
critical coupling $z_c^{(l+1)}$ as:
\beq
k_{n+l/N}(z) \, \simeq \, 
n \, + \, \frac{l \, + \, 1}{N} 
\, + \, n \left( z \, - \, z_c^{(l+1)} \right)
\qquad\quad (l>0).
\eeq
Because of linearity,
at the middle point ($m$)
of the coupling interval (\ref{coupl_interv_reson}) 
above, namely at
\beq
\qquad\qquad
z^{(l)}_m \, \equiv \, \frac{z_c^{(l)} \, + \, z_c^{(l+1)} }{2}
\, = \, \frac{l \, + \, 1/2}{n \, N}
\qquad\quad (l>0),
\eeq
the momentum takes the value
\beq
\qquad\qquad
k_{n+l/N}\left( z_m^{(l)} \right) 
\, \simeq \, 
n \, + \, \frac{l \, + \, 1/2}{N}
\qquad\quad (l>0).
\eeq
According to eq.(\ref{eq_loc_max}), the amplitude ratio takes a relative maximum close to $z = z^{(l)}_m$.
Therefore the non-resonant level $k_{n+l/N}(z)$ exhibits
an appreciable resonant behavior inside the coupling region
specified in eq.(\ref{coupl_interv_reson}).
By comparing with the resonant case,
we conclude that, in the coupling region (\ref{coupl_interv_reson}),
among all the momentum levels,
{\it only} the non-resonant level $k_{n+l/N}(z)$
exhibits a resonance
behavior.
\item
Finally, for larger values of the coupling
(a part of which can still be perturbative since $N$ is assumed to be very large),
\beq
\label{coupl_large_non_res}
z_c^{(l+1)} \,\,\, \lsim \,\,\, z \,\,\, 
< \, \,\, + \, \infty,
\eeq
the momentum level $k_{n+l/N}(z)$ roughly has a plateau at the
value
reached at the previous step:
\beq
k_{n+l/N}(z) \, \approx \, 
n \, + \, \frac{l \, + \, 1}{N}
\, \simeq \, n \, + \, \frac{l \, + \, 1}{N \, + \, 1}
\, = \, k_{n \, + \, l/N}(z=+\infty). 
\eeq
Because of eq.(\ref{eq_minima}),
the amplitude ratio in the coupling region 
(\ref{coupl_large_non_res}) is close to zero, exhibiting
then a non-resonant behavior.
We conclude that, in the
coupling region (\ref{coupl_large_non_res}),
neither the resonant level $k_n(z)$
nor the non-resonant level $k_{n+l/N}(z)$
do exhibit a resonance behavior.
\end{enumerate}
To summarize, 
similarly to the resonant levels, also the
non-resonant levels with $l>0$ basically have:
\begin{enumerate}
\item
A plateau at $k \simeq n + l/N$ for $z \, \lsim \, l/(n\,N)$,
with no resonant behavior;
\item
A linear growth with the coupling $z$ from $k=n+l/N$ 
up to $k=n+(l+1)/N$,
for $z$ going from $l/(nN)$ up to $(l+1)/(n\,N)$,
with a relatively strong resonant behavior;
\item
A plateau at $k = n+(l+1)/N$ for $z \gsim (l+1)/(nN)$,
with no resonant behavior.
\end{enumerate}
The range of a non-resonant level with $l>0$
is:
\beq
\mathrm{Range}\left[k_{n+l/N}\right]
\, = \, n \, + \, \frac{l \, + \, 1}{N \, + \, 1} -
\left( n \, + \, \frac{l}{N \, + \, 1}  \right) 
\, = \, \frac{1}{N \, + \, 1} \, \simeq \, \frac{1}{N} 
\qquad (l>0).
\eeq
Note that it is one-half of the range
of a resonant level (cfr. eq.(\ref{eq_range_res})).

Compared to a resonant level,
a non-resonant level with $l>0$ 
has its {\it linear growth region}
shifted to the right, 
\beq
0 \, \mapsto \, l>0
\quad \Rightarrow \quad
z \, \in \, 
\left( - \, \frac{1}{n\,N}, \, + \, \frac{1}{n\,N} \right)
\,\,\, \mapsto \,\,\,
z \, \in \, 
\left( \frac{l}{n\,N}, \,\, \frac{l \, + \, 1}{n\,N} \right),
\eeq
and reduced in size by a factor two
(the slope is approximately the same).

To understand the resonant behavior
of a non-resonant level $k_{n+l/N}(z)$, $l>0$, 
it is also
interesting to compute its range
(i.e. its variation) 
restricted to the positive half-line $(z \ge 0)$
and to the negative one $(z \le 0)$.
An elementary computation gives:
\bea
\mathrm{Range}^-\left[k_{n+l/N}\right]
&\equiv&
k_{n+l/N}(z=0) \, - \, k_{n+l/N}(z=-\infty)
\, = \, \frac{l}{N(N+1)};
\nonumber\\
\mathrm{Range}^+\left[k_{n+l/N}\right]
&\equiv&
k_{n+l/N}(z=+\infty) \, - \, k_{n+l/N}(z=0)
\, = \, \frac{1}{N+1} \left( 1 - \frac{l}{N}\right). 
\eea
For small $l$,
\beq
0 \, < \, \frac{l}{N} \, \ll \, 1,
\eeq
the negative range is approximately
\beq
\mathrm{Range}^-\left[k_{n+l/N}\right] 
\, \simeq \, \frac{l}{N^2},
\eeq
while the positive range is approximately
\beq
\mathrm{Range}^+\left[k_{n+l/N}\right] 
\, \simeq \, \frac{1}{N+1} \, \simeq \, \frac{1}{N}.
\eeq
Therefore the  variation of the level $k_{n+l/N}(z)$ 
is much larger in the positive half-line,
where the linear region occurs,
than in the negative one, as expected:
\beq
\frac{\mathrm{Range}^-\left[k_{n+l/N}\right]}{\mathrm{Range}^+ \left[k_{n+l/N}\right]}
\, \simeq \, \frac{l}{N} \, \ll \, 1.
\eeq
By increasing $l$, the negative range
gets progressively bigger, while the positive 
range gets smaller.
Assuming, for simplicity's sake, $N$ even, at
the largest possible value of $l$,
\beq
l \, = \, \frac{N}{2},
\eeq
the ranges become exactly equal:
\beq
\mathrm{Range}^-\left[k_{n+l/N}\right]  
\, = \, \mathrm{Range}^+\left[k_{n+l/N}\right]
\, = \, \frac{1}{2(N+1)}. 
\eeq


\subsubsection{Non-resonant levels with $l<0$}

Let us now consider the non-resonant levels 
$k_{n+l/N}(z)$ with $l<0$, $|l|\ll N$,
i.e. the levels right below the resonant one $k_n(z)$.
One finds the following behavior (see fig.$\,$\ref{fig_PkTOT}):
\begin{enumerate}
\item
For smaller couplings that the $(l-1)$-th
critical coupling,
\beq
\label{eq_non_res_small_coup_neg}
\qquad\quad
- \, \infty \,\,\, < \,\,\, z \,\,\, \lsim \,\,\, z_c^{(l-1)} 
\qquad\quad (l<0),
\eeq
there is a flat region
with a momentum value roughly equal to the
$z=-\infty$ limit:
\beq
k_{n+l/N}(z) \, \simeq \, n \, + \, \frac{l\,-\,1}{N}
\, \simeq \, n \, + \, \frac{l\,-\,1}{N\,+\,1}
\, = \, k_{n+l/N}(z=-\infty)
\qquad\quad (l<0).
\eeq
According to eq.(\ref{eq_minima}) 
(in which $l \mapsto l-1$),
in the coupling region above, the amplitude ratio
is much smaller than one, so there is not any 
resonant behavior;
\item
In the coupling window
\beq
\label{coupl_interv_reson_lneg}
\qquad\quad
z_c^{(l-1)} \,\,\, \lsim \,\,\, z 
\,\,\, \lsim \,\,\, z_c^{(l)}
\qquad\quad (l<0),
\eeq
the momentum level $k_{n+l/N}(z)$ rises roughly 
linearly with $z$, from
the value $k=n+(l-1)/N$ up to the next value $k=n+l/N$,
\beq
z: z_c^{(l-1)} \mapsto z_c^{(l)}
\qquad \Rightarrow \qquad
k_{n+l/N}(z): 
\,\, n \, + \, \frac{l\,-\,1}{N} \,\,\,\, \mapsto \,\,\,\,
n \, + \, \frac{l}{N}
\qquad\quad
(l-1 \mapsto l).
\eeq
To a first approximation,
we may write in this coupling interval:
\beq
\label{approx_beyond_pert}
k_{n+l/N}(z) \, \simeq \, 
n \, + \, \frac{l}{N} 
\, + \, n \left( z \, - \, z_c^{(l)} \right)
\, = \,
n \, + \, \frac{l \, - \, 1}{N} 
\, + \, n \left( z \, - \, z_c^{(l-1)} \right)
\qquad (l<0).
\eeq
At the midpoint of the coupling interval above,
\beq
\qquad\quad
z_m^{(l-1)} \, \equiv \, 
\frac{z_c^{(l-1)} \, + \, z_c^{(l)} }{2}
\, = \, \frac{l \, - \, 1/2}{n \, N}
\qquad\quad (l<0),
\eeq
according to eq.(\ref{approx_beyond_pert}), the momentum takes the value
\beq
k_{n+l/N}\left( z_m^{(l-1)} \right)
\, \simeq \, n \, + \, \frac{l \, - \, 1/2}{N}
\qquad (l<0),
\eeq
where, according to eq.(\ref{eq_loc_max}), the amplitude ratio approximately has
a (relative or secondary) maximum.
Therefore the non-resonant level $k_{n+l/N}(z)$,
$l<0$, exhibits
a relatively strong resonant behavior inside the coupling region
specified in eq.(\ref{coupl_interv_reson_lneg}).
By comparing with the resonant case,
we conclude that, in the region (\ref{coupl_interv_reson_lneg}),
{\it only} the non-resonant level $k_{n+l/N}(z)$
exhibits a resonance
behavior.
\item
Finally, for larger values of the coupling,
\beq
\qquad\quad
\label{coupl_large_non_res_lneg}
z_c^{(l)} \,\, \lsim \,\, z \,\, < \,\, + \, \infty
\qquad\quad (l<0),
\eeq
the momentum $k_{n+l/N}(z)$ has roughly a plateau at the
value
reached at the previous step:
\beq
\qquad\quad
k_{n+l/N}(z) \, \simeq \, 
n \, + \, \frac{l}{N}
\, \simeq \, n \, + \, \frac{l}{N \,  + \, 1}
\, = \, k_{n+l/N}(z=+\infty)
\qquad\quad (l<0). 
\eeq
Because of eq.(\ref{eq_minima}),
the amplitude ratio in the coupling region 
(\ref{coupl_large_non_res_lneg}) is close to zero, exhibiting
a non-resonant behavior.
We conclude that, in the
coupling region (\ref{coupl_large_non_res_lneg}),
neither the resonant level $k_n(z)$
nor the non-resonant level $k_{n+l/N}(z)$
do exhibit a resonance behavior.
\end{enumerate}
To summarize, 
quite similarly to the non-resonant levels
with $l>0$, also the
non-resonant levels with $l<0$ basically have:
\begin{enumerate}
\item
A plateau at $k = n+(l-1)/N$ for $z \lsim (l-1)/(nN)$,
with no resonant behavior;
\item
A linearly-rising behavior from $k=n+(l-1)/N$ 
up to $k=n+l/N$,
for $z$ going from $(l-1)/(nN)$ up to $l/(nN)$,
with a relatively strong resonant behavior;
\item
A plateau at $k=n+l/N$ for $z \gsim l/(nN)$,
with no resonant behavior.
\end{enumerate}
The range of the non-resonant level
with $l<0$ is:
\beq
\mathrm{Range}\left[k_{n+l/N}\right]
\, = \, n \, + \, \frac{l}{N+1} 
\, - \,
\left( n \, + \, \frac{l-1}{N+1}  \right) 
\, = \, \frac{1}{N\,+\,1} 
\, \simeq \, \frac{1}{N} 
\qquad\quad (l<0).
\eeq
Note that it is equal to the range
a non-resonant level with $l>0$.
Compared to the resonant level $k_n(z)$,
a non-resonant level $k_{n+l/N}(z)$
with $l < 0$
has its linear region
shifted to the left, 
\beq
0\mapsto l<0
\quad \Rightarrow \quad
z \, \in \, 
\left( - \, \frac{1}{nN}, \, + \, \frac{1}{nN} \right)
\,\,\, \mapsto \,\,\,
z \, \in \, \left( \frac{l\,-\,1}{nN}, \, \frac{l}{nN} \right),
\eeq
and reduced in size by a factor two
(the slope of the linearly-rising region with $z$
of the momentum levels is approximately the same for all
of them).

The computation of the positive and negative
ranges of a non-resonant level $k_{n+l/N}(z)$
with $l<0$ is similar to the one for $l>0$,
which we have presented at the end of the previous section,  
so we do not report it.
As expected, for $|l| \ll N$, the negative range, containing
the linear region, is much larger than the 
positive one.

Let us remark that
since, as we have seen, the so-called "non-resonant
levels" $k_{n+l/N}(z)$ actually do exhibit a relevant resonance behavior
in intermediate coupling regions,
they should be more properly called
"non-resonant levels around zero coupling"
or "resonant levels in intermediate
coupling regions". 
However, as in the previous cases, we will retain
the old terminology.


\subsubsection{General remarks}

In general, the levels $k=k(z)$
(both resonant and non-resonant)
have, in their flat regions,
\beq
\frac{dk}{dz} \, \approx \, 0,
\eeq
a very small amplitude ratio,
\beq
A_N(k) \, \ll \, 1
\qquad (\mathrm{flat\,\,regions}),
\eeq
and therefore they do not exhibit, 
as far as amplitudes are concerned, 
any resonant behavior.
On the contrary, the levels
in their linearly-rising regions,
\beq
\frac{dk}{dz} \, \approx \, 
\left( 1 \, + \, \frac{1}{N} \right) n 
\, \approx \, n,
\eeq
have large amplitude ratios, because the momenta
in this case are close to a 
(principal or secondary) maximum of $A_N(k)$:
\beq
A_N(k) \, \gg \, 1
\qquad (\mathrm{linearly-rising\,\,levels}).
\eeq
In concise form: All momentum levels have, 
in their respective linearly-rising regions, a resonant
behavior, while in the flat ones they do not.

In order to understand the resonant
behavior of the model "globally" in the coupling $z$,
let us begin considering a resonant level 
$k_n(z_{ini}) \approx n$
at an extremely small initial coupling, $0 \lsim z_{ini} \ll 1$,
and imagine to progressively increase the coupling.
When we cross the first critical coupling,
\beq
z_c^{(1)} \, = \, \frac{1}{n \, N},
\eeq
the resonant behavior of the system is transferred,
from the resonant level $k_n(z)$, up to the
first non-resonant level $k_{n+1/N}(z)$.
Let us remark that, in the "transition"
$k_n \mapsto k_{n+1/N}$  ($l=0\mapsto 1$),
the resonant behavior is softened,
because we go from the absolute maximum 
of the amplitude ratio 
\beq
\left( A_N \right)_{abs.\,max.} \, = \, N, 
\eeq
down to the
first relative maximum, which is roughly 
21\% of the former. 
By further increasing $z$, we hit 
at some moment the second
critical point  
\beq
z_c^{(2)} \, = \, \frac{2}{n \,N}.
\eeq
By crossing this point (from below),
the resonant behavior is transferred,
from the first non-resonant level 
$k_{n+1/N}(z)$, up to the second one
$k_{n+2/N}(z)$ ($l=1 \mapsto 2$).
Note that the second relative maximum is about 
13\% of the principal one.
Now, at this point of the discussion, the general
dynamical mechanism related to a resonance should be clear: 
By progressively increasing  the coupling $z$ from zero on,
the resonant behavior, which is initially taken up
by the resonant level $k_n(z)$, is transferred,
from the current non-resonant level $k_{n+l/N}$,
up to the next one $k_{n+(l+1)/N}$
and it is, at the same time, softened.
For negative couplings, $z<0$,
the mechanism  is similar.
By going from $z< 0$, $|z|\ll 1$, down 
to $z_c^{(-1)}=-1/(nN)$,
the resonant behavior is transferred
from $k_n(z)$ to $k_{n-1/N}(z)$.
By decreasing the coupling further down to
$z_c^{(-2)}=-2/(nN)$ and below,
the resonant behavior is transferred
from $k_{n-1/N}(z)$ down to $k_{n-2/N}(z)$,
and so on.

All these phenomena can be directly observed by
plotting the momentum levels $k=k_{n+l/N}(z)$
for a given value of the principal index $n$ 
and for different values of the sub-index $l$
around zero,
as functions of the coupling $z$. 
In fig.$\,$\ref{fig_PkTOT}
we have taken, as an example, $N=99$ and
we have plot the first resonant level $k_1(z)$
($n=1$),
together with a bunch of close non-resonant
levels close, $k_{1\pm 1/N}(z)$,
$k_{1\pm 2/N}(z)$, $k_{1\pm 3/N}(z)$, $\cdots$.

Let us note that the non-resonant levels $k=k_{n+l/N}(z)$
far from any resonant level, i.e. with large $l$,
\beq
l \, = \, \mathcal{O}(N),
\eeq
such as, let's say $l=\lfloor N/2  \rfloor$,
where $\lfloor \alpha \rfloor$ is the integer
part of $\alpha$,
do not exhibit a resonance behavior in any
coupling region. They never resonate.
These levels, as functions of $z$, 
do not present two approximately flat regions 
connected by a linear region, 
as we have seen in all the previous cases ($0 \le |l|\ll N$),
but a smooth and slow increase in a large coupling 
region (see fig.$\,$\ref{fig_PkTOT}).
These properties are in agreement with
simple physical intuition: momentum levels far
from any resonance (in momentum space)
should be very little affected by the latter.

Let us remark that, if we cut the plot in fig.$\,$\ref{fig_PkTOT}
with a vertical line,
of equation
\beq
z \, = \, \bar{z} \qquad (|\bar{z}| \ll 1),
\eeq
we find, in general,
that all the levels are quite flat
at their respective intersection points, with the exception
of a single level, which is linearly rising
at its intersection point.
We can summarize the physics above
by saying that, for any (small) fixed
value of the coupling $z$, there is one,
and only one, resonant-behaving level;
All the other levels are dormant.

We can also observe that
all the momentum levels $k_{n+l/N}(z)$,
for any given (and fixed) $n=1,2,3,\cdots$ and variable 
$l=0,\pm 1,\pm 2, \pm 3,\cdots$, restricted to
their respective resonant regions, "link together"
to form a segment of a straight line,
passing through the point $z=0$ and $k=n$,
of equation:
\beq
k_n^{res}(z) \, \simeq \, 
n \, + \, \left(1+\frac{1}{N}\right) n z.
\eeq
We can compare the above behavior with
a similar behavior in standard Winter model.
In the case of the latter,
if we expand in powers of $z$ the pole
in the complex $k$-plane below the real axis,
associated to the $n$-th resonance,
we find:
\beq
\label{eq_pole_stand_WM}
\tilde{k}_n^{res}(z) \, = \, n 
\, + \, n z \, + \, n z^2 \, - \, i \pi n^2 z^2
\, + \, \mathcal{O}\left(z^3\right).
\eeq
At first order in $z$, the slope of the 
finite-volume momentum level is:
\beq
\frac{dk_n^{res}}{dz} \, = \,
\left(1+\frac{1}{N}\right) n,
\eeq
while that of the infinite-volume level is
\beq
\frac{d\tilde{k}_n^{res}}{dz} \, = \, n. 
\eeq
In the limit $N \to \infty$, the slope
of the finite-volume momentum line converges to the 
infinite-volume one;
Furthermore, finite-volume corrections are of order $1/N$
(i.e. rather sizable).

In standard Winter model $(N=\infty)$, 
the properties
of a given resonance $n$ can be analyzed
by considering its corresponding complex pole 
$\tilde{k}_n^{res}(z)$
in the momentum plane, as a function of the coupling $z$.
In the corresponding finite-volume model,
the behavior of the resonant state $n$,
as function of the coupling $z$,
is studied instead by gluing together different
segments from many momentum levels $k_{n+l/N}(z)$,
with $l=0,\pm 1, \pm 2, \pm 3, \cdots$. 

When the coupling $z$ is very close to a critical one
(an exceptional or non-generic situation),
let's say
\beq
z \, \approx \, z_c^{(l)}
\qquad \mathrm{with} \,\,\, l \, \ge \, 1,
\eeq
the curves of the contiguous momentum levels 
$k_{n+(l-1)/N}(z)$ and $k_{n+l/N}(z)$,
are very close to each other, producing
an approximate degeneracy in the momentum
spectrum of the model.
Therefore the critical points $z_c^{(l)}$ constitute 
some sort of "transition points" of the model, 
where the resonant behavior
moves from one level to the next one.
As already remarked,
there points are naturally thought of as
non-generic or exceptional
values of the coupling.

An approximate resummation of the perturbative
corrections enhanced at large $N$ 
for $k_n(z)$ gives \cite{FVWMme}:
\beq
k_n(z)^{resum} \, \simeq \, n \,
\Big\{ 
1 \, + \, z \, + \, z^2
\big[ 
\pi n \cot(\pi n N z) \, + \, 1 
\big]
\Big\} .
\eeq
The function on the r.h.s. of the
above equation has simple poles
at all the critical couplings $z_c^{(l)}$,
$l \, = \, \pm \, 1, \, \pm \, 2, \, \pm \, 3, \, \cdots$,
which are its only singularities.
We may say that large-$N$ resummed perturbation
theory "signals" the transition
of the resonant behavior from one level 
to the next one,
by presenting singularities at all the
critical (transition) points.
Note also that the singularity 
of the above function for $z\to 0$
is an apparent one, and its power-series
expansion in $z$ up to first order, for example,  
gives eq.(\ref{eq_mom_reson_case})
--- as it should.
Furthermore, at the middle points
of the resonating intervals
of the non-resonant levels,
\beq
z_m^{(l)} \, \equiv \, \frac{l \, + \, 1/2}{n \, N},
\eeq
the cotangent function $\cot(\pi n N z)$ 
vanishes, so that the above
expression simplifies into:
\beq
\label{eq_reson_resum_mp}
k_n\left( z_m^{(l)} \right)^{resum} 
\, \simeq \,  
n \left[ 
1 \, + \, z_m^{(l)} \, + \, \left( z_m^{(l)} \right)^2
\right] .
\eeq
It is worth comparing the above equation
with eq.(\ref{eq_pole_stand_WM}):
The only difference is the imaginary
part of the second-order coefficient,
which in eq.(\ref{eq_reson_resum_mp}) is 
necessarily absent.


\subsection{Phase shift}

The phase shift of finite-volume
Winter model, as function of the particle
momentum $k$, reads \cite{FVWMme}:
\beq
\varphi_N(k) \, = \, - \, \pi (N+1) k 
\, \simeq \, - \, \pi N k
\qquad (\mathrm{mod} \,\, \pi).
\eeq
As well known, 
in the case of a resonance,
the above phase
quickly crosses the value $\pi/2$ (mod $\pi$),
as we cross the resonance momentum
(or energy).

Since the coupling $z$ is our control variable,
we expect a resonant behavior to occur
in our model when:
\beq
\left| \frac{d\varphi_N}{dz} \right|
\, = \,
\left| \frac{d\varphi_N}{dk} \right|
\frac{dk}{dz}
\, \simeq \, \pi N \, \frac{dk}{dz}
\, \gg \, 1
\qquad \left( \frac{dk}{dz} > 0 \right).
\eeq
Therefore, for $N \gg 1$, any momentum level $k$
in its linearly-rising region (if any), i.e. where
\beq
\frac{dk}{dz} \, \simeq \, n,
\eeq
gives rise to a resonant behavior,
as far as phase shift is concerned. 
On the contrary, any momentum level
in its (approximately) flat region/regions,
i.e. with
\beq
\frac{dk}{dz} \, \lsim \, \frac{1}{N},
\eeq
does not gives rise to a resonant
behavior, whatever (large) value of $N$
is assumed. 
 
To have a resonant behavior, 
the particle momentum $k$
also has to satisfy the equation
\beq
\label{eq_reson_fase_shift}
\qquad
k \, \simeq \, \frac{s \, + \, 1/2}{N} \, = \, 
n \, + \, \frac{l \, + \, 1/2}{N},
\qquad l \in \ZZ
\qquad
\mathrm{(phase-shift \,\, resonant \,\, momenta),}
\eeq
where $s$ is an integer; $n$ is the quotient and
$l$ is the remainder of the euclidean
division of $s$ by $N$ respectively,
$s=nN+l$. 
As already discussed, 
it is natural to assume for the index $l$
the range
\beq
- \, \frac{N}{2} \, < \, l \, \le \, + \, \frac{N}{2}.
\eeq

%
\begin{figure}[ht]
\begin{center}
\includegraphics[width=0.5\textwidth]{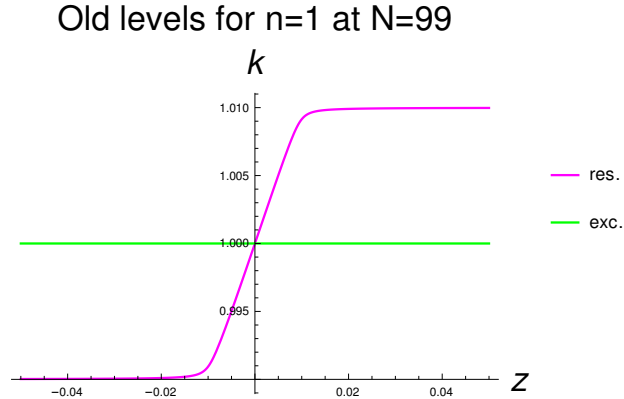}
\footnotesize
\caption{
\label{fig_Poldlev}
\it
First resonant momentum level $k_1(z)$ (magenta curve) and 
first exceptional level $k_{exc}\equiv 1$
(green horizontal line), in the case $N=99$.
These levels cross at the point $z=0$, $k=1$,
which is then, as expected, a singular point of the model.
}
\end{center}
\end{figure}
%


\subsubsection{Non-resonant case}

Let us first consider the simpler case
of the non-resonant levels $k_{n+l/N}(z)$
($l \ne 0$).
Eq.(\ref{eq_reson_fase_shift}) for $l \ne 0,-1$  exactly coincides
with eq.(\ref{eq_loc_max}),
the latter giving
the condition for having a local
maximum of the inside/outside 
amplitude ratio $A_N(k)$, namely:
\beq
A_N(k) \, \gg \, 1;
\qquad \frac{dA_N}{dk} \, = \, 0. 
\eeq
Therefore we  conclude that both signals
of resonance behavior simultaneously occur 
in all the non-resonant levels $k_{n+l/N}(z)$
with $l \ne -1$,
i.e. they occur for the same values of 
the momenta
\beq
k \, \simeq \, n \, + \, \frac{l \, + \, 1/2}{N},
\qquad l \ne 0,-1, 
\eeq
and therefore of the coupling
\beq
z \, \simeq \, z_m^{(l)} 
\, \equiv \, \frac{l \, + \, 1/2}{n \, N},
\qquad l \ne 0,-1.  
\eeq

%
\begin{figure}[ht]
\begin{center}
\includegraphics[width=0.5\textwidth]{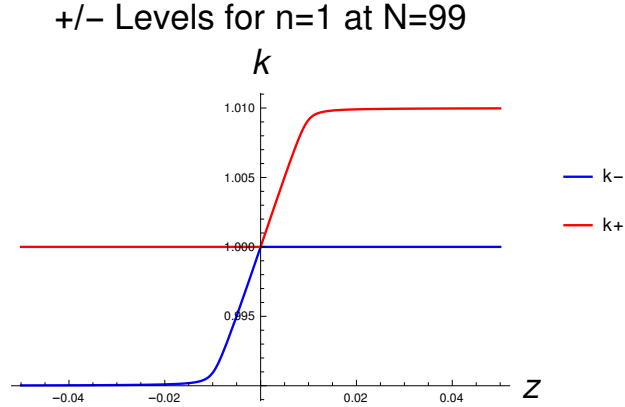}
\footnotesize
\caption{
\label{fig_Pnewlev}
\it
New $k_+(z)$ and $k_-(z)$ levels constructed
by gluing pieces of the first resonant and 
exceptional levels, as described in the main text,
for $N=99$ (compare with fig.\ref{fig_Poldlev}).
}
\end{center}
\end{figure}
%


\subsubsection{Resonant case}

As far as the resonant levels $k=k_n(z)$
are concerned, one has to 
explicitly consider their $z$ dependence.
By replacing their first-order
expansion, eq.(\ref{eq_mom_reson_case}),
inside eq.(\ref{eq_reson_fase_shift}),
one obtains for $l=0$ the relation:
\beq
z \, \simeq \, + \, \frac{1}{2nN}
\, = \, \frac{z_c^{(+1)}}{2},
\eeq
while,
 for $l=-1$, one obtains
\beq
z \, \simeq \, - \, \frac{1}{2nN}
\, = \, \frac{z_c^{(-1)}}{2}.
\eeq
Therefore,
by looking at the phase shift
$\varphi_N\left[k_n(z)\right]$,
we obtain two different  
values of $z$ (opposite to each other and 
very small in size),
for which a resonant behavior 
is expected to occur inside $k_n(z)$.
On the other hand, by looking
at the amplitude ratio $A_N\left[k_n(z)\right]$,
just one resonant behavior is expected
to occur in $k_n(z)$, inside the window
$\left( z_c^{(-1)}, \, z_c^{(+1)} \right)$,
namely at $z \approx 0$.

A possible solution to the above problem 
involves a careful re-consideration of the
properties of the resonant level $k_n(z)$.
As we know, the point $z=0$ is a singular point
of finite-volume Winter model, 
because the
resonant level $k_n(z) = n + \mathcal{O}(z)$
and the exceptional level $k_n^{exc}(z) \equiv n$ 
cross each other precisely at this point.
The same conclusion is reached
by looking at the differential equation for 
$dk/dz$,
which has a singularity at $z=0$:
\beq
\label{ODE1}
\frac{d k}{d z} \, = \, 
\frac{k}{
z \, \Big\{ z \, (\pi k)^2
\big[
\csc^2\left(\pi k\right) 
\, + \, 
N \, \csc^2\left(\pi N k\right) 
\big] 
\, - \, 1
\Big\}
};
\eeq
where
\beq
\csc(y) \, \equiv \, \frac{1}{\sin(y)} 
\eeq
is the standard cosecant of the angle $y$.
It is immediate to check that the above equation 
is non-singular in the (simpler) repulsive case, $z<0$.
Furthermore, the Hamiltonian
of finite-volume Winter model, eq.(\ref{H_Winter}),
has a simple pole at $z=0$.

All the facts above imply that $k_n(z)$,
for real $z$,
basically consists of two independent
branches, namely its restriction to $z>0$,
$k_n(z)|_{z>0}$,
and its restriction to $z<0$,
$k_n(z)|_{z<0}$.
By means of similar considerations, one
concludes that
also the restrictions of the exceptional level
$k_n^{exc}(z) \equiv n$ 
to $z>0$ and to $z<0$ are quite natural.
Let us also note that the ranges of 
the restrictions of the resonant level,
namely $k_n(z)|_{z>0}$ and of $k_n(z)|_{z<0}$, 
evaluated on the half-lines where
they are defined, are both equal to $1/(N+1)$.
They are equal to the ranges (evaluated on the entire real line) of all the non-resonant levels $k_{n+l/N}(z)$, $l \ne 0$.
As we have shown, the resonant eigenfunction
$\psi_{k_n(z)}(x)$ has a resonant
behavior for $z \mapsto 0$, as $A_N[k_n(z)]\mapsto N \gg 1$
in this limit.
The exceptional
eigenfunction $\psi_n^{exc}(x) \equiv \sqrt{2/L} \sin(n x)$,
$0 \le x \le L$, has instead
an amplitude ratio identically
equal to one for any value of $z$, 
so it has neither 
a resonant behavior nor an anti-resonant one.

Since, as we have just shown in different ways, 
the point $z=0$ is a singular point of the model,
we are allowed to construct
a new resonant level, let's call it
$k_{n+}(z)$, by gluing 
the exceptional level, restricted to $z \le 0$,
to the resonant level, restricted to $z>0$
(compare fig.$\,$\ref{fig_Poldlev} with 
fig.$\,$\ref{fig_Pnewlev}):
\beq
\label{eq_def_knplus}
k_{n+}(z) 
\, \equiv \,
\left\{
\begin{array}{cc}
n,  & \,\, z \, \le \, 0 ;
\\
k_n(z), & \,\, z \, > \, 0 ;
\end{array}
\right.
\qquad n = 1,2,3,\cdots.
\eeq
In a similar way, we construct the level
$k_{n-}(z)$ as:
\beq
\label{eq_def_knminus}
k_{n-}(z) 
\, \equiv \,
\left\{
\begin{array}{cc}
k_n(z),  & \,\, z \, \le \, 0 ;
\\
n, & \,\, z \, > \, 0 ;
\end{array}
\right.
\qquad n = 1,2,3,\cdots.
\eeq
The new level $k_{n+}(z)$ has a rising linear
behavior in $z$ (and therefore a resonant
behavior as far as amplitudes are concerned)
in the interval
\beq
0 \, < \, z \, \lsim \, z_c^{(1)}.
\eeq
The level $k_{n-}(z)$ instead has a rising linear
behavior in the interval
\beq
z_c^{(-1)} \, \lsim \, z \, < \, 0  .
\eeq
Let us now compare the range of
all the momentum levels.
The exceptional level $k_n^{exc}(z) \equiv n$
has trivially zero range (it is constant!) while,
as we have shown, the resonant level $k_n(z)$
has range $2/(N+1)$ and the non-resonant
levels $k_{n+l/N}(z)$, $l \ne 0$, have range $1/(N+1)$.
Therefore, the resonant level $k_n(z)$
and the exceptional level $k_n^{exc}(z) \equiv n$
have different ranges from each other,
as well as from the non-resonant levels.
On the contrary, the new levels $k_{n+}(z)$ and $k_{n-}(z)$
both have range $1/(N+1)$,
equal to the range of all the non-resonant levels.
Furthermore, the new levels $k_{n\pm}(z)$ have shapes, 
in the $(z,k)$ plane,
which are qualitatively more similar
to the shapes of the non-resonant levels $k_{n+l/N}(z)$,
$l\ne 0$, with respect to the old levels $k_n(z)$
and $k_n^{exc}(z) \equiv n$.
Because of the increased regularity (or symmetry),
the new constructed levels seem therefore
to be more natural than the old ones.

Let us now study the properties of the new levels.
The level $k_{n-}(z)$ has an approximate
degeneracy at the point $z=z_c^{(-1)}$
with the lower non-resonant level $k_{n-1/N}(z)$
($l=-1$),
while it has an exact degeneracy at
the origin, $z=0$, with the level $k_{n+}(z)$ as:
\beq
k_{n+}(z=0) \, = \, k_{n-}(z=0) \, = \, n.
\eeq
The minus level $k_{n-}(z)$ has a resonant
behavior for $z<0$, $|z|\ll 1$, let's
say inside the region
\beq
\label{eq_reson_minus}
z_c^{(-1)} \,\, \lsim \, \, z \, \, < \,\, 0.
\eeq
The level $k_{n+}(z)$, in addition to the
exact degeneracy at $z=0$ with the level $k_{n-}(z)$
just discussed, has an approximate
degeneracy at $z=z_c^{(1)}$ with the next level
$k_{n+1/N}(z)$ ($l=+1$).
The plus level has a resonant
behavior for $0 < z \ll 1$, more specifically
in the region
\beq
\label{eq_reson_plus}
0 \,\, < \,\, z \,\, \lsim \,\, z_c^{(+1)}.
\eeq
As a consequence of the construction
of the new levels,
we may consider also the point $z=0$
as a critical point, with zero index:
\beq
z_c^{(0)} \, \equiv \, 0.
\eeq
With the above definition, it  turns out
that the critical couplings form
a regular sequence of points 
in the $z$-axis,
with the constant spacing $1/(nN)$
(a simple one-dimensional lattice):
\beq
z_c^{(l)} \, \equiv \, \frac{l}{n \, N};
\qquad l \, = \, 0, \, \pm \, 1, \, \pm \, 2, \, \pm \, 3, \, \cdots.
\eeq
Let us remark that,
unlike the old levels $k_n(z)$ and 
$k_n^{exc}(z)$, which are smooth at the
origin, the new levels $k_{n\pm}(z)$
have a cusp at this point.
That also implies that the amplitude ratio
of the level $k_{n-}(z)$ discontinuously jumps
from $N$ to one as we cross the point $z=0$
from below ($z<0\mapsto z>0$).
Similarly, the amplitude ratio
of the level $k_{n+}(z)$ discontinuously jumps
from $1$ to $N$, again as we cross the origin
from below.
These losses of regularity are however
acceptable, because $z=0$, as we have
discussed, is a singular point, 
where a smooth behavior is not to be expected.
Once we have "accepted" the new plus and minus levels,
we may think that the coupling value
\beq 
z_+ \, \equiv \, \frac{z^{(+1)}_c}{2}, 
\eeq 
where the phase shift quickly crosses
the value $\pi/2$ (modulo $\pi$),
is related to the resonant behavior
of the plus resonant state 
$k_{n+}(z)$.
Indeed, the coupling $z_+$
is right at the middle of the interval (\ref{eq_reson_plus})
where, as we have seen, the plus state
exhibits a resonant behavior as
far as amplitudes are concerned.
The coupling value
\beq
z_- \, \equiv \, \frac{z_c^{(-1)}}{2}
\, = \, - \, z_+, 
\eeq
having similar properties to $z_+$,
is at the middle of the interval (\ref{eq_reson_minus}),
where the minus state
exhibits a resonant behavior as
far as amplitudes are concerned.

We may conclude the present discussion
by saying that, by introducing
the plus and minus states
in place of the resonance
and exceptional states, 
it is possible to completely
reconcile the resonance
properties coming from the
amplitude analysis with the resonance
properties coming from
phase analysis.


\section{The Symmetric Case $N=1$}
\label{sec_symm_case}

\begin{figure}[ht]
\begin{center}
\includegraphics[width=0.5\textwidth]{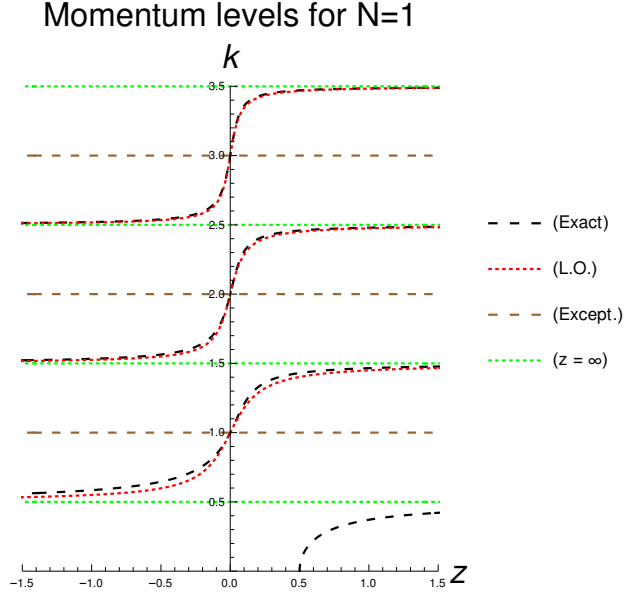}
\footnotesize
\caption{
\label{fig_plotNeq1}
\it
Lowest six momentum levels $k$ (excluding the 
level related to the bound state)  
as functions of the coupling $z\equiv-g$, 
namely $k=k(z)$, for the case $N=1$.
The three lowest resonant levels $k_1(z)$, $k_2(z)$
and $k_3(z)$ are represented by black dashed lines;
The three lowest exceptional levels $k_{exc} \equiv 1,2,3$
are represented by brown dashed lines.
The Leading Order (L.O.) resonant levels are given by 
red dotted lines.
The accuracy of the approximation
quickly increases with the order of the level $n=1,2,3$, 
uniformly in $z \in \RR$; Already at $n=3$, the difference
between the exact level and the L.O. approximation
is barely visible.
The infinite-coupling limits $(z \to \pm \infty)$ of the resonant levels are
represented by horizontal green dotted lines.
The level below the lowest horizontal line, 
i.e. the level with  $k<1/2$, is related to the bound state 
of the model (the latter existing only for $0<z<1/2$) and  cannot be described by
our high-energy expansion.
}
\end{center}
\end{figure}

In this section we consider Winter model
at finite volume in the case $N=1$,
in which the two resonant cavities have the
same length $l=L/2=\pi$ \cite{flugge2}. 
Because of that, the two cavities
have the same resonant momenta.
In other words, there are no non-resonant
momenta in this case, but only resonant
and exceptional momenta 
(see fig.$\,$\ref{fig_plotNeq1}).

Eq.(\ref{eq_basic}) for the normal
momentum spectrum of the model simplifies, in the
case $N=1$, to the transcendental equation
\beq
\qquad\qquad
\label{eq_Neq1_momenta}
\tan(w) \, = \, \zeta \, w
\qquad\qquad (N=1);
\eeq
where we have defined the quantities:
\bea
\zeta &\equiv& 
- \, \left( 1 \, + \, \frac{1}{N} \right) \, g  
\, = \, - \, 2 \, g
\qquad (N=1);
\nonumber\\ 
w &\equiv& + \, \pi \, k.
\eea
The factor $-2$ in front of the coupling $g$,
as well as the factor $\pi$ in front of $k$,
are purely conventional;
They are inserted just to simplify the forthcoming
formulae.
%
%


\subsection{Resummed perturbation theory for high-energy states}
\label{sec_resum_method_one}

The ordinary perturbative expansion for 
the $\pi$-re-scaled momenta $w$ of the particle
reads:
\beq
\label{w_Neq_1_HE}
w \, = \, \nu \, + \, \nu \, \zeta \, + \, \nu \, \zeta^2
\, + \, \left( - \, \frac{\nu^3}{3}  \, + \, \nu \right) \zeta^3
\, + \, \mathcal{O}\left(\zeta^4\right),
\eeq
where, to simplify the formula, we have defined
the re-scaled index
\beq
\nu \, \equiv \, \pi \, n;
\qquad
n = 1, 2, 3, \cdots.
\eeq
Let us consider eigenfunctions of 
the Hamiltonian operator $\hat{H}$
of the model (eq.(\ref{H_Winter}) with $L=2\pi$) 
with a high energy 
$E = k^2 = (w/\pi)^2 \gg 1$, i.e.
let us restrict ourselves to the case
\beq
\nu \, \gg \, 1.
\eeq
The third-order coefficient (i.e. the term multiplied by $\zeta^3$) on the r.h.s. of eq.(\ref{w_Neq_1_HE}) contains
the term $- \nu^3/3$, proportional to $\nu^3$,
as well as the term $\nu$, down by two powers of $\nu$.
In the above, large-$\nu$ region,
this coefficient is clearly dominated by the first term,
so that we may write, to a first approximation:
\beq
\left( - \, \frac{\nu^3}{3}  \, + \, \nu \right) \zeta^3
\, \approx \, - \, \frac{1}{3} \, 
\left( \nu \, \zeta \right)^3.
\eeq
In general, we find that the leading terms
in the perturbative expansion of the particle
momentum $w=w_\nu(\zeta)$ are,
for $\nu \gg 1$, of the form:
\beq
\left(\nu \, \zeta \right)^h; \qquad h = 1,2,3,\cdots.
\eeq
These are the terms which, for any given power
$\zeta^h$ of the coupling $\zeta$,
contain the leading power of $\nu$, namely $\nu^h$.
Because of the occurrence of the above "secular
terms" at high energy, we expect ordinary perturbation theory
to be convergent only for:
\beq
\left| \nu \, \zeta \right| \, \lsim \, 1,
\eeq
i.e. only for:
\beq
\label{PT_strong_restr}
\left| \zeta \right| \, \lsim \, \frac{1}{\nu}.
\eeq
Even in the (lucky) case in which the perturbative series
has a non-zero radius of convergence 
$R = R_\nu > 0$,
the latter is expected to vanish,
according to the above relation, as $1/\nu$
for $\nu\to\infty$.
Therefore $R_\nu$ is 
expected to be very small for large $\nu$,
i.e. for high-energy states.

The crucial point is that relation (\ref{PT_strong_restr}) implies a very strong limitation on the range of
ordinary perturbation theory --- an additional and
stronger limitation with respect to the usual 
weak-coupling condition
\beq
\left| \zeta \right| \, \ll \, 1.
\eeq
If we take for example 
$\nu = \mathcal{O}\left(10^4\right)$,
we have to assume that:
\beq
\left| \zeta \right| \, \lsim \, 10^{-4},
\eeq
while we would like to use perturbation
theory for couplings much smaller than
one, but much larger than the above ones,
such as for example:
\beq
\zeta \, \approx \, 10^{-2},
\eeq
or even:
\beq
\zeta \, \approx \, 10^{-1}.
\eeq
The conclusion of this analysis is simply
that ordinary perturbation theory is not the 
right approximation scheme for studying high-energy states
of the model.
Our aim is to construct a new perturbative scheme, 
which still
assumes weak coupling (which is, by the way, 
the only coupling domain relevant to resonances), 
but in which the quantity $\nu \, \zeta$
is not restricted anymore to be smaller than one.
In other words, we are interested to the dynamical situation
in which:
\beq
\left| \zeta \right| \, \ll \, 1, \qquad \nu \, \gg \, 1 , 
\qquad \left| \nu \, \zeta \right| \, \gsim \, 1.
\eeq
Formally, that means to consider the 
combined/correlated limit
\beq
\label{new_scheme_pert}
\zeta \, \to \, 0, 
\qquad \nu \, \to \, \infty, 
\qquad \omega \, \equiv \, \nu \,\zeta \, \to \, \mathrm{constant},
\eeq
where the constant is assumed to be different from zero 
(and infinity).
A similar limit has been considered in 
ref.\cite{Aglietti:2016zzm} to compute the poles
of standard resonances in various $\delta$-shell models.

We have already said that the maximal power of $\nu$
in the coefficient of $\zeta^h$ is just $h$,
i.e. the maximal power of the index $\nu$ is equal to
the current power of the coupling $\zeta$.
In general, the coefficient $C_h=C_h(\nu)$ of each power $\zeta^h$ 
of $\zeta$ is a polynomial in $\nu$ of degree $\le h$:
\beq
\qquad
C_h(\nu) \, \zeta^h \, = \,
\Big(
c_{h,h} \, \nu^h \, + \, 
c_{h,h-1} \, \nu^{h-1} \, + \,
\cdots \, + \, c_{h,1} \, \nu \, + \, c_{h,0}
\Big) \zeta^h;
\qquad h \, = \, 1,2,3,\cdots;
\eeq
where the $c_{h,j}$'s are real coefficients
independent of $\nu$ and assumed to be of order one
(some of them may actually vanish).
In the high-energy scheme we are
constructing,
the above term is approximated, in Leading Order (L.O.), 
as we have seen, by the leading power of $\nu$:
\beq
C_h(\nu) \, \zeta^h \, \approx \,
c_{h,h} \, \nu^h \, \zeta^h
\, = \, c_{h,h} \, \omega^h;
\eeq
where, on the last member, we have introduced the new variable $\omega$, defined in eq.(\ref{new_scheme_pert}).
Therefore, the expansion of the particle momentum $w=w_\nu$
reads at the L.O. of the new perturbative scheme (\ref{new_scheme_pert}):
\beq
w_\nu \, \simeq \, \nu \, + \, 
\sum_{h=1}^\infty c_{h,h} \, \omega^h
\qquad (\mathrm{L.O.\,\,in\,\,the\,\,new\,\,scheme}).
\eeq
Note that the variable $\omega$
plays the role of an effective, state-dependent 
coupling of the model.
Let us also remark that, since the quantity 
$\omega$ is not assumed to be small,
we are not allowed to truncate 
the above series to any finite order
in $\omega$: Terms of all orders
in $\omega$ have to be consistently included
in the new perturbative scheme.

Let us observe that, in the new scheme,
the expansion of $w_\nu$ at L.O. involves an 
{\it infinite series} 
of terms of the form $c_{h,h} \, \omega^h$
while, in ordinary perturbation theory,
the lowest-order approximation involves the term
proportional to $\zeta$ only, namely 
\beq
\qquad\qquad\qquad
w_\nu \, = \, \nu \, + \, \nu \, \zeta 
\, + \, \mathcal{O}\left( \zeta^2 \right)
\qquad 
\mathrm{
(L.O.\,\,in\,\,ordinary\,\,perturbation\,\,theory).
}
\eeq
By expressing for example the index $\nu$ 
in terms of $\omega$ and $\zeta$,
\beq
\nu \,\, \mapsto \,\, \frac{\omega}{\zeta},
\eeq
the term $C_h \, \zeta^h$, occurring  
in the expansion of $w_\nu$ at order $h$ in $\zeta$, 
is rewritten:
\beq
C_h \, \zeta^h \, = \,
c_{h,h} \, \omega^h \, + \, 
c_{h,h-1} \, \omega^{h-1} \, \zeta \, + \,
c_{h,h-2} \, \omega^{h-2} \, \zeta^2 \, + \,
\cdots \, + \, c_{h,1} \, \omega \, \zeta^{h-1} 
\, + \, c_{h,0} \, \zeta^h.
\eeq
Since by assumption $\left| \zeta \right| \ll 1$,
at the Next-to-Leading Order (N.L.O.),
we include, in the perturbative
expansion of $w_\nu$, all the terms in which
{\it one} power of $\omega$ is replaced by {\it one} power
of $\zeta$, i.e. we include all the terms of the form
\beq
\zeta \, c_{h,h-1} \, \omega^{h-1};
\qquad h = 1,2,3,\cdots.
\eeq
Note that, in the new scheme, 
an eventual term proportional to $\zeta$
with a coefficient independent of $\nu$,
is a N.L.O. one.

\noindent
The new expansion of the particle momentum $w_\nu$
can be written, at N.L.O., in the form:
\beq
w_\nu \, \simeq \, \nu \, + \, 
\varphi_1(\omega) \, + \, \zeta \, \varphi_2(\omega)
\qquad (\mathrm{N.L.O.});
\eeq
where we have defined the above functions of $\omega$
as the following power series in $\omega$:
\beq
\varphi_1(\omega) \, \equiv \, 
\sum_{h=1}^\infty c_{h,h} \, \omega^h
\eeq
and
\beq
\varphi_2(\omega) \, \equiv \, 
\sum_{h=1}^\infty c_{h,h-1} \, \omega^{h-1}.
\eeq
Now, at this point of the discussion,
it should be clear to the reader how to 
construct the next-order approximation 
in the new scheme;
It  should also be clear how the
general scheme works. 
The momentum $w=w_\nu$ of the particle in a high-energy 
state, i.e. with the index $\nu \gg 1$,
is written as the formal sum of
a function series of the form:
\bea
\label{w_new_scheme_exact}
w_\nu &=& \nu \, + \, 
\sum_{h=0}^\infty \zeta^h \, \varphi_{h+1}(\omega) \, =
\nonumber\\
&=& \nu \, + \, \varphi_1(\omega) \, + \, \zeta \, \varphi_2(\omega) \, + \, \zeta^2 \, \varphi_3(\omega)
\, + \, \zeta^3 \, \varphi_4(\omega)
\, + \, \cdots. 
\eea
From a mathematical point of view,
the construction of the new perturbative scheme 
is legitimate, only if we can rearrange terms
in the double series in $\zeta$ and $\omega$
representing the particle momentum:
\beq
w_\nu \, = \, \nu  \, + \, \sum_{h=1}^\infty 
\left( \sum_{j=0}^h c_{h,j} \, \nu^j \right) \zeta^h
\, = \, \nu  \, + \, 
\sum_{h'=0}^\infty 
\left( \sum_{j=0}^\infty c_{h'+j,j} \, \omega^j \right)
\zeta^{h'}.
\eeq
In the last equality we have changed index
according to $h \mapsto h' \equiv h-j$.
Note that the internal sum (over $j$) at the last member 
of the above equation is infinite, 
while the internal sum at the second member is finite.
As well-known in mathematics, a sufficient condition 
for terms rearrangement --- the so-called unconditional
convergence  --- is that
the double series is absolutely convergent,
a property therefore which we assume from now on.

Let us remark that, in the new scheme, 
even though we are resumming
terms of all orders in $\zeta$ even at the L.O.,
the all-order resummation of the perturbative series
which we have realized, is an approximate one,
as the coefficients $C_h(\nu)$ are not evaluated
exactly.
That is because, in practice, we are always forced to truncate the function
series above to some finite order $P$ (possibly large):
\beq
\label{eq_funz_ser_sch_P}
\qquad\qquad
w_\nu \, \cong \, \nu  \, + \, 
\sum_{h=0}^P \zeta^h \sum_{j=0}^\infty c_{h+j,j} \, \omega^j;
\qquad 0 \, \le \, P \, < \, \infty. 
\eeq 
It is also important to stress that the 
resummation scheme which we have formally constructed
is relevant --- i.e. is useful in "real life" --- 
if, and only if, we are able to compute the 
series of the coefficients $c_{h+j,j}$
in closed analytic form for any value of the index 
$j=0,1,2,3,\cdots$, i.e. in symbolic form in $j$,
for some values of $h=0,1,2,\cdots$.
If we do not possess such a knowledge,
we are forced to truncate the series in $\omega$, defining
the functions $\varphi_i(\omega)$,
at some finite order in $\omega$ --- an approximation which
is not legitimate in the new scheme,
as already noted.

To explicitly compute the functions $\varphi_i(\omega)$'s,
we have to go back, from the ordinary (truncated) perturbative expansion of the momenta,  to the exact equation determining the momentum spectrum of the $N=1$ model.
We simply insert the function-series expansion
given in eq.(\ref{w_new_scheme_exact}) inside the $N=1$
momentum spectrum equation (\ref{eq_Neq1_momenta}), 
conveniently rewritten as:
\beq
\Delta w_\nu \, \equiv \, w_\nu \, - \, \nu 
\, = \, \arctan\left( \zeta \, w_\nu \right).
\eeq
The master equation is obtained:
\bea
&& \varphi_1(\omega) \, + \, \zeta \, \varphi_2(\omega)
\, + \, \zeta^2 \, \varphi_3(\omega) 
\, + \, \zeta^3 \, \varphi_4(\omega)
\, + \, \zeta^4 \, \varphi_5(\omega)
\, + \, \cdots \, = \,
\nonumber\\
&=&
\arctan\Big[
\omega \, + \, \zeta \, \varphi_1(\omega) 
\, + \, \zeta^2 \, \varphi_2(\omega)
\, + \, \zeta^3 \, \varphi_3(\omega)
\, + \, \zeta^4 \, \varphi_4(\omega)
\, + \, \cdots
\Big] \, =
\nonumber\\
&=&
\arctan(\omega) \, + \, \zeta \,
\frac{1}{1 \, + \, \omega^2} \,
\Big[
\varphi_1(\omega) \, + \, \zeta \, \varphi_2(\omega)
\, + \, \zeta^2 \, \varphi_3(\omega)
\, + \, \cdots
\Big] \, +
\nonumber\\
&&
\qquad\qquad\,\,\,
\, - \, \zeta^2 \, \frac{\omega}{(1+\omega^2)^2} \,
\Big[
\varphi_1(\omega) \, + \, \zeta \, \varphi_2(\omega)
\, + \, \zeta^2 \, \varphi_3(\omega)
\, + \, \cdots
\Big]^2
\, + \, \cdots
\nonumber.
\eea
In order to isolate
the different orders in $\zeta$
in the above equation,
in the last member 
we have made a Taylor
expansion of the function
$\arctan(\omega+\Delta\omega)$
around the point $\omega$.
We expand then the last member of the above 
equation in powers of $\zeta$, up to the required 
order.
By equating the coefficients of same powers
of $\zeta$ at the first and the last member, 
we obtain a sequence of equations of the form:
\bea
\label{syst_eqs_ini}
\varphi_1(\omega) &=& \arctan(\omega) ;
\nonumber\\
\varphi_2(\omega) &=& \frac{\varphi_1(\omega)}{
1 \, + \, \omega^2} ;
\nonumber\\
\varphi_3(\omega) &=&
\frac{1}{1 \, + \, \omega^2}
\left[
\varphi_2(\omega)
\, - \, \frac{\omega}{1 \, + \, \omega^2} 
\, \varphi_1^2(\omega)
\right];
\nonumber\\
\cdots &\cdots& \cdots.
\eea
The first equation directly gives the L.O.
function:
\beq
\label{eq_varphi1}
\varphi_1(\omega) \, = \, \arctan(\omega).
\eeq
By substituting the above equation (in principle) in
all the equations of the system (\ref{syst_eqs_ini}) and solving the second equation with
respect to the N.L.O. function $\varphi_2(\omega)$, one obtains for the latter:
\beq
\varphi_2(\omega) \, = \, 
\frac{\arctan(\omega)}{1 \, + \, \omega^2}.
\eeq
In a similar way, by substituting the above equation
in the system and solving the third equation
with respect to the Next-to-Next-to-Leading Order 
(N.N.L.O. or N$^2$LO for brevity) function 
$\varphi_3(\omega)$, one obtains for the latter:
\beq
\label{eq_Neq1_varphi3}
\varphi_3(\omega) \, = \, 
\frac{\arctan(\omega)}{(1 \, + \, \omega^2)^2}
\, \Big[
1 \, - \, \omega \, \arctan(\omega) 
\Big] .
\eeq 
At this point, it should be clear to the reader
how to explicitly compute higher-order functions 
$\varphi_i(\omega)$'s, to any required order,
as well as the general structure of the scheme.
By means of a symbolic manipulation program
(and a computer, of course!), 
the explicit evaluation of large number of functions 
$\varphi_i(\omega)$'s 
is trivial and immediate.
Actually, with enough computing resources, one can evaluate
an arbitrarily large number of such functions. 
By computing the first twenty functions
by means of the Mathematica system on a standard PC, 
we obtained analytic approximations
of the particle momentum $w_\nu$ with
a relative error of order $\mathcal{O}\left(10^{-15}\right)$.


\subsection{Comparison of resummed perturbation theory with ordinary one}

In this section we compare our resummed
formulae with ordinary perturbation theory
ones;
That provides a consistency check of the resummation 
scheme which we have constructed in the previous section.

According to eqs.(\ref{coef_ris_straight})
with $N=1$,
the momentum $w=w_\nu$ of the particle has the
following ordinary perturbative expansion:
\beq
\label{eq_PT_ord5}
w_\nu \, = \, \nu \, + \, \nu \, \zeta \, + \, \nu \, \zeta^2
\, + \, \left( - \, \frac{\nu^3}{3}  \, + \, \nu \right) \zeta^3
\, + \,  
\left( - \, \frac{4}{3} \nu^3 \, + \, \nu \right) \zeta^4
\, + \,  
\left(
\frac{1}{5} \nu^5- \frac{10}{3} \nu^3 + \nu
\right) \zeta^5
\, + \, \mathcal{O}\left(\zeta^6\right).
\eeq
By means of a Taylor expansion 
of the arctangent function at the origin,
\beq
\arctan(x) \, = \, x \, - \, \frac{x^3}{3} 
\, + \, \frac{x^5}{5}
\, + \, \mathcal{O}\left(x^7\right)
\quad\qquad (|x|<1),
\eeq
the expansion of the L.O. resummed formula for $w_\nu$
in powers of $\zeta$ (eq.(\ref{eq_funz_ser_sch_P}) with $P=0$), up to fifth order included,
reads:
\beq
\label{recursion1}
w_\nu^{LO}
\, = \, 
\nu \, + \, \varphi_1(\omega) \, = \, 
\nu \, + \, \arctan\left( \nu \, \zeta \right)
\, = \, \nu \, + \, \nu \, \zeta 
\, - \, \frac{1}{3} \nu^3 \zeta^3
\, + \, \frac{1}{5} \nu^5 \zeta^5
\, + \, \mathcal{O}\left(\zeta^7\right).
\eeq
By comparing the last member of the above equation
with the r.h.s. of eq.(\ref{eq_PT_ord5}), we find
that:
\begin{enumerate}
\item
The first-order term in $\zeta$, namely $\nu \, \zeta$, is correctly reproduced (as it should)
by the L.O. resummed formula
(let us remark that any good resummation scheme
should reproduce, at least, the lowest-order correction); 
\item
The second-order term, $\nu \, \zeta^2$, is 
totally missing in the resummed formula 
(the last member of eq.(\ref{recursion1})).
However, that is consistent, because this term is a N.L.O.
correction,
having an additional power of $\zeta$
with respect to $\nu$.
Formally, the absence of this term
may be seen as a trivial consequence of the fact 
that $\arctan(\nu \zeta)$ is an odd function of $\zeta$,
so that only odd powers of $\zeta$ enter
the expansion of the L.O. resummed formula;
\item
Only the contribution $-\,\nu^3 \zeta^3/3$
of the third-order term is reproduced,
while the remaining contribution, $\nu \, \zeta^3$, 
is missing.  
That is (again) consistent, because the latter
term is a N.N.L.O. correction, having two additional
powers of $\zeta$ with respect to $\nu$;
\item
The fourth-order term is completely missing in the
L.O. resummed formula for the same reason
of the second-order term;
\item
Only the contribution $\nu^5 \, \zeta^5/5$
to the fifth-order term
is reproduced by the L.O. resummed formula. 
The terms $-10 \, \nu^3 \, \zeta^5/3$ 
and $\nu \, \zeta^5$ are missing,
as they are N$^2$LO and N$^4$LO 
corrections respectively.
\end{enumerate}
We may conclude that,
up to fifth order in $\zeta$ included,
our L.O. resummed formula (\ref{recursion1}) 
contains {\it all} the L.O. terms, 
i.e. all the terms of the form 
$\nu^j \, \zeta^j$, $j=1,2,3,\cdots$. 

To convince ourselves that our resummed scheme
also works at higher orders,
let us explicitly consider also the N.L.O.
resummed formula.
The expansion in powers of $\zeta$ of the latter
reads:
\bea
\label{momenta_NLO}
w_\nu^{NLO} &=&
\nu \, + \, \arctan\left( \nu \, \zeta \right)
\, + \, \zeta \, \frac{\arctan(\nu \, \zeta)}{
1 \, + \, (\nu \, \zeta)^2}
\,\, = 
\nonumber\\
&=& \nu \, + \, \nu \, \zeta \, + \, \nu \, \zeta^2
\, - \, \frac{1}{3} \, \nu^3 \zeta^3
\, - \, \frac{4}{3} \, \nu^3 \zeta^4
\, + \, \frac{1}{5} \, \nu^5 \zeta^5
\, + \, \mathcal{O}\left(\zeta^6\right).
\eea
By comparing the expansion in powers of $\zeta$
of the N.L.O. resummed formula 
(the last line above)
with the ordinary perturbative
formula, the r.h.s. of eq.(\ref{eq_PT_ord5}), we find that:
\begin{enumerate}
\item
The first-order term in $\zeta$, 
namely $\nu \, \zeta$,
is correctly reproduced in N.L.O. resummation,
as was already the case with L.O. resummation;
\item
The second-order term, $\nu \, \zeta^2$,
is also correctly reproduced in N.L.O. resummation,
as it should, such term being a N.L.O. correction.
Note that this term was instead completely missing 
in the L.O. resummation.
We see therefore a first improvement
by going from L.O. to N.L.O. resummation;
\item
Only the contribution $-\nu^3 \zeta^3/3$
of the third-order term is reproduced,
while the remaining contribution, $\nu \, \zeta^3$, 
is missing, as it is a N.N.L.O. term.
Therefore there is no improvement, at this order
in $\zeta$, of the N.L.O. resummation
with respect to the L.O. one;  
\item
At fourth order, only the N.L.O. term
$-4/3 \, \nu^3 \, \zeta^4$ is reproduced,
while the N$^3$LO term $\nu\,\zeta^4$ 
is missing.
Note that the fourth-order term,
as well as all the even-order terms, 
was instead completely missing
in L.O. resummation;
\item
At fifth order, only the L.O. term
$\nu^5 \, \zeta^5/5$ is reproduced,
while the N$^2$LO term $- 10 \, \nu^3 \zeta^5 /3$ 
and the N$^4$LO term $\nu\, \zeta^5$
are missing.
Therefore, at $\mathcal{O}\left(\zeta^5\right)$ 
there is no improvement 
with respect to LO resummation.
\end{enumerate}
The above comparison of perturbative corrections
for the particle momenta
explicitly shows that the resummation scheme 
we have constructed, also works at N.L.O..
Even though we do not provide a formal proof,
it is quite natural to conjecture
that this scheme also works at higher orders.
%
\begin{figure}[ht]
\begin{center}
\includegraphics[width=0.5\textwidth]{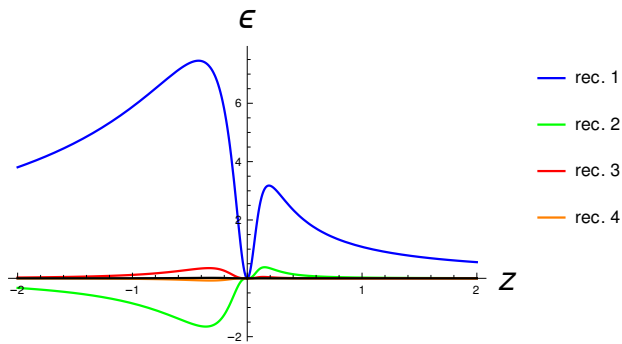}
\footnotesize
\caption{
\label{fig_plotNeq1b}
\it
Percent relative error 
$\epsilon^{(i)} \equiv 
100 \times \big( k_1^{ex} - k_1^{(i)} \big)/k_1^{ex}$
for the first resonance $(n=1)$ in the case $N=1$,
as function of the coupling $z \equiv -g$.
More explicitly, $k_1^{ex}=k_1^{ex}(z)$ is the exact 
first momentum level,
while $k_1^{(i)}=k_1^{(i)}(z)$ is the
recursive approximation to $k_1^{ex}(z)$
of order $i=1,2,3,4$.
The error $\epsilon^{(1)}=\epsilon^{(1)}(z)$ 
of the first recursion
(blue line), providing the lowest-order non-trivial approximation,
is uniformly below $8\%$ in the coupling 
$z \in \RR$.
The errors of the second recursion (green line),
third recursion (red line) 
and fourth recursion (barely visible orange line)
are uniformly below $2\%$, $0.35\%$ and $0.08\%$ 
respectively.
The error of the fifth recursion
(not drawn in the plot) is uniformly below $0.02\%$.
Note that we are considering the fundamental
resonance, i.e. the "worst" situation;
The errors in the calculation of higher momenta
$(n>1)$ turn out to be, as expected, much smaller.
}
\end{center}
\end{figure}
%


\subsection{Resummation of the perturbative series by recursion}

In this section we present a different resummation
scheme for the perturbative series
of particle momenta, which is 
based on a sequence of functions defined
recursively.
As we are going to show,
the two resummation schemes give
the same results in Leading Order (L.O.) 
and begin to differ from Next-to-Leading Order 
(N.L.O.) on.

The idea behind the representation of 
the particle momentum $w=w_\nu(\zeta)$ which we are going to derive, is the following.
Suppose that the variable $w$ on the r.h.s of the momentum equation (\ref{eq_Neq1_momenta}),
namely
\beq
\label{eq_w_Neq1_repeat}
\tan(w) \, = \, \zeta \, w,
\eeq
is replaced by some constant, let's call it $w_0$
(which is non zero but is, for the moment, arbitrary). 
The above equation then simplifies to
the equation
\beq
\tan(w) \, = \, \zeta \, w_0, 
\eeq
which can be immediately solved to give:
\beq
w \, = \, \arctan\left( w_0 \, \zeta \right).
\eeq
The function $\arctan(x)$ is the branch of the  
multi-valued arctangent function,
having image, for a real argument $x$, 
in the interval $(-\pi/2,+\pi/2)$:%
\footnote{
As function of a complex argument
(complex function), $\arctan(w)$
is a multi-valued analytic function,
with logarithmic branch points at 
$w = \pm i$. 
To make $\arctan(w)$ single-valued,
i.e. to obtain well-defined
analytic branches,
we can cut the complex $w$ plane, 
for example, along the imaginary axis, 
from $+i$ to $+ i \infty$
and from from $-i$ to $-i \infty$.
By writing $w=u+iv$, $u,v\in \RR$,
the above cuts have equations $u=0, \,\, |v|\ge 1$
%
%
Since these cuts do not cross the real
axis of the $w$-plane, i.e. the $u$-axis,
the branches of $\arctan(w)$ so constructed
are continuous for real $w$.
}
\beq
- \, \infty \, < \, x \, < \, + \, \infty, 
\qquad 
- \, \frac{\pi}{2} \, < \, \arctan(x) \, < \, + \, \frac{\pi}{2}.
\eeq
Sometimes, to stress such choice, we will denote this branch 
by $\arctan_0(x)$.

We can reduce ourselves to the above 
"desirable" situation 
($w \, \zeta \mapsto w_0 \, \zeta$)
by means of the following reasoning.
Eq.(\ref{eq_w_Neq1_repeat}) involves a
periodic function, namely $\tan(w)$, 
of period $T=\pi$ (which is a quantity 
of order one), and a monotonic function, $f(w) \equiv w$. 
Inside any period $(a,a+\pi)$, $a \in \RR$, 
such as for example
the fundamental period $(-\pi/2,+\pi/2)$ 
(corresponding to the choice $a=-\pi/2$), 
the function $\tan(w)$ takes all its values, 
ranging from $-\infty$ to $+\infty$.
Let's now think to $w$ as an independent variable.
If we consider a large starting value of this variable,
\beq
w \, \gg \, 1,
\eeq
and make an increment of the latter of order one 
(or of order $\pi$, if preferred),
\beq
w \, \mapsto \, w \, + \, \Delta w, 
\qquad \Delta w \, = \, \mathcal{O}(1),
\eeq
the relative (or fractional) change of 
the function $f(w)=w$ will be small as,
by definition,
\beq
\frac{ \left| \Delta w \right| }{w} \, \ll \, 1,
\eeq
while the relative change of the function $\tan(w)$
will be, in general, very large.
Now, eq.(\ref{eq_Neq1_momenta}) can be trivially
rewritten as:
\beq
\label{eq_Neq1_momenta_bis}
\frac{\tan(w)}{w} \, = \, \zeta.
\eeq
In the weak-coupling regime, 
\beq
\left| \zeta \right| \, \ll \, 1,
\eeq
we can consider the situation in which:%
\footnote{
In ordinary (or straightforward) perturbation theory,
one assumes instead that $w=\mathcal{O}(1)$,
while $\tan(w)=\mathcal{O}(\zeta)$.}
\beq
w \, \approx \, \frac{1}{ \left| \zeta \right| }, 
\eeq
so that:
\beq
w \, \gg \, 1,
\eeq
while, generically:
\beq
\tan(w) \, = \, \mathcal{O}(1). 
\eeq
Therefore let us consider a
high-energy expansion for the
particle momentum $w$ of the form:
\beq
\label{basic_expan_large_n}
w \, = \, \nu \, + \, \Delta w_\nu;
\eeq
where  we have defined the convenient index
\beq
\nu \, \equiv \, \pi \, n.
\eeq
The index $\nu$ is assumed to be large, 
while $\Delta w_\nu$ is assumed to be a 
generic quantity of order one
(as will be checked {\it a posteriori}):
\beq
\nu \, \gg \, 1; 
\qquad \Delta w_\nu \, = \, \mathcal{O}(1).
\eeq 
Since the modulus of their ratio is small,
\beq
 \frac{\left| \Delta w_\nu \right|}{\nu}  \, \ll \, 1,
\eeq
the quantity
\beq
\frac{\Delta w_\nu}{\nu}
\eeq 
provides us the required expansion parameter.
By substituting the expansion for $w$ 
given in eq.(\ref{basic_expan_large_n})
inside eq.(\ref{eq_w_Neq1_repeat}), one obtains:
\beq 
\label{eq_Neq1_reduce}
\tan\left(\Delta w_\nu \right) \, = \, 
\zeta \left( \nu \, + \, \Delta w_\nu \right)
\, = \, \nu \, \zeta \, + \, \zeta \, \Delta w_\nu;
\eeq
where we have explicitly taken into account the periodicity of $\tan(w)$, so that:
\beq
\tan\left(\nu \, + \, \Delta w_\nu\right) \, = \, 
\tan\left(\Delta w_\nu\right).
\eeq
The crucial point is that the first member of 
eq.(\ref{eq_Neq1_reduce}) only depends on the correction term
$\Delta w_\nu$, as the leading term $\nu$ has disappeared.
Instead, the last member of eq.(\ref{eq_Neq1_reduce})
depends on both $\nu$ and $\Delta w_\nu$,
which however are both multiplied by 
the small coupling $\zeta$.
The idea now is to analyze the size of the
various terms which enter 
the first and the last member of eq.(\ref{eq_Neq1_reduce}),
namely the basic equation
\beq 
\label{eq_Neq1_reduce2}
\tan\left( \Delta w_\nu \right)
\, = \, 
\nu \, \zeta \, + \, \zeta \, \Delta w_\nu.
\eeq
It is easily found that $\zeta \, \Delta w_\nu$ is the 
smallest quantity entering eq.(\ref{eq_Neq1_reduce2}) because:
\begin{enumerate}
\item
It is much smaller than the 
momentum correction term $\Delta w_\nu$,
appearing on the r.h.s. of eq.(\ref{eq_Neq1_reduce2}), 
because it is multiplied by the small coupling $\zeta$:
\beq
\left| \zeta \, \Delta w_{\nu} \right|
\, \ll \, 
\left| \Delta w_{\nu} \right|
\quad \mathrm{as}
\,\,\, \left| \zeta \right| \, \ll \, 1;
\eeq
\item
It is much smaller than the quantity $\nu \, \zeta$,
\beq
\left| \zeta \, \Delta w_{\nu} \right| 
\, \ll \, \nu \left| \zeta \right|,
\eeq
because, by assumption:
\beq
\left| \Delta w_{\nu} \right| \, \ll \, \nu .
\eeq
\end{enumerate}
Therefore, to a first approximation,
one can safely neglect the term
$\zeta \, \Delta w_{\nu}$  
in eq.(\ref{eq_Neq1_reduce2}), obtaining
the much simpler equation
\beq 
\tan(\Delta w_\nu) \, \simeq \, \nu \, \zeta
\qquad (\mathrm{leading\,\,order}).
\eeq
The latter equation can be immediately solved to give:
\beq
\label{eq_lo_heuristic}
\qquad\qquad
w \, \equiv \, \nu \, + \, \Delta w_\nu 
\, \simeq \, \nu \, + \, \arctan(\omega)
\qquad\qquad (\mathrm{leading\,\,order}).
\eeq
Note that the result obtained, the last member of 
eq.(\ref{eq_lo_heuristic}),
is not trivial at all, and does
not coincide with the corresponding ordinary perturbation
expression previously obtained.
Note also that, 
as anticipated at the beginning of this section, 
eq.(\ref{eq_lo_heuristic}) 
exactly coincides with the L.O. resummed momentum $w=w_\nu$ 
derived in the previous section
(involving the function $\varphi_1(\omega)$ only),
given in eq.(\ref{eq_varphi1}).

Arrived at this point, we can also check the correctness of our initial small-parameter assumption 
$\left| \Delta w_{\nu} \right| \ll \nu$.
For any real argument $\omega$,
the uniform bound holds
\beq
\left| \arctan(\omega) \right| \, < \, \frac{\pi}{2},
\qquad \omega \in \RR.
\eeq
It follows that:
\beq
\frac{ \left| \Delta w_{\nu} \right| }{\nu} 
\, \lsim \, \frac{\pi}{2 \, \nu} \, \ll \, 1
\qquad \mathrm{for} \quad \nu \, \gg \, 1.
\eeq
Now, to evaluate sub-leading corrections to $w=w_\nu$
in this scheme and, in general, to derive a systematic expansion,
the idea is simply to think to the above,
leading approximation,
as the first step of a recursion process.
Therefore let's  go back to the exact momentum equation (\ref{eq_Neq1_reduce2}).
By thinking to the small term $\zeta \, \Delta w_\nu$
as a known quantity,
we can formally solve eq.(\ref{eq_Neq1_reduce2}) with 
respect to the momentum correction term $\Delta w_{\nu}$
appearing at its l.h.s.:
\beq
\Delta w_\nu \, = \,
\arctan\big( \nu \, \zeta \, + \, \zeta \, \Delta w_\nu \big).
\eeq
We solve the above equation recursively in the following way.
We generate a sequence
of functions 
\beq
\Big\{ \Delta w^{(h)}_\nu(\zeta); \,\,\, h = 0,1,2,3,\cdots 
\Big\}
\eeq 
by means of the following one-step recursion equation:
\beq
\label{map_fixed_point}
\qquad\qquad\qquad
\Delta w_\nu^{(h+1)}(\zeta) \, = \,
\arctan\left[ \nu \, \zeta 
\, + \, \zeta \, \Delta w_\nu^{(h)}(\zeta) \right];
\qquad
h = 0, 1, 2, 3, \cdots.
\eeq
The initial condition is given at $h=0$ 
and coincides with the free limit $(\zeta \to 0)$,
i.e. a vanishing value
of the momentum correction $\Delta w_\nu(\zeta)$:
\beq
\Delta w^{(0)}_\nu(\zeta) \, = \, 0.
\eeq
Therefore the momentum shift $\Delta w_\nu(\zeta)$
is the fixed point of the map given in eq.(\ref{map_fixed_point}).
The momentum level $w_{\nu}(\zeta)$ is given
by the limit of the above function sequence:
\beq
w_\nu(\zeta) \, = \, \nu \, + \, \Delta w_\nu(\zeta)
\, = \, \nu \, + \, \lim_{h\to\infty} \Delta w^{(h)}_\nu(\zeta).
\eeq
We have checked numerically the convergence of the function series 
$\left\{ \Delta w^{(h)}_{\nu}(\zeta) \right\}$ 
to $\Delta w_{\nu}(\zeta)$ in many cases 
(see fig.$\,$\ref{fig_plotNeq1b}),
which turns out to be quite fast with the recursion order
and uniform in $z \in \RR$. 
We admit that we do not have an analytic proof of the convergence yet, 
which is, in any case, well beyond the aims of the present paper.

There is an equivalent formulation of the
recursive solution for $w=w_\nu$, involving 
the complete momentum of the particle
in place of the momentum correction 
$\Delta w_\nu \equiv w_\nu - \nu$.
One constructs the sequence of functions 
\beq
\Big\{ w^{(h)}_\nu(\zeta); \,\,\, 
h = 0,1,2,3,\cdots \Big\} 
\eeq
by means of the following recursion equation:
\beq
w^{(h+1)}_\nu(\zeta) \, = \, \nu \, + \,
\arctan\Big[ \zeta \, w_\nu^{(h)}(\zeta) \Big];
\qquad
h = 0,1,2,3,\cdots.
\eeq
The initial condition reads:
\beq
w^{(0)}_{\nu}(\zeta) \, = \, \nu.
\eeq
The complete particle momentum 
(including the $0$-th order term $\nu$)
is simply the limit
of the above sequence:
\beq
w_{\nu}(\zeta) 
\, = \, 
\lim_{h \to \infty} w_\nu^{(h)}(\zeta).
\eeq
Let us work out explicitly the first few recursions,
in the first scheme constructed.
At the lowest, zero order:
\beq
\Delta w_\nu^{(0)} \, = \, 0;
\eeq
so that:
\beq
\label{eq_zero_ord_trivial}
\qquad w_\nu^{(0)} \, = \, \nu \, + \, \Delta w_\nu^{(0)} 
\, = \, \nu.
\eeq
The zero order is, by definition, trivial.
The first recursion gives:
\beq
\label{recursion1_bis}
w_\nu^{(1)}(\zeta) \, = \, \nu \, + \, \Delta w_\nu^{(1)}(\zeta)
\, = \, \nu \, + \,
\arctan\left( \zeta \, w_\nu^{(0)} \right)
\, = \, \nu \, + \,
\arctan\left( \omega \right).
\eeq
The function on the last member of the above equation 
exactly coincides with the approximation
to the particle momentum $w_\nu$
obtained in eq.(\ref{eq_lo_heuristic}).
Therefore, with the first recursion, 
we obtain the lowest-order non-trivial
approximation to the particle momentum.

\noindent
The second recursion reads:
\beq
w_\nu^{(2)}(\zeta) 
\, = \, \nu \, + \, 
\arctan\left[ \zeta \, w_\nu^{(1)}(\zeta) \right]
\, = \, \nu \, + \,
\arctan\left[ \omega 
\, + \,
\zeta \arctan\big(\omega \big) 
\right].
\eeq
In a similar way, the third recursion 
is written:
\beq
w_\nu^{(3)}(\zeta) 
\, = \, \nu \, + \, 
\arctan\left[ \zeta \, w_\nu^{(2)}(\zeta) \right]
\, = \, \nu \, + \,
\arctan\Big\{ \omega 
\, + \,
\zeta \arctan\big[\omega + \zeta \arctan(\omega) \big] 
\Big\}.
\eeq
We may notice that a recursion step is obtained,
in general, by substituting $\omega$ in the "innermost place"
with $\omega + \zeta \arctan(\omega)$:
\beq
\label{non_basic_repl_rule}
\omega \,\, \mapsto \,\, \omega \, + \, \zeta \arctan(\omega).
\eeq
We may also notice that the above rule
is obtained by multiplying, 
on both sides by the coupling $\zeta$, 
the basic replacement rule
\beq
\nu \,\, \mapsto \,\, \nu \, + \, \arctan(\omega),
\eeq
which represents the first recursion step ($h:0\mapsto 1$,
compare eq.(\ref{eq_zero_ord_trivial}) with
eq.(\ref{recursion1_bis})).
The observation leading to the substitution rule (\ref{non_basic_repl_rule}) can be formalized in the 
following way.
Let us define the function
\beq
f(x) \, \equiv \, \omega \, + \, \zeta \arctan(x).
\eeq
Note that $f = f_{\omega,\zeta}$, 
i.e. $f$ depends on both $\omega$ and $\zeta$
parametrically.
In terms of the function $f$,
the second recursion can be written:
\beq
w_\nu^{(2)}(\zeta) 
\, = \, \nu \, + \,
\arctan\big[ f(\omega) 
\big];
\eeq
and the third recursion:
\beq
w_\nu^{(3)}(\zeta) 
\, = \, \nu \, + \,
\arctan\Big\{ f\big[f(\omega)\big] \Big\}.
\eeq
In general, by iterating the recursion, 
one evaluates the composition of $f$ with itself a progressively
larger number of times.
Therefore 
let us define $f_h$ as the composition of $f$
with itself $h=2,3,4,\cdots$ times:
\beq
f_h \, \equiv \, 
\overbrace{ f \circ f \circ \cdots \circ f }^{h\,\mathrm{times}},
\eeq
i.e.:
\beq
f_h(x) \, = \, f(f(...f(x))) \,\,\,\, h\,\mathrm{times.}
\eeq
For example, by taking  $h=2$,
we obtain the function
\beq
f_2 \, = \, f \circ f,
\eeq
meaning, as well known, that:
\beq
f_2(x) \, = \, \big(f \circ f \big)(x) \, \equiv \, f\big[f(x)\big].
\eeq
It is natural to define $f_1(x) \equiv f(x)$ and
$f_0(x) \equiv x$ 
(i.e. $f_1$ is simply the initial function $f$, while $f_0$ is the identity function).
In terms of the sequence of functions 
$\big\{ f_h \big\}$,
the $h$-th recursion for the momentum $w_\nu$ can be written
in the following compact way:
\beq
w_\nu^{(h)} \, = \, \nu \, + \,
\arctan\left[ f_{h-1}(\omega) \right];
\qquad h = 1,2,3,\cdots. 
\eeq
Let us end this section by noting that
the momentum level $w_\nu(\zeta)$,
obtained as the limit of $w_\nu^{(h)}(\zeta)$
for $h\to \infty$, contains, roughly speaking, an infinite
number of compositions of the function $f$
with itself.

%
\begin{figure}[ht]
\begin{center}
\includegraphics[width=0.5\textwidth]{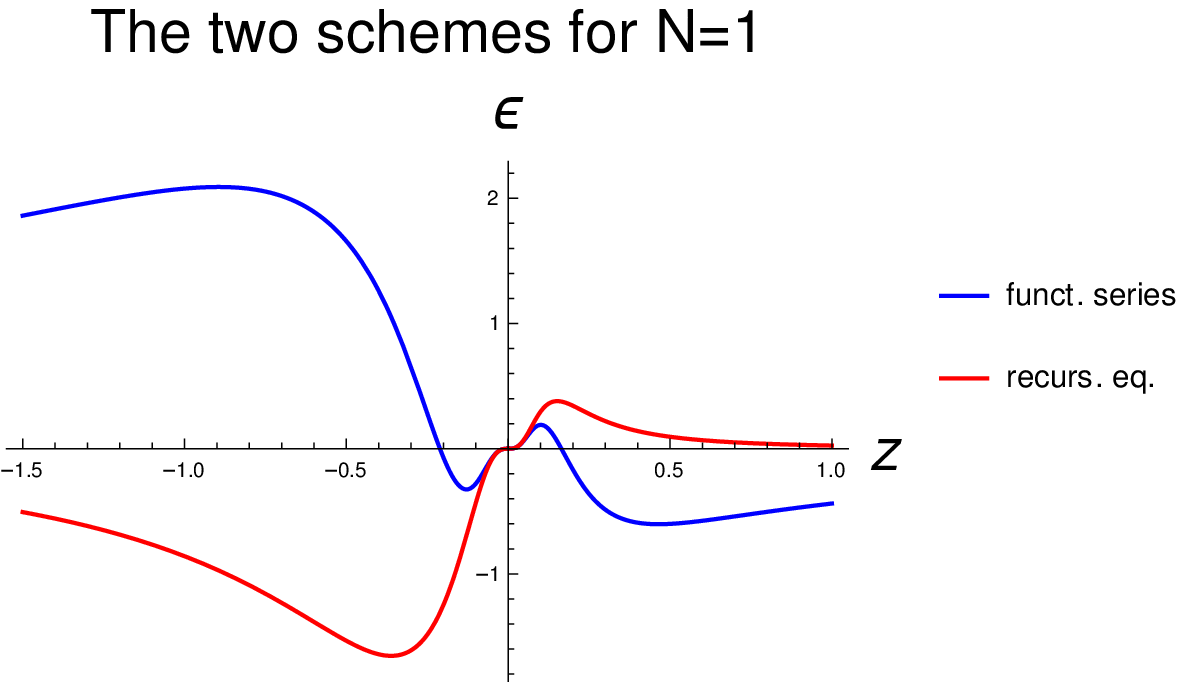}
\footnotesize
\caption{
\label{fig_plotNeq1c}
\it
Comparison of the percent relative errors 
$\epsilon$ as functions
of the coupling $z \equiv -g$,
namely $\epsilon=\epsilon(z)$,
in the two resummation schemes for $N=1$.
The blue line is the error for the fundamental
momentum level $(n=1)$ of the Next-to-Leading Order (N.L.O.) 
approximation in the function-series scheme;
The red line is the error of the second recursion
in the recursive-equation scheme.
As expected,
the two schemes basically coincide for small
coupling, let's say $|z| \lsim 0.1$.
For larger couplings, they differ more in the
negative-coupling region (repulsive potential)
than in the positive one.
The overall error is smaller in the recursion 
scheme, implying that the latter 
scheme is, at least in this case, better.
}
\end{center}
\end{figure}
%


\subsection{Comparison between the two resummation schemes}

In the previous sections, we presented
two different (both approximate) resummation
schemes for the perturbative series
of the particle momenta.
As we have discussed in detail, the first scheme
is based on a (truncated) function series,
while the second scheme is based on 
the approximate (i.e. finite order) solution of a 
recursion equation.
It is quite natural, arrived at this point, 
to compare the two schemes,
in order to find their common properties,
as well as their differences. 
That is indeed the aim of the present section.

\noindent
As already discussed, in the recursive scheme,
the first recursion,
\beq
w_\nu^{(1)}(\omega) \, = \, \nu \, + \, \arctan(\omega),
\eeq
exactly coincides with the function series truncated 
at first order, which we have called the Leading Order (L.O.)
approximation ($P=0$ in eq.(\ref{eq_funz_ser_sch_P})):
\beq
w_\nu^{\mathrm{LO}}(\omega) 
\, = \,  
\, \nu \, + \, \varphi_1(\omega).
\eeq
The leading-order agreement is related to the
fact that it turns out that: 
\beq
\varphi_1(\omega) \, = \, \arctan(\omega).
\eeq
At second order, i.e. at N.L.O.,
the two schemes begin to (slightly) differ from each other 
(see fig.$\,$\ref{fig_plotNeq1c}).
The scheme based on the function series 
gives, at N.L.O.:
\beq
\label{equat_sec_ord_i}
w_\nu^{NLO}(\zeta,\omega) \, = \, 
\nu \, + \, \varphi_1(\omega)
\, + \, \zeta \, \varphi_2(\omega)
\, = \,
\nu \, + \, \arctan(\omega) \, + \,
\zeta \, \frac{\arctan(\omega)}{1 \, + \, \omega^2}.
\eeq
The scheme based on the recursion equation
gives instead, for the second recursion,
as we have seen:
\beq
w_\nu^{(2)}(\zeta,\omega) 
\, = \, \nu \, + \,
\arctan\left[ \omega 
\, + \,
\zeta \arctan\big(\omega \big) 
\right].
\eeq
Note that the latter formula,
unlike the previous one,
contains terms of all orders in $\zeta$.
By expanding the r.h.s. of the above equation
in powers of $\zeta$ up to second order, 
one obtains:
\beq
\label{equat_sec_ord_ii}
w_\nu^{(2)}(\zeta,\omega) \, = \,
\nu \, + \, \arctan(\omega) \, + \,
\zeta \, \frac{\arctan(\omega)}{1 \, + \, \omega^2}
\, - \, \zeta^2 \, \frac{\omega \arctan^2(\omega)}{\left( 1 \, + \,\omega^2 \right)^2} 
\, + \, \mathcal{O}\left( \zeta^3 \right).
\eeq
The second-order term in $\zeta$ on the r.h.s. of
the above equation
is naturally compared with the N.N.L.O. function 
$\varphi_3(\omega)$ which, according to eq.(\ref{eq_Neq1_varphi3}), 
is given by:
\beq
\varphi_3(\omega) \, = \, 
\, - \, 
\frac{\omega \, \arctan^2(\omega)}{(1 \, + \, \omega^2)^2}
\, + \,
\frac{\arctan(\omega)}{(1 \, + \, \omega^2)^2}.
\eeq
By comparing the last member of eq.(\ref{equat_sec_ord_i}) 
with the r.h.s. of eq.(\ref{equat_sec_ord_ii}), 
one finds
that the second recursion $w_\nu^{(2)}(\zeta,\omega)$
contains the first contribution to $\varphi_3(\omega)$,
namely the term
\beq
\, - \, \frac{\omega \arctan^2(\omega)}{\left( 1 \, + \,\omega^2 \right)^2}.
\eeq
Note that the above term is the dominant one 
in $\varphi_3(\omega)$ for $\omega \gg 1$, 
because of the additional factor $\omega$ at the numerator.
By pushing the expansion of $w_\nu^{(2)}(\zeta,\omega)$ 
in powers of $\zeta$ to even higher
orders, one finds that the second
recursion also contains the leading
contributions to $\varphi_4(\omega)$,
$\varphi_5(\omega)$, $\varphi_6(\omega)$,
$\cdots$, for $\omega \to \infty$.
With a suggestive language,
we may say that the scheme based on the
recursion equation "resums the resummation"
realized by the function-series scheme.
Such a "resummation of a resummation" is,
of course, again an approximate one 
(we are not able to solve exactly the $N=1$
momentum-spectrum equation!)
and is relevant in the case $|\omega| \gg 1$.

Let us also note that the functions
$\varphi_1(\omega)$ and $\varphi_2(\omega)$
can also be computed, in addition to the
direct evaluation described at the end of
sec.$\,$\ref{sec_resum_method_one},
also by expanding the second recursion 
$w_\nu^{(2)}=w_\nu^{(2)}(\zeta,\omega)$
in powers of $\zeta$ up to first order.
In general, the functions 
\beq
\varphi_1(\omega), \,\,\, \varphi_2(\omega),
\,\,\, \cdots, \,\,\, 
\varphi_h(\omega)
\eeq
can be computed by expanding $w_\nu^{(h)}(\zeta,\omega)$, 
the recursion of order $h$, 
in powers of $\zeta$ up to the order $h-1$ included.

A natural question at this point is which
resummation scheme, among the two, is preferable.
It is fair to say that each scheme
has its own virtues and shortcomings.
The scheme based on the function series
--- let's call it the first scheme ---
is conceptually simpler, as it involves
truncated expansions in the small
parameter $\zeta$.
The scheme based on the recursion equation
--- let's call it the second scheme ---
includes higher-order terms in $\zeta$,
which are missed in a truncated power series 
in $\zeta$, and which are the dominant ones
for very large $\omega$.
In view of this latter fact,
we may conclude that the second resummation scheme 
is perhaps better than the first one 
in the intermediate and in the large coupling regions
($|z|\gsim 1$), or in the description
of states with very high energies.
Because of that, in the following sections we will mostly
consider the second or recursive scheme.


\subsection{Discussion}

Finite-volume Winter model in the symmetric case $N=1$,
in which the two resonant cavities have the same length,
is an oscillating, rather than a decaying system. 
Amplitude is constantly transferred from one cavity 
to the other, back and forth.
Similar oscillation phenomena occur in classical mechanics,
by considering for example two weakly-coupled harmonic oscillators,
where energy is constantly transferred
from one mode to the other \cite{nayfeh}.

To fully understand the dynamics of the model,
we may consider the explicit time evolution
of a wavefunction initially contained
in the "small" cavity $[0,\pi]$, 
i.e. identically vanishing in the "large" cavity
$[\pi,2\pi]$, which is equal
to a $[0,\pi]$ box eigenfunction:
\beq
\label{eq_wf_initial}
\psi(x; \, t=0) \, \equiv \, \psi_{\mathrm{ini}}(x)
\, = \, \sqrt{\frac{2}{\pi}} \, \theta(\pi-x) \, \sin(j x);
\qquad 0 \, \le \, x \, \le \, 2 \, \pi;
\eeq
where $j=1,2,3,\cdots$ is a positive integer and 
$\theta(y) \equiv 1$ for $y>0$
and zero otherwise is the usual Heaviside
step function.
Since the two cavities have the same length
or, equivalently,
because of the absence of non-resonant levels,
we expect the no-escape probability,
for example,
\beq
P(t)^{no\,esc} \, \equiv \, 
\int\limits_0^\pi 
\left| \psi(x,t) \right|^2 \, dx,
\eeq
to not exhibit an approximate exponential decay in 
any time region.
Heuristically, we may consider the exponential decay
of an unstable state with time a statistical phenomenon, 
in which the initial amplitude, contained in
the small cavity, is transferred to the large number of
available states of the system.
Since the large cavity has the same density of states
(in momentum space)
of the small one, the former is not expected to be able 
to "exponentially absorb", for some time, a large part of the initial amplitude.   

We may conclude that the symmetric case $N=1$
is qualitatively different from the
quasi-continuum one ($N\gg 1$) and describes
oscillation, rather than decay, phenomena. 


\section{The Double Case $N=2$}
\label{sec_double_case}

In this section we consider Winter model
at finite volume in the particular case $N=2$,
in which the right cavity $[\pi,3\pi]$ (the large one) 
is two times larger than the left one $[0,\pi]$
(the small one). 

\begin{figure}[ht]
\begin{center}
\includegraphics[width=0.5\textwidth]{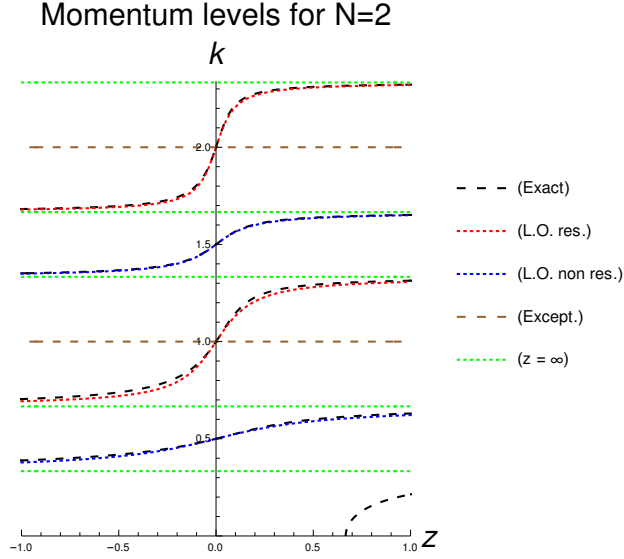}
\footnotesize
\caption{
\label{fig_plotNeq2}
\it
Lowest six momentum levels for the case $N=2$.
The lowest four normal (or non-exceptional) momenta, 
$k_{1/2}(z), \, k_1(z), \, k_{3/2}(z), \,
k_2(z)$, are represented by black dashed lines, 
while the two lowest exceptional momenta, $k_{exc}\equiv 1,2$,
are represented by brown dashed lines.
The Leading Order (L.O.) resonant levels 
$k_1^{LO}(z)$ and $k_2^{LO}(z)$ 
are given by red dotted lines, while
the L.O. non-resonant levels $k_{1/2}^{LO}(z)$ and
$k_{3/2}^{LO}(z)$ are
given by blue dotted lines.
The infinite-coupling limit of the non-exceptional levels 
$k(z=\pm\infty) = 1/3, \, 2/3, \, 4/3, \, 5/3, \, 7/3$
(integer values are absent) are
represented by horizontal green dotted lines.
The range of the resonant levels, 
i.e. the difference $k_n(z=+\infty) - k_n(z=-\infty)$,
is $2/3$ and it is
two times larger than the range of the non-resonant
ones, which is $1/3$.
The accuracy of the L.O. approximation
quickly increases with the order, i.e. with the energy 
of the level, as expected.
Quite remarkably, the L.O. approximation  is rather good also for 
the lowest non-resonant level $(k_{1/2}(z))$.
In general, the L.O. approximation seems to work slightly 
better for non-resonant levels than for
resonant levels of close energy (contiguous levels).
}
\end{center}
\end{figure}

In this case, resonant levels $k_n(z)$, 
having integer limit $n=1,\,2,\,3,\,\cdots$
in the free limit, 
alternate with non-resonant levels
$k_{n+1/2}(z)$, having instead semi-integer limit
$n+1/2=1/2,\,\,3/2,\,\,5/2,\,\,\cdots$ for $z \to 0$.
In other words, resonant levels and non-resonant
ones separate each other (see fig.$\,$\ref{fig_plotNeq2}).
The equation for the normal part of the momentum
spectrum explicitly reads:
\beq
\label{equat_Neq2_begin}
\qquad\qquad
\frac{1}{\tan(w)} \, + \, \frac{1}{\tan(2w)} 
\, = \, \frac{1}{z \, w}
\qquad\qquad(N=2);
\eeq
where:
\beq
z \, \equiv \, - \, g.
\eeq
Note that eq.(\ref{equat_Neq2_begin}) is considerably more complex than eq.(\ref{eq_Neq1_momenta}), the latter
describing the physically and mathematically much simpler
$N=1$ case.

Now it comes one of the main technical ideas
of this work, which will be used also in the
larger $N$ cases: 
we transform $\tan(2w)$,
the tangent of the double momentum $2w$ above,
into a rational function (of degree two)
in $\tan(w)$, the tangent of the momentum $w$.
The general tangent reduction formula is derived
in Appendix \ref{app_tang_mult_ang}, which, 
in the case $N=2$, reduces to the first of 
eqs.(\ref{eq_tan_red_partic}).
By means of the latter, a non-linearity of transcendental type
is converted into a non-linearity of algebraic
type, which is much more tractable.
By means of the first of eqs.(\ref{eq_tan_red_partic}),
the momentum equation above is rewritten:
\beq
\qquad\qquad
\label{eq_mom_spec_Neq2_pol}
\tan^2(w) \, + \, \frac{2}{z w} \, \tan(w) 
\, - \, 3 \, = \, 0
\qquad\qquad (N=2).
\eeq
By assuming $z w$ to be a known quantity, 
we can solve the above equation with respect to $\tan(w)$,
obtaining:
\beq
\tan(w) \, = \, 
\frac{ \pm \, \Big[ 1 \, + \, 3(w z)^2 \Big]^{1/2} \, - \, 1 }{w z}.
\eeq
The two sign determinations in front of the arithmetic (i.e. positive) square root above give rise to two different sequence of levels.
Indeed, in the weak-coupling regime, $|z| \ll 1$, the
plus/minus equations read respectively:
\bea
\label{eq_plus}
1. \,\,\,\, \tan(w) &=& + \,
\frac{ \sqrt{ 1 \, + \, 3(w \, z)^2 } \,\, - \, 1 }{w \, z}
\, \approx \, + \, \frac{3}{2} \, w \, z;
\\
\label{eq_minus}
2. \,\,\,\, \tan(w) &=& - \,
\frac{ \sqrt{ 1 \, + \, 3(w \, z)^2 } \, + \, 1 }{w \, z}
\, \approx \, - \, \frac{2}{w \, z}.
\eea
In the first case 1., by taking the free limit $z \to 0$, 
the function $\tan(w)$ vanishes, implying that $w$ tends to an integer $n$, times $\pi$.
Therefore the "plus equation" above describes the resonant levels of the system.

In the second case 2., for $z \to 0$, $\tan(w)$
instead  diverges, so that $w$ tends to $n+1/2$,
i.e. to a half-integer, times $\pi$.
Therefore the "minus equation" above  
describes the non-resonant levels.

%
%


\subsection{Resonant levels}

Let us first consider the simpler, resonant case 1.
The particle momentum $w=w_\nu$ is naturally
decomposed, as in the previous $N=1$ case, as:
\beq
\label{eq_basic_decompos_Neq2}
w_\nu \, = \, \nu \, + \, \Delta w_\nu
\eeq
where:
\beq
\nu \, \equiv \, \pi \, n;
\qquad n = 1,2,3,\cdots. 
\eeq
Eq.(\ref{eq_plus}) is then rewritten:
\beq
\label{eq_Neq2_specialized_rescase}
\Delta w_\nu \, = \,
\arctan 
\left[
\frac{ \sqrt{ 1 \, + \, 3( z \, w_\nu )^2 } \,\, - \, 1 }{z \, w_\nu}
\right] .
\eeq
Just as in the case $N=1$, there are two possible
resummation schemes for the perturbative series
of the particle momenta, which are discussed
separately in the next sections.


\subsubsection{Resummation by means of a function series}

The perturbative series for the
resonant particle momentum $w_\nu$
can be (approximately) resummed 
to all orders in
\beq
\eta \, \equiv \, \nu \, z
\eeq
and to finite order in $z$, by means 
of a function series. 
Similarly to the previous $N=1$ case,
the particle momentum is written:
\beq
\label{eq_w_reson_Neq2}
\qquad
w_\nu \, \cong \, \nu \, + \,
\sum_{h=0}^P z^h \, \phi_{h+1}(\eta)
\, = \, \nu \, + \, \phi_1(\eta) 
\, + \, z \, \phi_2(\eta)
\, + \, \cdots
\, + \, z^P \, \phi_{P+1}(\eta)
\qquad\quad\qquad (N=2); \qquad
\eeq
where the non-negative integer $0 \le P < \infty$ specifies the order of the approximation.
The Leading-Order (L.O.) function 
$\phi_1(\eta)$
(corresponding to the choice $P=0$) and 
the Next-to-Leading Order (N.L.O.) 
function $\phi_2(\eta)$ 
($P=1$) have the following explicit expressions:
\bea
\phi_1(\eta) &=&
\arctan\left[\frac{\Big(1 + 3 \, \eta^2\Big)^{1/2} \, - \, 1}{\eta}\right];
\qquad\qquad\qquad\qquad\qquad\qquad\qquad\qquad (N=2)
\nonumber\\
\phi_2(\eta) &=& 
\left( 1 + \frac{1}{2 \sqrt{1 + 3 \eta^2}} \right)
\frac{1}{ 1 + 4 \eta^2}
\arctan\left[\frac{\Big(1 + 3 \, \eta^2\Big)^{1/2} \, - \, 1}{\eta}\right].
\eea
The evaluation of higher-order functions is,
in principle, straightforward.
However, by increasing $P$, i.e. the order of the
approximation, the complexity of the
functions $\phi_i(\eta)$ grows quite
quickly.


\subsubsection{Resummation by recursion}

Similarly to the previous $N=1$ case, 
the momentum $w_\nu$ of the particle is written:
\beq
\label{eq_Neq2__rescase_exact}
w_\nu \, = \, \nu \, + \, \Delta w_\nu;
\qquad \nu \, \gg \, 1.
\eeq 
According to the above decomposition, 
eq.(\ref{eq_plus}) is conveniently rewritten:
\beq
\Delta w_\nu \, = \, \arctan
\left[
\frac{ \sqrt{ 1 \, + \, 3 
\Big(\eta \, + \, z \, \Delta w_\nu \Big)^2 } \,\, - \, 1 }{
\eta \, + \, z \, \Delta w_\nu }
\right].
\eeq
We construct a function sequence
$\left\{ w_\nu^{(h)}(z); \,\, h=0,1,2,3,\cdots \right\}$
by means of the following one-step recursion in 
the index $h$:
\beq
\label{eq_rec_Neq2}
w_\nu^{(h+1)}(z) \, = \,
\nu \, + \, \arctan
\left[
\frac{ \sqrt{ 1 \, + \, 3 \, z^2 \, w_\nu^{(h)}(z)^2 } 
\,\, - \, 1 }{ z \, w_\nu^{(h)}(z) }
\right];
\qquad
h \, = \, 0, 1, 2, 3, \cdots;
\eeq
with the initial condition (at $h=0$):
\beq
w_\nu^{(0)}(z) \, = \, \nu .
\eeq
The function $w = w_\nu(z)$ is obtained as
the limit of the above function sequence:
\beq
w_\nu(z) \, = \, \lim_{h\to\infty} w_\nu^{(h)}(z).
\eeq
In practice, it is sufficient to compute few 
recursions only, as they already provide a good, uniform approximation in $z \in \RR$, to the exact result.
In other words, convergence 
of the function sequence $\left\{ w_\nu^{(h)}(z) \right\}$
to the exact momentum level $w_\nu(z)$ (for $h\to\infty$) is rather fast.

The first recursion ($h=1$) gives the lowest-order non-trivial approximation for the resonant particle momenta
and explicitly reads:
\beq
\label{eq_Neq2_first_rec}
\qquad\qquad\qquad\qquad\qquad
w_\nu^{(1)}(\eta) \, = \,
\nu \, + \, \arctan
\left(
\frac{ \sqrt{ 1 \, + \, 3 \, \eta^2 } 
\,\, - \, 1 }{ \eta }
\right)
\qquad\qquad (N=2).
\eeq
Let us remark that it is not legitimate
to expand the square root above in powers of $\eta$, 
as $\eta$ is not, by assumption, a small variable.

By means of eq.(\ref{eq_rec_Neq2}) for $h=2$,
the second recursion is written:
\beq
\qquad\qquad\qquad
w_\nu^{(2)}(z,\eta) \, = \,
\nu \, + \, \arctan
\left[
\frac{ \sqrt{ 1 \, + \, 3 \, z^2 \, w_\nu^{(1)}(\eta)^2 } 
\,\, - \, 1 }{ z \, w_\nu^{(1)}(\eta) }
\right]
\qquad\qquad (N=2);
\eeq
with $w_\nu^{(1)}(\eta)$ explicitly given in 
the previous equation (\ref{eq_Neq2_first_rec}).
To give the reader an idea about the kind of functions
coming out of the recursions,
let us give the second recursion for $w_\nu$
in completely explicitly form:
\beq
w_\nu^{(2)}(z,\eta) \, = \,
\nu \, + \, \arctan
\left(
\frac{  \sqrt{  1 \, + \,
3 \, \bigg\{ 
\, \eta \, + \, z \, \arctan
\Big[
\big( \sqrt{ 1 \, + \, 3 \, \eta^2 } 
\,\, - \, 1 
\big) / \eta
\Big]  
\, \bigg\}^2 } 
\, - \, 1 }{ \eta \, + \, z \, \arctan
\left[
\big( \sqrt{ 1 \, + \, 3 \, \eta^2 } 
\,\, - \, 1 \big) / \eta
\right] }
\right) .
\eeq
Similarly to the previous $N=1$ case,
we may notice that a recursion step is obtained 
by replacing $\eta$,
in its innermost place, according to the rule:
\beq
\eta \,\, \mapsto \,\, \eta \, + \, 
z \arctan\left[\frac{\big(1 + 3 \, \eta^2 \big)^{1/2} \, - \, 1}{\eta}\right]. 
\eeq
The above rule, in turn, is obtained by multiplying on both sides by $z$, the basic substitution rule
\beq
\nu \,\,\, \mapsto \,\,\, \nu \, + \, 
\arctan\left[\frac{\Big(1 + 3 \, \eta^2\Big)^{1/2} \, - \, 1}{\eta}\right]. 
\eeq
The recursions are easily written by means of the function
\beq
g(x) \, \equiv \, \eta \, + \, z 
\arctan\left[\frac{\big( 1 + 3 \, x^2 \big)^{1/2} \, - \, 1}{x}\right].
\eeq
The treatment is similar to the $N=1$ case, so
we omit the details.
The resonant momentum $w_\nu$ at the step $h$
is written:
\beq
w_{\nu}^{(h)} \, = \,
\nu \, + \, 
\arctan\left\{
\frac{\Big[1 + 3 \, g_{h-1}(\eta)^2 \Big]^{1/2} \, - \, 1}{g_{h-1}(\eta)}
\right\};
\qquad h=1,2,3,\cdots;
\eeq
where $g_h$ is the (function) composition of $g$
with itself $h$ times, with $g_{h=1}(\eta)\equiv g(\eta)$
and $g_{h=0}(\eta) \equiv \eta$.


\subsubsection{Comparison with ordinary perturbation theory}

The comparison of the perturbative resummed formulae
derived in the above two sections with the fixed-order ones
is completely similar to the $N=1$ case, so we 
just sketch the procedure;
The actual computation is left to the reader as an exercise.
We expand in powers of $z$ eq.(\ref{eq_w_reson_Neq2}),
with a selected value of $P$, 
or eq.(\ref{eq_rec_Neq2}), 
with a selected value of $h$.
We then compare the coefficients of the various
powers of $z$ with the coefficients given
in eqs.(\ref{coef_ris_straight}), 
in which we have set $N=2$.


\subsection{Non-resonant levels}

In this section we consider the 
non-resonant levels of the system, 
described by eq.(\ref{eq_minus})
(the case $2.$).
For the weak-coupling expansion, $|z|\ll 1$,
it is convenient to write the particle
momentum $w = w_{\hat{\nu}}$ in the form
\beq
\label{eq_nonres_decomp_natur_Neq2}
w_{\hat{\nu}} \, = \, \hat{\nu} \, + \, \delta w_{\hat{\nu}};
\eeq
where we have defined the "ad-hoc" index,
involving semi-integer numbers,
\beq
\hat{\nu} \, \equiv \, 
\pi \left( n \, + \, \frac{1}{2} \right).
\eeq
Note that:
\beq
\Delta w_\nu \, = \, \delta w_{\hat{\nu}} \, + \, \frac{\pi}{2}
\eeq
and that $\delta w = \delta w_{\hat{\nu}}(z)$
vanishes in the free limit:
\beq
\lim_{z \to 0} \delta w_{\hat{\nu}}(z)
\, = \, 0.
\eeq
By means of the formula
\beq
\tan\left( x \, + \, \frac{\pi}{2} \right)
\, = \, - \, \frac{1}{\tan(x)},
\eeq
the following equation for the non-resonant momentum shift 
$\delta w_{\hat{\nu}}$ is obtained
from eq.(\ref{eq_minus}):
\beq
\delta w_{\hat{\nu}} \, = \,
\arctan\left[ \frac{ \sqrt{ 1 \, + \, 3 \, 
\left( z \, w_{\hat{\nu}}  \right)^2 } 
\,\, - \, 1 }{
3 \, z \, w_{\hat{\nu}}}     \right]
\qquad\qquad\qquad (N=2).
\eeq
The above equation can also be obtained
by transforming eq.(\ref{eq_minus})
according to the formula
\beq
\label{eq_arctan_toinv}
\arctan(x) \, + \, \arctan\left( \frac{1}{x} \right) 
\, = \, \pm \, \frac{\pi}{2},
\qquad \mathrm{for}\,\, x>0 
\,\,\ \mathrm{and}\,\,\, x<0  \,\,\, \mathrm{respectively}.
\eeq
We consider the simpler repulsive case $z<0$,
implying $z\, w<0$ (the attractive case, $z>0$,
is obtained later, from the case $z<0$, 
by means of analytic continuation;
That way, continuity of particle momenta at $z=0$
is guaranteed). 

The complete particle momentum $w_{\hat{\nu}}(z)$, 
according to eq.(\ref{eq_nonres_decomp_natur_Neq2}),
satisfies the following equation:
\beq
\label{eq_Neq2_nonres_exact}
\qquad\qquad\qquad\qquad
w_{\hat{\nu}}(z) \, = \, \hat{\nu}
\, + \,
\arctan\left[ \frac{ \sqrt{ 1 \, + \, 
3 \, \left[ z \, w_{\hat{\nu}}(z) \right]^2 } 
\,\, - \, 1 }{
3 \, z \, w_{\hat{\nu}}(z)  }  \right]
\qquad\qquad\qquad (N=2).
\eeq
Note that the above equation is still exact
and that the only differences with respect
to the corresponding resonance equation
(\ref{eq_basic_decompos_Neq2}) are: 
\begin{enumerate}
\item
In the lowest-order term (coinciding with the free limit
$z\to 0$),
the index $\nu \equiv \pi n$ is replaced by the new index 
$\hat{\nu} \equiv \pi (n+1/2)$,
the latter containing an additional, half-integer contribution;
\item
The argument of the arctangent function
contains an additional factor three
at the denominator.
That implies, in particular, that,
in the weak-coupling region,
the growth of the non-resonant momentum with $z$ 
is softer with respect to the resonant case.
\end{enumerate}
Just like in the resonant case, also in the
non-resonant case two resummation schemes
are available, which are discussed in the
next sections.


\subsubsection{Resummation by function series}

The non-resonant particle momentum $w_{\hat{\nu}}$
is written as the following function series:
\beq
w_{\hat{\nu}} \, \cong \, \hat{\nu}
\, + \, \sum_{j=0}^P 
z^j \, \psi_{j+1}\left( \hat{\eta} \right)
\, = \, \hat{\nu}
\, + \, \psi_1\left( \hat{\eta}\right)
\, + \, z \, \psi_2\left( \hat{\eta}\right)
\, + \, \cdots
\, + \, z^P \, \psi_{P+1}\left( \hat{\eta}\right);
\eeq
where $0 \le P < \infty$ specifies the order of the 
approximation and we have defined the convenient 
$\eta$-type variable
\beq
\hat{\eta} 
\, \equiv \, \hat{\nu} \, z
\, = \, \pi \left( n \, + \, \frac{1}{2} \right) z.
\eeq
The evaluation of the functions 
$\psi_j\left( \hat{\eta} \right)$
is straightforward.
For concreteness sake,
let's just give explicitly the L.O. function:
\beq
\label{eq_Neq2_nonres_scheme1}
\qquad\qquad
\psi_1\left( \hat{\eta} \right)
\, = \, 
\arctan
\left(
\frac{ \sqrt{ 1 \, + \, 3 \, {\hat{\eta}}^2 } 
\,\, - \, 1 }{ 3 \, \hat{\eta} }
\right)
\qquad\qquad (N=2). 
\eeq


\subsubsection{Resummation by recursion}

Similarly to the previous cases,
a function sequence 
$\left\{ w_{\hat{\nu}}^{(h)} \right\}$
converging to the desired
function $w_{\hat{\nu}}$ is generated 
by means of the following recursion equation:
\beq
\label{eq_Neq2_nonres_recurs}
\qquad\qquad\qquad
w_{\hat{\nu}}^{(h+1)}(z) \, = \, \hat{\nu}
\, + \,
\arctan\left\{ \frac{ 
\sqrt{ 1 \, + \, 3 \, z^2 
\left[ w_{\hat{\nu}}^{(h)}(z) \right]^2 } \, - \, 1 }{
3 \, z \, w_{\hat{\nu}}^{(h)}(z)  }     
\right\}
\qquad\qquad (N=2);
\eeq
with
\beq
h \, = \, 1,2,3,\cdots.
\eeq
The initial condition, as usual, coincides
with the free limit of the non-resonant levels:
\beq
w_{\hat{\nu}}^{(0)} \, = \, \hat{\nu}.
\eeq
The first recursion ($h=1$) provides
the lowest-order, non-trivial approximation 
for the non-resonant momenta:
\beq
\label{eq_Neq2_nonres_firstord}
\qquad\qquad\qquad\qquad\qquad
w_{\hat{\nu}}^{(1)}(\hat{\eta}) \, = \,
\hat{\nu} \, + \, \arctan
\left(
\frac{ \sqrt{ 1 \, + \, 3 \, {\hat{\eta}}^2 } 
\,\, - \, 1 }{ 3 \, \hat{\eta} }
\right)
\qquad\qquad (N=2).
\eeq
Note that the correction term in the above equation,
i.e. the second term at its r.h.s., exactly
coincides with the L.O. function 
$\psi_1\left(\hat{\eta}\right)$
given in eq.(\ref{eq_Neq2_nonres_scheme1}).
Therefore the two resummation schemes
exactly coincide, at the first non-trivial order,
also in this case.
Note also that the first-order momentum correction
\beq
\label{delw_non reson_Neq2_ord1}
\delta w_{n+1/2}^{(1)}(z) \, = \,
\arctan
\left(
\frac{ \sqrt{ 1 \, + \, 3 \, \pi^2 \, (n+1/2)^2 \, z^2 } 
\,\, - \, 1 }{ 3 \, \pi \, (n+1/2) \, z }
\right)
\eeq
is an odd function of $z$.
The second recursion ($h=2$) is written
in terms of the first one ($h=1$) as:
\beq
\label{eq_Neq2_nonres_secondord}
\qquad\qquad\qquad
w_{\hat{\nu}}^{(2)}(z,\hat{\eta}) \, = \,
\hat{\nu} \, + \, \arctan
\left[
\frac{ \sqrt{ 1 \, + \, 3 \, z^2 \, 
w_{\hat{\nu}}^{(1)}(\eta)^2 } 
\,\, - \, 1 }{ 3 \, z \, w_{\hat{\nu}}^{(1)}(\eta) }
\right]
\qquad\qquad (N=2).
\eeq
As observed before by looking at the exact equation,
because of the factor three
at the denominator of the argument of the
arctangent function at the r.h.s. of 
eq.(\ref{eq_Neq2_nonres_firstord})
and (\ref{eq_Neq2_nonres_secondord}), 
the growth of the
particle momentum $w$ with the coupling $z$ 
is softer in the
non-resonant case with respect to
the resonant one.
This observation \cite{FVWMme}
has also been made in section$\,$\ref{Winter_Phys} on the
basis of ordinary perturbation-theory formulae.
This factor three is also responsable of the smaller
range of the non-resonant levels (by a factor two)
with respect to the resonant ones 
(see fig.$\,$\ref{fig_plotNeq2}).


\subsubsection{Comparison with ordinary perturbation theory}

Since the resummation of the perturbative
series of a non-resonant momentum 
--- eq.(\ref{eq_Neq2_nonres_firstord})
for the first order and eq.(\ref{eq_Neq2_nonres_secondord})
for the second order in the recursive scheme ---
constitutes a new kind of resummation,
let us discuss in this section, in some detail, 
the comparison with ordinary 
perturbation-theory formulae.

By evaluating the coefficients 
of ordinary perturbation theory
--- eqs.(\ref{eq_Neq2_nonris}) for $N=2$ ---,
one obtains, up to third order in $z$ included:
\beq
\qquad
w_{\hat{\nu}}(z) \, = \, \hat{\nu}
\, + \, \frac{1}{2} \, \hat{\nu} \, z 
\, + \, \frac{1}{4} \, \hat{\nu} \, z^2
\, - \, \frac{5}{12} \, \hat{\nu}^3 \, z^3
\, + \, \frac{1}{8} \, \hat{\nu} \, z^3
\, + \, \mathcal{O}\left(z^5\right)
\qquad\qquad (N=2).   
\eeq
The expansion in powers of $z$ of 
the first recursion, eq.(\ref{eq_Neq2_nonres_firstord}),
reads instead:
\beq
w_{\hat{\nu}}^{(1)}(z) \, = \, \hat{\nu}
\, + \, \frac{1}{2} \, \hat{\nu} \, z 
\, - \, \frac{5}{12} \, \hat{\nu}^3 \, z^3
\, + \, \mathcal{O}\left(z^4\right)
\qquad\qquad\qquad\qquad (N=2). 
\eeq
As in previous cases, the first recursion 
correctly reproduces the first-order term in $z$,
completely misses the second-order term and
gives only the leading, large-$\hat{\nu}$ contribution 
at third order. 


\subsection{Discussion}

Finite-volume Winter model in the double case, $N=2$, 
in which the large cavity, $[\pi,3\pi]$, is two times
larger than the small one, $[0,\pi]$, is, in a sense, 
the lowest-$N$ non-trivial case. 
Even though $N=2$ cannot be reasonably considered a large number, by going from the case $N=1$ to the case $N=2$, 
we begin to see, 
in the small-coupling domain $|z| \ll 1$, 
some hints of resonance dynamics of the
quasi-continuum $(N \gg 1)$.
The density of states of the large cavity
(in momentum space) is indeed two times larger
than that of the small cavity, so that
non-resonant levels occur in the spectrum.

We may say that the main change, 
by going from $N=1$ to $N=2$,
is indeed the appearance of non-resonant levels
in the momentum spectrum of the model.
The latter are rather symmetrical
under a change of sign of the coupling, 
i.e. for $z \mapsto -z$, so they cannot
be naturally associated to the 
contiguous upper or lower resonant level.
As already noted, the L.O. approximation
to the momentum shift $\delta w_{\hat{\nu}}^{(1)}(z)$, 
eq.(\ref{delw_non reson_Neq2_ord1}),
is indeed an odd function of the coupling $z$.
Furthermore, as we have shown again with
analytic computations,
non-resonant levels have a smaller slope 
around $z=0$ than resonant ones.
Unlike what happens in the
quasi-continuum case (see sec.\ref{Winter_Phys}),
non-resonant levels
do not exhibit a resonant behavior
in any coupling region.  
The main role of non-resonant levels
is simply that of separating the resonant levels
from each other.


\section{General method}
\label{sect_gen_meth}

The transcendental equation (\ref{eq_basic}) for the
ordinary momentum spectrum of the model can be written:
\beq
\qquad
\frac{1}{\tan(w)} \, + \, \frac{1}{\tan(N w)}
\, = \, \frac{1}{z\, w};
\qquad\qquad N \in \NN \equiv \left\{ 1,2,3,\cdots\right\}.
\eeq
By using the tangent reduction equation (\ref{eq_TanNwExpansion}), derived in 
Appendix$\,\,$\ref{app_tang_mult_ang},
inside the momentum equation above, 
one obtains an equation of the
form:
\beq
\label{eq_after_red}
S_N\big( t \big) \, = \, z \, w;
\eeq
where:
\beq
\label{eq_def_S_N}
S_N(t) \, \equiv \, \frac{t \, R_N(t)}{ t \, + \, R_N(t) }
\, = \, \frac{t \, P_{2\lfloor(N-1)/2\rfloor}(t)}{P_{2\lfloor(N-1)/2\rfloor}(t)\, + \, 
Q_{2\lfloor N/2 \rfloor }(t)}.
\eeq
To simplify notation, we have defined
\beq
t \, \equiv \, \tan(w).
\eeq
In the last equality above we have used eq.(\ref{eq_R_fromPandQ})
in Appendix$\,\,$\ref{app_tang_mult_ang}.
The explicit expressions of the polynomials
$P_{2\lfloor(N-1)/2\rfloor}(t)$ and
$Q_{2\lfloor N/2 \rfloor }(t)$ 
are given by eqs.(\ref{eq_P_polin})
and (\ref{eq_Q_polin})
respectively.
For even $N$, the numerator of the rational function (in $t$)
on the last member of
eq.(\ref{eq_def_S_N}) has degree $N-1$,
while the denominator has degree $N$.
For odd $N$, the degrees are reverted:
The numerator has degree $N$,
while the denominator has degree $N-1$.
Therefore, for any $N \in \NN$,
also $S_N(t)$ is a rational function of degree $N$ in 
its argument $t$, like $R_N(t)$.

Eq.(\ref{eq_after_red}) can be rewritten as a general polynomial
equation in $t$ of order $N$,
with coefficients $c_i$ dependent on $z \, w$
(the latter quantity is assumed to be known):
\beq
p_N(t) \, \equiv \, 
z w \left[ P_{2\lfloor(N-1)/2\rfloor}(t) 
\, + \, Q_{2\lfloor N/2\rfloor }(t) \right]
\, - \, t \, P_{2\lfloor(N-1)/2\rfloor}(t)
\, = \,
\sum_{i=0}^N c_i \, t^i \, = \, 0.
\eeq
All the coefficients $c_i=c_i(zw)$ of 
$p_N$ are real for $z w\in\RR$, i.e. for a real coupling $z$
(the physical case).
It is immediate to check that,
for any $N \in \NN$, the polynomial $p_N=p_N(t)$
has degree $N$.
For $N \le 4$, i.e. in practice in the cases $N=2,3,4$
(the case $N=1$ is, of course, trivial),
the above equation can be explicitly solved 
with respect to $t$
in terms of nested radicals 
--- involving square roots and cubic roots ---
and rational operations.
For $N>4$, because of Galois theory, it is instead
in general impossible to solve the above equation
in terms of radicals;
In the latter case, one is forced to use special 
functions.
That makes the cases $N>4$, in practice, 
much less elementary and tractable than 
the $N \le 4$ ones.

Because of the fundamental theorem
of algebra, the above equation, namely
\beq
p_N(t) \, = \, 0
\eeq
has in general $N$ solutions, usually called zeroes
or roots, $t_0,t_1,\cdots,t_{N-1}$,
for any specified value of the coefficients 
$c_0(zw),$ $c_1(zw), \cdots, c_N(zw)$.
By varying $zw$, 
the zeroes, $t_0=t_0(zw)$, $t_1=t_1(zw)$, $\cdots$, 
$t_{N-1}=t_{N-1}(zw)$, also vary,
so that we may think that the
above equation implicitly defines
a multi-valued function $t=t(zw)$.
Let us choose one branch among its $N$ branches,
$t=F_0(zw),t=F_1(zw), \cdots, t=F_{N-1}(zw)$,
let's say $F_l(zw)$, so that:
\beq
\label{eq_formal_inverse}
\tan(w) \, = \, F_l (z \, w).
\eeq
Formally, we may write:
\beq
F_l \, \in \, \left\{ \, S_N^{-1} \, \right\};
\qquad
l = 0,1,2,\cdots,N-1.
\eeq
For $N \le 4$, the branches 
$F_0, \, F_1, \, \cdots$, $F_{N-1}$ originate from 
properly cutting the complex planes where
the square roots and cubic roots of the formulae
for the zeroes, are defined.

Eq.(\ref{eq_formal_inverse}) can also be
derived  by taking the formal inverse of $S_N$
on both sides of eq.(\ref{eq_after_red}),
i.e. by applying the function $S_N^{-1}$
to both sides of this equation.
Since, in general, the function $S_N(t)$ is
not one-to-one, the inverse
function $S_N^{-1}$ is multi-valued.
Generally, $S_N$ is a $N \mapsto 1$ map,
i.e. $N$, in general distinct, $\tan(w)$'s 
have the same image under $S_N$.
That implies that $S_N^{-1}$ has 
$N$ different branches, corresponding
to the $N-1$ non-resonant levels for each
resonant (and exceptional) level.
The branch $F_l$ of $S_N^{-1}$, 
which we have selected, 
is fixed from now on.


\subsection{Resummation by recursion}
\label{eq_gen_meth_recursion}

Now it comes the fundamental decomposition
of the particle momentum $w$, which allows
substantial simplifications due to the 
periodicity of the tangent:
\beq
\label{eq_decomp_w}
w \, = \, w_\nu \, \equiv \, 
\nu \, + \, \Delta \, w_\nu;
\eeq
where:
\beq
\nu \, \equiv \, \pi \, n;
\qquad n \, = \, 1, 2, 3, \cdots.
\eeq
By taking indeed into account the periodicity of the tangent, 
eq.(\ref{eq_formal_inverse}) is transformed into
the equation
\beq
\label{eq_formal_inverse_bis}
\tan(\Delta w_\nu)
\, = \, F_l\left( \eta \, + \, z \, \Delta w_\nu \right);
\eeq
where we have defined the effective coupling
\beq
\eta \, \equiv \, \nu \, z.
\eeq
Note that eq.(\ref{eq_formal_inverse_bis})
does not explicitly depend on $\nu$ which,
as we are going to show later in this section,
is the largest quantity in the problem.

By taking the formal inverse of the tangent
on both sides of eq.(\ref{eq_formal_inverse_bis}), 
i.e. by applying the arctangent function
on both sides, one obtains the equation:
\beq
\label{eq_Deltaw_ini}
\Delta w_\nu \, = \, 
H_{0,l}\left( \eta \, + \, z \, \Delta w_\nu \right);
\eeq
where we have defined the function
\beq
H_{0,l}(x) \, \equiv \, 
\arctan_0 \left[ F_l(x) \right].
\eeq
To obtain a well-defined equation,
one has to select an (arbitrary) branch
of the arctangent function.
We have assumed the branch with image,
for a real argument,
in the interval $(-\pi/2,+\pi/2)$,
which we have denoted $\arctan_0$,
and which is also fixed from now on.
Note that we have already implicitly taken into account 
the multi-valued character
of the arctangent function by writing $w$
in the form (\ref{eq_decomp_w}) and also by going from 
eq.(\ref{eq_formal_inverse})
to eq.(\ref{eq_formal_inverse_bis}).
To simplify the notation,
let us drop both indices $0$ and $l$ in $H_{0,l}$, 
so that, from now on, $H_{0,l} \mapsto H$.

Let us observe that, in the momentum levels
$k = k_{n+l/N}(z)$, 
the principal index $n=1,2,3,\cdots$
is related to the infinite multivaluedness
of the general arctangent function,
while the sub-index $l=-\lfloor (N-1)/2 \rfloor, \cdots,
-1,0,+1,\cdots, + \lfloor N/2 \rfloor$
is related
to the finite multivaluedness, "$N$-valuedness", 
of the function $S_N^{-1}$.

According to eq.(\ref{eq_decomp_w}),
the total momentum $w_\nu$ of the particle
is written:
\beq
\label{eq_ricor_Heq0}
w_\nu \, = \, 
\nu \, + \, 
H\left( \eta \, + \, z \, \Delta w_\nu \right).
\eeq
Let us now assume to be at high energies, i.e.
at high momenta:
\beq
n \, \gg \, 1,
\eeq
or, equivalently:
\beq
\nu \, \gg \, 1,
\eeq
as well as in the weak-coupling regime:
\beq
|z| \, \ll \, 1.
\eeq
We also assume that $\eta$, the product of the above
variables $\nu$ and $z$, is of order one, 
or somewhat larger than one:
\beq
\left| \eta \right| \, \gsim \, 1.
\eeq
In the above region, 
if the function $H$ vanishes at the origin, i.e.
\beq
H(0) \, = \, 0,
\eeq
as it happens for the resonant levels,
it makes sense to 
construct the following recursion equation
out of eq.(\ref{eq_Deltaw_ini}):
\beq
\Delta w_\nu^{(h+1)}(z) \, = \, 
H\big[ 
\eta \, + \, z \, \Delta w_\nu^{(h)}(z) 
\big],
\eeq
with the initial condition (at $h=0$): 
\beq
\label{eq_recur_initial_cond}
\Delta w_\nu^{(0)}(z) \, \equiv \, 0.
\eeq
We had to assume $H(0)=0$, because the initial 
condition, accompanying the recursion equation 
in $h\mapsto h+1$,
assigns to the initial momentum shift 
$\Delta w_\nu^{(0)}(z)$, at $h=0$, its free-theory value
($\Delta w_\nu(z=0)=0$).
If the function $H$ does not vanish
at the origin, 
as it happens for the non-resonant levels,
we redefine $H$
by subtracting its value in zero:
\beq
\qquad\qquad
H(x) \,\,\, \mapsto \,\,\, \tilde{H}(x) 
\, \equiv \, H(x) \, - \, H(0)
\qquad\qquad (H(0)\ne 0).
\eeq
Eq.(\ref{eq_ricor_Heq0}) for the particle momentum
$w = w_{\tilde{\nu}}$ is then rewritten:
\beq
\label{eq_ricor_Hnoneq0}
w_{\tilde{\nu}} \, = \, 
\tilde{\nu} \, + \, \delta w_{\tilde{\nu}}
\, = \, \tilde{\nu} \, + \,
\tilde{H}\left( \tilde{\eta} \, + \, z \, \delta w_{\tilde{\nu}} \right),
\eeq
where:
\beq
\tilde{\nu} \, \equiv \, \nu \, + \, H(0)
\eeq
and
\beq
\tilde{\eta} \, \equiv \, z \, \tilde{\nu}
\, = \, z \left[ \nu \, + \, H(0) \right].
\eeq
Let us look at the first few recursions
(we do not write the tilde's, but we consider
both cases $H(0)=0$ and $H(0) \ne 0$).
The first recursion gives:
\beq
\label{eq_first_recur_GT}
\Delta w_\nu^{(1)} \, = \, 
H\left( \eta \, + \, z \, \Delta w_\nu^{(0)} \right)
\, = \, H\left( \eta \right), 
\eeq
where, on the last member,
we have imposed the vanishing initial condition, 
eq.(\ref{eq_recur_initial_cond}).

If we consider the total particle momentum $w_\nu$,
we see that the first recursion simply adds 
$H(\eta) = \mathcal{O}(1)$
to the free-theory momentum $\nu \gg 1$:
\beq
h: 0 \, \mapsto \, 1
\qquad \Rightarrow \qquad
w_\nu^{(0)} \, = \, \nu 
\quad \mapsto \quad
w_\nu^{(1)} \, = \, \nu 
\, + \, H\left(\eta \right).
\eeq
Therefore the first recursion is equivalent
to the following replacement rule:
\beq
\label{rule_substitue_0}
\nu
\quad \mapsto \quad
\nu \, + \, H\left(\eta \right).
\eeq
Let's now consider the second recursion,
which is given by:
\beq
\label{eq_second_recursion_expl}
\Delta w_\nu^{(2)} \, = \, 
H\left(\eta \, + \, z \, \Delta w_\nu^{(1)} \right)
\, = \, 
H\big[ \,
\eta \, + \, z \, H\left( \eta \right) 
\big];
\eeq
where, on the last member, we have replaced
the first recursion, i.e. we have used 
eq.(\ref{eq_first_recur_GT}).
Therefore, going from the first recursion to the second one
involves the "transition"
\beq
\Delta w_\nu^{(1)} \, = \, H\left( \eta \right)
\quad \mapsto \quad
\Delta w_\nu^{(2)} \, = \, 
H\left[\eta \, + \, z \, H \left(\eta \right) \right].
\eeq
The second recursion step ($h=1 \mapsto 2$)
is then equivalent to the following substitution
inside the argument of the function $H(\,\cdot\,)$:
\beq
\label{eq_basic_replace}
\eta
\quad \mapsto \quad
\eta \, + \, z \, H \left(\eta \right).
\eeq
Note that the above rule (\ref{eq_basic_replace}) simply
comes from multiplying by $z$
on both sides the basic
replacement rule (\ref{rule_substitue_0}).

In order to understand the functional character
of the general recursion equation above,
let us define the function
\beq
\label{eq_def_sigma}
\sigma(x) \, \equiv \, 
\eta \, + \, z \, H(x) .
\eeq
It is immediate to check that
the second recursion can be written 
in terms of $\sigma(x)$ evaluated at the point
$x=\eta$, as:
\beq
\Delta w_\nu^{(2)} \, = \, H\big[ \sigma(\eta) \big].
\eeq
In order to find some general rules in the
recursions, i.e. some mathematical structures produced
by the repeated recursions --- if any ---,
let us work out the third recursion:
\beq
\label{eq_Delw_3_Gen}
\Delta w_\nu^{(3)} \, = \, 
H\left(\eta \, + \, z \, \Delta w_\nu^{(2)} \right)
\, = \, 
H\Big\{ \eta \, + \, z \, H\big[\eta \, + \, z \, H\left(\eta \right) \big] \Big\}.
\eeq
Therefore, with the third recursion step
$(h=2 \mapsto 3)$, $\Delta w_\nu$, the momentum shift 
from the free value, is modified as:
\beq
\label{eq_compare_gen_h2to3}
\Delta w_\nu^{(2)} \, = \, 
H\left[ \eta \, + \, z \, H \left(\eta \right) \right]
\quad \mapsto \quad
\Delta w_\nu^{(3)} \, = \,
H\Big\{ 
\eta \, + \, z \, H\big[\eta \, + \, z \, H\left( \eta \right) \big] 
\Big\}.
\eeq
We may notice that, 
in the expression for $\Delta w_\nu^{(2)}$ 
above on the left,
the variable $\eta$ occurs two times, namely
inside the argument of the external function $H$,
\beq
\label{eq_first_occur}
\Delta w_\nu^{(2)} \, = \,  
H\left[ \eta \, + \, \cdots \right],
\eeq
and as the argument of the internal $H$ function, 
\beq
\label{eq_second_occur}
\Delta w_\nu^{(2)} \, = \,  
H\big[ \cdots H(\eta) \big].
\eeq
By going from  the second recursion
to the third one, one may note that 
the first occurrence
of $\eta$ on the left of
(\ref{eq_compare_gen_h2to3})
(see eq.(\ref{eq_first_occur}))
is not modified.
In the second occurrence,
(see eq.(\ref{eq_second_occur})),
the variable $\eta$ is instead transformed 
according to rule
(\ref{eq_basic_replace}),
which we have already used 
in going from $h=1$ to $h=2$.
In other words, by going from the second
recursion to the third one, the variable
$\eta$ is replaced according to the 
rule (\ref{eq_basic_replace})
{\it only} in the innermost place.

In terms of the function $\sigma$
defined in eq.(\ref{eq_def_sigma}),
it is easy to check that the third 
recursion can be written:
\beq
\Delta w_\nu^{(3)} \, = \, 
H\Big\{ \sigma \big[ \sigma(\eta) \big] \Big\}.
\eeq
The composition of $\sigma$
with itself, evaluated at the point $x=\eta$, 
is the argument of $H$. 
 
We may generalize the above observations
by conjecturing that any recursion 
step ($h \mapsto h+1$) is obtained
by shifting $\eta$ 
according to the rule (\ref{eq_basic_replace})
in the innermost place,
i.e. inside the innermost function $H$.
By explicitly looking at higher order recursions,
we find that the above conjecture is true, namely:
\beq
h \, \mapsto \, h + 1 
\quad
\Rightarrow
\quad
H\Big(
\cdots,H\big(\cdots, H(\eta) \big) 
\Big) 
\, \mapsto \, 
H\Big(
\cdots,H\big(
\cdots, H(\eta \, + \, z \, H\left(\eta \right)) 
\big) 
\Big) .
\eeq
Furthermore, the momentum correction 
$\Delta w_{\nu}^{(h)}$
can be written, in general,
in terms of the function $\sigma$ as:
\beq
\label{eq_sigma_to_Deltaw_general}
\Delta w_\nu^{(h)} \, = \, 
H\big[ \sigma^{(h-1)}(\eta) \big],
\qquad h = 2,3,4, \cdots;
\eeq
where $\sigma^{(i)}$ is the composition of
the function $\sigma$ with itself $i$ times.
By convention,
\beq
\sigma^{(1)}(x) \, \equiv \, \sigma(x).
\eeq
By defining the zero-index function 
$\sigma^{(0)}$
as the identity function,
\beq
\sigma^{(0)}(x) \, \equiv \, x,
\eeq
one may write also the first recursion 
using the general formula (\ref{eq_sigma_to_Deltaw_general}):
\beq
\Delta w_\nu^{(1)} \, = \, 
H\big[ \sigma^{(0)}(\eta) \big]
\, = \,
H(\eta).
\eeq


\subsection{Resummation by function series}
\label{eq_gen_meth_func_series}

The perturbative series for the particle
momentum $w=w_\nu(z)$ can also be (approximately)
resummed to all orders in $z$,
by means of a (truncated) series of
functions of $\eta$ multiplied
by progressively higher powers of $z$:
\beq
w_\nu^{N^{P}LO} \, = \, \nu \, + \, 
\sum_{i=0}^P z^i \, \varphi_{i+1}(\eta).
\eeq
In general, by expanding 
$\Delta w_\nu^{(P+1)} = \Delta w_\nu^{(P+1)}(z,\eta)$,
i.e. the solution of order $P+1$ of the recursive equation,
in powers of $z$ up to order $h=P$ included, 
and comparing with the r.h.s.
of the above equation, one obtains
the following expression for the $\varphi_i$'s:
\beq
\varphi_{i+1}(\eta)
\, = \,
\frac{1}{i!} 
\left[
\frac{\partial^{i}}{\partial z^{i}}
\Delta w_\nu^{(P+1)}(z,\eta)
\right]_{z=0},
\qquad i=0,1,2,\cdots,P.
\eeq
Note that the variables $z$ and $\eta$
are independent from each other.
By considering, for example, the third-order recursive
solution $w^{(3)}_\nu = w^{(3)}_\nu(z,\eta)$ 
given in eq.(\ref{eq_Delw_3_Gen}), 
we obtain for the first three functions, according to the above equation (with $P=2$):
\bea
\label{eq_gen_meth_func_series_low_ord}
\varphi_1(\eta) &=& H(\eta);
\nonumber\\
\varphi_2(\eta) &=& H(\eta) \, \frac{d H}{d\eta}(\eta)
\, = \, \frac{1}{2} \, \frac{d H^2}{d\eta}(\eta);
\\
\varphi_3(\eta) &=& 
\frac{1}{2} \, H^2(\eta) \, \frac{d^2 H}{d\eta^2}(\eta) 
\, + \, H(\eta) \left( \frac{dH}{d\eta}(\eta) \right)^2.
\nonumber
\eea


\section{The Triple Case $N=3$}
\label{sec_triple_case}

In this section we consider Winter model
at finite volume in the case $N=3$,
in which the right cavity $[\pi,4\pi]$ (the large one) 
is three times larger than the left one $[0,\pi]$
(the small one).
As we are going to show, this case can be reasonably 
considered a "large-$N$" case.

\begin{figure}[ht]
\begin{center}
\includegraphics[width=0.5\textwidth]{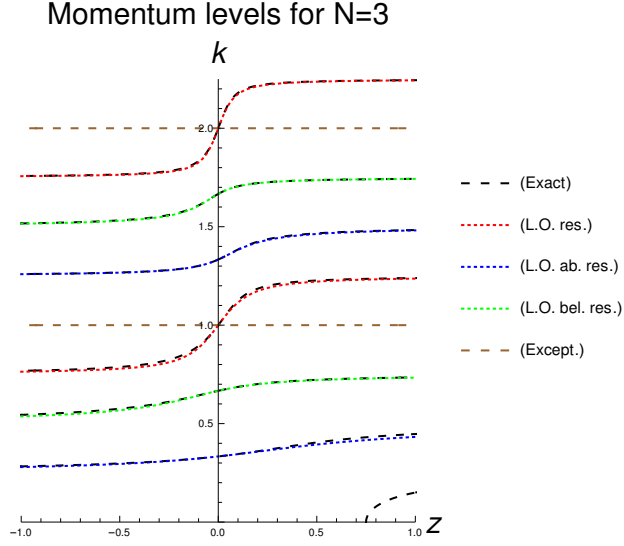}
\footnotesize
\caption{
\label{fig_plotNeq3}
\it
Lowest eight momentum levels for the case $N=3$.
The six exact normal momenta, 
$k_{1/3}(z)$, $k_{2/3}(z)$, $k_1(z)$,
$k_{4/3}(z)$, $k_{5/3}(z)$, $k_2(z)$,
are plotted as black dashed lines;
The two exceptional momenta $k_{exc} \equiv 1,2$
are plotted as brown dashed lines.
The L.O. resonant levels $k_1^{LO}(z)$ and $k_2^{LO}(z)$
are given by red dotted lines;
The L.O. non-resonant levels above the resonant ones,
$k_{1/3}^{LO}(z)$ and $k_{4/3}^{LO}(z)$, by
blue dotted lines;
Finally, the L.O. non-resonant levels below the resonant ones,
$k_{2/3}^{LO}(z)$ and $k_{5/3}^{LO}(z)$,
by green dotted lines.
The L.O. approximation works pretty well
for all the levels but the lowest-momentum level
$k_{1/3}(z)$ (the lower blue line), for which
there is a small discrepancy from the
exact level for $z \gsim 0.3$.
Taking into account that the latter is really a 
low-energy state, the overall agreement is quite 
satisfactory.
}
\end{center}
\end{figure}

In the free limit $z \to 0$, the momenta $k=k(z)$
of the particle become integer multiples of $1/N=1/3$:
\beq
\lim_{z\to 0} k(z) \,\,\, = \,\,\,
\frac{1}{3}, \,\, \frac{2}{3}, \,\,\, 1, \,\,\, \frac{4}{3}, 
\,\, \frac{5}{3}, \,\,\, 2, \,\,\, \frac{7}{3}, 
\,\, \frac{8}{3}, \,\, 3, \,\, \cdots.
\eeq
Therefore there are two contiguous non-resonant levels 
before (or after) each resonant (or exceptional) level
(see fig.$\,$\ref{fig_plotNeq3}).

The equation for the $\pi$-rescaled momenta
$w \equiv \pi \, k$ of the particle 
is simply obtained by setting $N=3$
in the general momentum equation (\ref{eq_basic}): 
\beq
\qquad\qquad\qquad\qquad
\frac{1}{\tan(w)} \, + \, \frac{1}{\tan(3w)} 
\, = \, \frac{1}{z \, w}
\qquad\qquad\qquad (N=3).
\eeq
By expressing the function $\tan(3w)$ as a rational
function (of third order) in $\tan(w)$,
as shown in appendix$\,$\ref{app_tang_mult_ang}, 
the above equation can be 
transformed into the equation:
\beq
\label{mom_eq_Neq3_rat_form}
\qquad\qquad\qquad
\frac{\tan(w) \big[ \tan^2(w) \, - \, 3 \big] }{
4 \, w \big[ \tan^2(w) \, - \, 1 \big]} \, = \, z
\qquad\qquad\qquad (N=3).
\eeq
In the free limit $z \to 0$, 
if the above equation is to be satisfied,
one of the factors
at the numerator of the rational function in $\tan(w)$ at its l.h.s., has to vanish.
There are three different possibilities:
\bea
1. && \tan(w) \, \to \, 0, \,\, w \neq 0 \,\, 
\Rightarrow \, \quad 
w \, = \, \pi \, n \, + \, \cdots
\,\,\,\qquad\qquad (n > 0);
\nonumber\\
2. && \tan(w) \, \to \, + \, \sqrt{3} 
\qquad \Rightarrow \quad 
w \, = \, \pi \left( n \, + \, \frac{1}{3} \right) 
\, + \, \cdots
\quad (n \ge 0);
\nonumber\\
3. && \tan(w) \, \to \, - \, \sqrt{3} 
 \qquad \Rightarrow \quad 
w \, = \, \pi \left( n - \frac{1}{3} \right) \, + \, \cdots
\,\,\quad (n > 0).
\eea
We have taken into account that, with our choice
of the arctangent branch, 
$\arctan\left(\pm\sqrt{3}\right) \, = \, \pm \, \pi/3$.
In the first case $(1.)$, 
we find the sequence of the resonant levels,
while in the second case $(2.)$ and in the third one $(3.)$,
we find the two sequences of non-resonant levels.

The momentum equation (\ref{mom_eq_Neq3_rat_form})
can be trivially transformed into the following
algebraic equation: 
\beq
\label{eq_mom_Neq3_pol_form}
t^3 \, - \, 3 \, \Omega \, t^2 
\, - \, 3 \, t \, + \, 3 \, \Omega \, = \, 0;
\eeq
where, to make the formulas more compact, we have defined:
\beq
t \, \equiv \, \tan(w).
\eeq
We have also introduced the convenient variable:
\beq
\Omega \, \equiv \, \zeta \, w;
\eeq
involving the effective coupling
\beq
\zeta \, \equiv \, - \, \left( 1 \, + \, \frac{1}{N} \right) g
\, = \, - \, \frac{4}{3} \, g
\qquad (N=3).
\eeq
In terms of the coupling $z\equiv -g$:
\beq
\zeta \, = \, \frac{4}{3} \, z.
\eeq
Note that eq.(\ref{eq_mom_Neq3_pol_form})
is a general, third-order algebraic equation
in $t$ with real coefficients.
By means of standard Cardano's formula
for third-order algebraic equations
(see Appendix \ref{app_Card_ord3}),
the explicit solutions of the above equation 
for $w=w_\nu$ can be written:
\beq
\label{Cardano_eq}
w_\nu \, = \, \nu \, + \,
\arctan
\bigg\{
\Omega \, + \, 
\exp\left(
- \, i \pi \, \frac{4 \, j \, + \, 1}{6}  \, \right) 
\, C_+(\Omega) \, + \, 
\exp\left(
 + \, i \pi \, \frac{4 \, j \, + \, 1}{6} 
\right) \, C_-(\Omega)
\bigg\}; \qquad
\eeq
where: 
\beq
\nu \, \equiv \, \pi \, n
\eeq 
and we have defined the (nested) radicals:
\beq
C_\pm(\Omega) \, \equiv \, 
\Big[
R(\Omega) \, \pm \, i \, \Omega^3 
\Big]^{1/3};
\eeq
with:
\beq
R(\Omega) \, \equiv \,
\Big( 1 \, + \, 3 \, \Omega^2 \, + \, 3 \, \Omega^4 \Big)^{1/2}.
\eeq
Note that the index $j$ is defined modulo three,
so we can limit ourselves, for example,
to consider the values:
\beq
j \, = \, 0, \, 1, \, 2.
\eeq
Note also that, with our branch of the cubic root:
\beq
\overline{C_\pm(\Omega)} \, = \, C_\mp(\Omega),
\qquad \Omega \in \RR,
\eeq
where a bar above a quantity denotes
its complex conjugation.
Different multi-valued functions
do appear in eq.(\ref{Cardano_eq})
so, in order to avoid ambiguities, 
let us specify in detail their branches:
\begin{enumerate}
\item
The square root in the function $R(\Omega)$
is the arithmetical (i.e. positive)
one: $1^{1/2} \, \equiv \, 1$.
That is consistent because, for any real value of $\Omega$ (the physical case),
the argument of the above square root is always positive 
(actually, it is greater than one).
In general, we consider the following branch of the 
complex square root. By writing a (non-zero) complex number $\xi$
in the polar form 
\beq
\label{eq_xi_pol_form}
\xi \, = \, \rho \, e^{i\theta}
\quad \mathrm{with} \quad
\rho \, > \,  0 
\quad \mathrm{and} \,\,\,
- \, \pi \, < \, \theta \, \le \, + \, \pi; 
\eeq
our square-root branch is defined by:
\beq
\xi^{1/2} \, \equiv \, 
\sqrt{\rho} \, e^{i\theta/2};
\eeq
\item
The choice of the branch of the cubic root
appearing in the definition of the functions $C_\pm(\Omega)$,
i.e. of the function $f(\xi) \equiv \xi^{1/3}$,
is similar to the one of the square root. 
We cut the complex $\xi$-plane along the negative 
part of the real axis, namely for $\xi \le 0$, 
and we select the branch which is real
on the positive axis: $1^{1/3} \equiv 1$.
By writing the independent variable $\xi$ 
in polar form as in eq.(\ref{eq_xi_pol_form}),
the cubic root branch in defined by:  
\beq
\xi^{1/3} \, \equiv \,
\sqrt[3]{\rho} \, e^{i\theta/3}.
\eeq
\item
We select the usual branch of
the $\arctan$ function, 
which is continuous and odd for real argument,
vanishes at the origin, $\arctan(0)=0$, 
and has image, again for real argument, in the
interval $(- \pi/2,\, + \pi/2)$.
\end{enumerate}
Let us now comment on our main
result, namely eq.(\ref{Cardano_eq}).
For each value of the index $j=0,1,2$,
the solution $w_\nu$ is real for any real value of
$\Omega$.
That is  because the second and third
terms inside the curly bracket on the r.h.s. of 
eq.(\ref{Cardano_eq}), are complex conjugate
of each other.
Therefore there are always three real and distinct roots. 
That is in complete agreement with physics:
We know that, for any choice of the integer 
$n = 1,2,3,\cdots$ in eq.(\ref{Cardano_eq}), 
there must be three real
momentum values $w = w_\nu^j(\zeta)$
for any value of the coupling
$\zeta \in \RR$.
Note also that all the radicals 
we have introduced are normalized
to one at $\Omega=0$, i.e. in the free limit:
\beq
C_\pm(0) \, = \, R(0) \, = \, 1.
\eeq
For $j=2$, the phase factors
in front of the cubic roots in eq.(\ref{Cardano_eq})
\beq
\exp\left(\pm \, i \pi \, \frac{4 \, j \, + \, 1}{6} \right) 
\,\, \mapsto \,\, 
\exp\left( \pm \, i \, \frac{3 \, \pi}{2} \right) 
\, = \, \mp \, i
\qquad (j=2),
\eeq
i.e. they become opposite of each other.
That implies that: 
\beq
\lim_{\Omega \to 0} w_\nu|_{j=2} \, = \, \nu,
\eeq
so that the index choice $j=2$ gives the resonant levels.
The remaining index choices $j=0,1$ give the non-resonant levels, as:
\bea
\lim_{\Omega \to 0} w|_{j=0} &=& 
\pi \left( n \, + \, \frac{1}{3} \right) 
\quad \Rightarrow \quad l \, = \, + \, 1;
\nonumber\\
\lim_{\Omega \to 0} w|_{j=1} &=& 
\pi \left( n \, - \, \frac{1}{3} \right)\quad \Rightarrow \quad l \, = \, - \, 1. 
\eea
The high-energy expansions of the resonant and non-resonant momenta are discussed in the following sections.

%
%


\subsection{Resonant levels}

By setting $j=2$ in the general eq.(\ref{Cardano_eq}), one obtains the following
equation for the resonant particle momenta: 
\beq
\label{Cardano_eq_2}
w_\nu \, = \, \nu \, + \, H\left( \Omega_\nu \right);
\eeq
where we have defined the function
\beq
\label{eq_def_fun_H_reson_Neq3}
H(x) \, \equiv \,
\arctan\Big[ x \, + \, i \, C_+\left( x \right) 
\, - \, i \, C_-\left( x\right) \Big].
\eeq
The variable $\Omega_\nu$ is the value taken by
$\Omega$ for $w=w_\nu$:
\beq
\Omega_\nu \, \equiv \, \zeta \, w_\nu \, = \, 
\nu \, \zeta \, + \, \zeta \, \Delta w_\nu;
\eeq
as $w_\nu = \nu + \Delta w_\nu$.
Note that:
\beq
H(0) \, = \, 0,
\eeq
as it should. Similarly to previous cases,
we can solve the above equation by
means of a function-series expansion
or by means of a recursion equation, 
as described in the next two sub-sections.


\subsubsection{Resummation by means of a function series}

The resonant levels $w=w_\nu(\zeta,\omega)$ posses a
function-series expansion of the form:
\beq
w_\nu(\zeta,\omega) \, = \, \nu \, + \, \sum_{j=0}^\infty
\zeta^j \, \varphi_{j+1}(\omega) 
\, = \,  \nu \, + \, \varphi_1(\omega) \, + \, 
\zeta \, \varphi_2(\omega)
\, + \, \cdots;
\qquad \nu = \pi, \, 2 \pi, \, 3\pi, \, \cdots;
\eeq
where we have defined:
\beq
\omega \, \equiv \, \nu \, \zeta. 
\eeq
The Leading-Order (L.O.) function reads:
\beq
\label{eq_LO_Neq3_funt_series}
\varphi_1(\omega) \, = \, H(\omega);
\eeq
where the function $H$ has been defined in 
eq.(\ref{eq_def_fun_H_reson_Neq3}).
The computation of higher-order functions is 
straightforward and is easily made, for example, by
following the general method described in 
section$\,$\ref{eq_gen_meth_func_series}.
By substituting $\eta$ with $\omega$ (as $z$ is
replaced by $\zeta$)
in eq.(\ref{eq_gen_meth_func_series_low_ord}), one obtains for the second-order function and the third-order one respectively:
\bea
\varphi_2(\omega) &=& 
\frac{1}{2} \, \frac{d H^2}{d\omega}(\omega);
\nonumber\\
\varphi_3(\omega) &=& 
\frac{1}{2} \, H^2(\omega) \, \frac{d^2 H}{d\omega^2}(\omega) 
\, + \, H(\omega) \left[ \frac{dH}{d\omega}(\omega) \right]^2.
\eea


\subsubsection{Resummation by recursion}

In the recursive scheme, the resonant
particle momentum is written:
\beq
w_\nu \, = \, \nu \, + \, \Delta w_\nu(\zeta,\omega).
\eeq
The momentum correction 
$\Delta w_\nu(\zeta,\omega)=\mathcal{O}(1)$ to the (large) free-theory value $\nu \gg 1$
is obtained as the limit of a function sequence
$\left\{\Delta w_\nu^{(h)}(\zeta,\omega); \,\, h=0,1,2,3,\cdots \right\}$:
\beq
\Delta w_\nu(\zeta,\omega) \, = \, 
\lim_{h\to\infty} \Delta w_\nu^{(h)}(\zeta,\omega).
\eeq
The function sequence is constructed by
means of the following recursion equation:
\beq
\Delta w_\nu^{(h+1)}(\zeta,\omega)
\, = \, H\left[  
\omega \, + \, \zeta \, \Delta w_\nu^{(h)}(\zeta,\omega)
\right];
\eeq
with the initial condition, at $h=0$:
\beq
\Delta w_\nu^{(0)}(\zeta,\omega) \, \equiv \, 0.
\eeq
The first recursion ($h=0\mapsto 1$)
gives for the complete momentum:
\beq
w_\nu^{(1)}(\zeta,\omega) \, \equiv \, \nu \, + \, 
\Delta w_\nu^{(1)}(\zeta,\omega) 
\, = \, \nu \, + \, H(\omega);
\eeq
where the function $H$ has been defined in 
eq.(\ref{eq_def_fun_H_reson_Neq3}).
The above first-order result, which is $\zeta$ independent,
is in complete agreement
with the L.O. approximation for $w_\nu$ in the
function-series scheme, 
eq.(\ref{eq_LO_Neq3_funt_series}). 
By substituting $\eta$ with $\omega$ and $z$ with $\zeta$
in eq.(\ref{eq_second_recursion_expl}) 
and (\ref{eq_Delw_3_Gen}), one obtains for the second and third recursions respectively:
\bea
w_\nu^{(2)}(\zeta,\omega) &=& \nu \, + \, 
H\big[ \omega \, + \, \zeta \, H(\omega) \big];
\nonumber\\
w_\nu^{(3)}(\zeta,\omega) &=& \nu \, + \, 
H\Big\{ \omega \, + \, \zeta \, 
H\big[\omega\, + \, \zeta \, H(\omega) \big] 
\Big\}.
\eea


\subsection{Non-resonant levels above the resonant ones $(l=+1)$}

In this section we consider the (approximate) resummation,
to all orders in the coupling $\zeta$, of the perturbative series of the
non-resonant momentum levels right above the resonance ones.
These are the momentum levels with the principal index
$n = 0,1,2,3,\cdots$ and the sub-index $l=+1$,
of the form:
\beq
w \, = \, w_{\hat{\nu}}(\zeta,\hat{\omega}) 
\, = \, 
\hat{\nu}
\, + \, \delta w_{\hat{\nu}}(\zeta,\hat{\omega});
\eeq
where we have defined the new variable
\beq
\hat{\omega} \, \equiv \, \zeta \, \hat{\nu}
\eeq
and the shifted index
\beq
\hat{\nu} \, \equiv \,
\pi \left( n \, + \, \frac{1}{3} \right).
\eeq
As in the non-resonant case of the $N=2$ model
previously considered, resummation is made
either by means of a function series expansion
or by means of a recursion equation.


\subsubsection{Resummation by function series $(l=+1)$}

For the non-resonant levels right
above the resonant ones, i.e. with $l=+1$,
the function-series expansion holds:
\beq
w_{\hat{\nu}}( \zeta,\hat{\omega} ) \, = \, 
\hat{\nu}
\, + \, \sum_{j=0}^\infty 
\zeta^j \, \psi_{j+1}\left( \hat{\omega} \right)
\, = \, \hat{\nu}
\, + \, \psi_1\left( \hat{\omega} \right) 
\, + \, \zeta \, \psi_2\left( \hat{\omega} \right)
\, + \, \cdots.
\eeq
The Leading-Order (L.O.) function explicitly reads:
\beq
\psi_1(\hat{\omega}) \, = \, 
\hat{H}\left( \hat{\omega} \right);
\eeq
where we have defined the function
\beq
\hat{H}\left( x \right)
\, \equiv \,
\arctan\left[ \, x \, + \, 
\frac{\sqrt{3} \, - \, i}{2} \, C_+\left(x\right) \, + \,
\frac{\sqrt{3} \, + \, i }{2} \, C_-\left(x\right)
\right]
\, - \, \frac{\pi}{3}.
\eeq
Note that $\hat{H}\left( x \right)$ is real for a real 
argument $x$ (the physical case) and that:
\beq
\hat{H}\left( 0 \right) \, = \, 0.
\eeq
Higher-order functions 
$\psi_2(\hat{\omega})$, $\psi_3(\hat{\omega})$, $\cdots$
are easily calculated by following the general method 
described in section$\,$\ref{eq_gen_meth_func_series}.
In the formulae in the latter section, one
has to make the following replacements:
$\nu \mapsto \hat{\nu}$, $z \mapsto \zeta$,
$\eta \mapsto \hat{\omega}$ and $H \mapsto \hat{H}$. 

It is remarkable that the above, L.O. formula
provides a fairly good approximation to the
exact momentum levels also at $n=0$, i.e. to $k_{1/3}(\zeta)$
(the lower blue dotted line in fig.$\,$\ref{fig_plotNeq3}).


\subsubsection{Resummation by recursion $(l=+1)$}

In the recursive scheme, the $l=+1$ non-resonant 
particle momenta are written:
\beq
w \, = \, w_{\hat{\nu}}\left(\zeta,\, \hat{\omega}\right) 
\, = \, \hat{\nu} 
\, + \, \delta w_{\hat{\nu}}\left(\zeta,\, \hat{\omega}\right).
\eeq
The non-resonant momentum correction,
$\delta w_{\hat{\nu}}\left(\zeta,\hat{\omega}\right)$,
is obtained as the limit of a function sequence
$\left\{\delta w_{\hat{\nu}}^{(h)}\left(\zeta,\hat{\omega}\right); 
\,\, h=0,1,2,3,\cdots \right\}$:
\beq
\delta w_{\hat{\nu}}\left( \zeta, \hat{\omega} \right) 
\, = \, 
\lim_{h\to\infty} \delta w_{\hat{\nu}}^{(h)}
\left( \zeta, \hat{\omega} \right).
\eeq
The function sequence is constructed by
means of the following recursion equation:
\bea
\delta w_{\hat{\nu}}^{(h+1)}\left( \zeta, \hat{\omega} \right)
\, = \, \hat{H}
\left[
\hat{\omega} \, + \, \zeta \, 
\delta w_{\hat{\nu}}^{(h)}\left( \zeta, \hat{\omega} \right)
\right];
\eea
with the initial condition, at $h=0$:
\beq
\delta w_{\hat{\nu}}^{(0)}\left( \zeta, \hat{\omega} \right) 
\, = \, 0.
\eeq
Higher-order recursions are easily computed by means
of the general method described in 
section$\,$\ref{eq_gen_meth_recursion}.
One has to change the symbols in the equations
of the latter section as described at the end
of the previous section.


\subsection{Non-resonant levels below the resonant ones $(l=-1)$}

In this section we consider the resummation
of the perturbative series of the non-resonant momentum levels 
right below the resonance ones, i.e.
the levels with the sub-index $l=-1$:
\beq
w \, = \, w_{\check{\nu}}(\zeta, \check{\omega}) \, = \, 
\check{\nu}
\, + \, \delta w_{\check{\nu}}(\zeta, \check{\omega});
\eeq
where we have defined the new variable 
\beq
\check{\omega} \, \equiv \, \zeta \, \check{\nu}
\eeq
and the new shifted index
\beq
\check{\nu} \, \equiv \,
\pi \left( n \, - \, \frac{1}{3} \right).
\eeq
Unlike the previous case $(l=+1)$,
in this case the principal-index value  $n=0$
has to be discarded, so that $n=1,2,3,\cdots$.
Apart from this detail,
this case closely parallels 
the previous one, so the discussion will be
very concise.


\subsubsection{Resummation by means of a function series $(l=-1)$}

The resummation of the non-resonant levels
below the resonant ones, i.e. with  $l=-1$, involves 
a function series-expansion of the form:
\beq
w_{\check{\nu}}(\zeta,\check{\omega}) \, = \, 
\check{\nu} \, + \, 
\sum_{j=0}^\infty 
\zeta^j \, \chi_{j+1}(\check{\omega}) .
\eeq
The L.O. function explicitly reads:
\beq
\chi_1\left( \check{\omega} \right) 
\, = \, \check{H}\left( \check{\omega} \right);
\eeq
where we have defined the function:
\beq
\check{H}\left( x \right)
\, \equiv \,
\arctan\left[ \, x \, - \, 
\frac{
\sqrt{3} \, + \, i}{2} \, 
C_+\left( x \right)
\, - \, \frac{\sqrt{3} \, - \, i}{2} \, 
C_-\left( x \right)
\right] \, + \, \frac{\pi}{3}.
\eeq


\subsubsection{Resummation by recursion $(l=-1)$}

In the recursive scheme, the non-resonant 
particle momenta with $l=-1$ are written:
\beq
w \, = \, w_{\check{\nu}} \left( \zeta, \check{\omega} \right)
\, = \, 
\check{\nu} \, + \, 
\delta w_{\check{\nu}}\left( \zeta, \check{\omega} \right) .
\eeq
The non-resonant momentum correction,
$\delta w_{\check{\nu}}\left( \zeta, \check{\omega} \right)$,
is obtained as the limit of a function sequence
$\left\{\delta w_{\check{\nu}}^{(h)}\left( \zeta, \check{\omega} \right); \,\, h=0,1,2,3,\cdots \right\}$:
\beq
\delta w_{\check{\nu}}\left( \zeta, \check{\omega} \right) \, = \, 
\lim_{h\to\infty} 
\delta w_{\check{\nu}}^{(h)}\left( \zeta, \check{\omega} \right).
\eeq
The function sequence is constructed by
means of the following recursion equation:
\bea
\delta w_{\check{\nu}}^{(h+1)}\left( \zeta, \check{\omega} \right)
\, = \, 
\check{H}\left[ \check{\omega} + \zeta \, \delta w_{\check{\nu}}^{(h)}\left( \zeta, \check{\omega} \right)  \right].
\eea
The initial condition to the non-resonant momentum shift, assigned at $h=0$, is, as usual, given by the vanishing, free-theory value:
\beq
\delta w_{\check{\nu}}^{(0)}\left( \zeta, \check{\omega} \right) 
\, = \, 0.
\eeq
 

\subsection{Limits of analytical computations}

In order to analytically evaluate the momentum spectrum
of finite-volume Winter model in the case $N=3$,
we had to solve a general, third-order
algebraic equation with real coefficients,
finding three real and distinct roots.
In algebra, this case is called
the irreducible one 
\cite{libroPC},
because it is not possible to express
the real roots in terms of radicals
involving real numbers only.
Even though the final results are all real,
it is necessary to
go through the complex numbers.
Any momentum is written indeed 
as the arctangent of a real quantity plus 
(two times) the real part of a truly 
complex number (see for example eq.(\ref{Cardano_eq_2})).
This situation is to be contrasted with
the one encountered in the previous case $N=2$,
where much simpler second-order equations
were involved and the
real roots could be written in terms
of radicals of real numbers.

Because of this mathematical irreducibility
phenomenon, the formulae for the particle momenta
in the case $N=3$ are much less readable than
the formulae in the case $N=2$.
As we have shown in the previous sections, 
in the case $N=3$ one has to (arbitrarily
but necessarily) fix branches of various
multi-valued functions, so that at the end
the intuitive content of the formulae is
drastically reduced with respect
to the case $N=2$.

As we are going to show in the next section,
the new mathematical situation which we have found by 
going from the case $N=2$ to the case $N=3$
persists --- and actually gets much worse --- in the case $N=4$,
where general algebraic equations 
with real coefficients of fourth order
have to be solved. 


\subsection{Discussion}

People say that in physics there is one, two, three,
infinity. By that, they mean that relatively small-$N$
cases, such as the case $N=3$, offer the possibility of 
describing some qualitative properties of
the $N=\infty$ case as well as, approximately,
some quantitative properties.

In the $N=3$ case, the density of states of the large cavity
is, in momentum space, three times bigger than that
of the small cavity, implying that
two contiguous resonant levels,
with indices $(n,l=0)$ and $(n+1,l=0)$, with $n=1,2,3,\cdots$,
are separated by a pair of
non-resonant levels, with indices $(n,l = +1)$ and $(n+1,l=-1)$.
A non-resonant $(n,l = +1)$ level, right above a 
resonant one, $(n,l = 0)$, has most of its variation
at positive couplings, $z>0$, where it exhibits 
a mild resonant behavior.
Therefore, as far as resonance dynamics is concerned,
a $(n,l=+1)$ level is naturally related to this 
lower, contiguous resonant level $(n,l=0)$.
Similarly, a non-resonant $(n,l=-1)$ level, right below 
a resonant one, $(n,l=0)$, has most of its variation at $z<0$, 
where it exhibits a mild resonant behavior.
Therefore the $(n,l-1)$ level is related to the contiguous,
upper resonant level, $(n,l=0)$.

The conclusion is that,
as far as resonance behavior is concerned, 
the momentum levels of the $N=3$ model 
naturally group into triplets, centered
around a resonant level.

If the case $N=3$ can be considered
a large-$N$ case, as we conjecture,
an initial box eigenfunction 
contained in the small cavity
--- the analog of the wavefunction 
in eq.(\ref{eq_wf_initial}) ---
should exhibit an approximate exponential decay 
with time in some intermediate temporal window.


\section{The Quadruple case $N=4$}
\label{sec_quadr_case}

In this section we consider Winter model
at finite volume for $N=4$, in which
the left cavity $[\pi,5\pi]$ (the large one) is
four times larger than the right one $[0,\pi]$
(the small one).
This is the largest-$N$ case which we can treat
using the fundamental algebraic simplification 
introduced in the case $N=2$
(involving the reduction
of the tangent of a multiple argument),
by means of elementary functions.

In the free-theory limit, $z \to 0$, 
the allowed momenta $k=k(z)$
of the particle tend to integer multiples of $1/N=1/4$,
\beq
k(z \to 0) \,\, = \,\, 
\frac{1}{4}, \,\, \frac{1}{2}, \,\, \frac{3}{4}, 
\,\, 1, \,\, \frac{5}{4}, 
\,\, \frac{3}{2}, \,\, \frac{7}{4}, \,\, 2, \,\, 
\frac{9}{4}, \,\, \frac{5}{2}, \,\, \frac{11}{4}, 
\,\, 3, \,\, \frac{13}{4}, \,\, \cdots.
\eeq
Therefore there are three contiguous non-resonant levels for
each resonant level (and exceptional level).

The equation in $\tan(w)$ determining the normal part of the momentum spectrum of the $N=4$ model reads:
\beq
\label{eq_mom_ord_Neq4}
\qquad\qquad
\frac{4 \, \tan(w) \, \Big[1 \, - \, \tan^2(w)\Big]}{w \,
\Big[\tan^4(w) \, - \, 10 \, \tan^2(w) \, + \, 5 \Big]}
\, = \, z \qquad\qquad (N=4).
\eeq
The denominator of the fraction on the r.h.s. of the above equation vanishes for:
\beq
\tan(w) \, = \, \pm \, \sqrt{5 \, \pm \, 2 \sqrt{5}},
\eeq
where the numerator does not vanish
(the two sign determinations in the above equation are 
independent from each other because a bi-quadratic 
equation has been solved).
Furthermore, the numerator vanishes at
$\tan(w)=0,\pm 1$, where the denominator does
not vanish.
Therefore the zeroes of the numerator
on the r.h.s. of eq.(\ref{eq_mom_ord_Neq4})
are zeroes of the complete expression.
In the free limit $z \to 0$, eq.(\ref{eq_mom_ord_Neq4})
is satisfied in the following four cases:
\bea
&& 1. \,\,\, \tan(w) \, \to \, 0, 
\,\, w \ne 0 \,\,\, \Rightarrow \quad 
w \, \to \, n \, \pi
\qquad\qquad\qquad (n \ge 1);
\nonumber\\
&& 2. \,\,\, \tan(w) \, \to \, \infty 
\,\quad\qquad\, \Rightarrow \quad 
w \, \to \, 
\left(n \, + \, \frac{1}{2}\right) \pi
\qquad\,\, (n \ge 0);
\nonumber\\
&& 3. \,\,\, \tan(w) \, \to \, + \, 1 
\qquad\quad \Rightarrow \quad w \, \to \,
\left(n \, + \, \frac{1}{4}\right) \pi
\qquad\,\, (n \ge 0);
\nonumber\\
&& 4. \,\,\, \tan(w) \, \to \, - \, 1 
\qquad\quad \Rightarrow \quad w \to 
\left(n \, - \, \frac{1}{4}\right) \pi
\qquad\,\,\,\, (n \ge 1).
\eea
In general, for even $N$, i.e. for 
\beq
N \, = \, 2 \, K, \qquad K \, \in \, \NN,
\eeq
there is a non-resonant momentum level
with $l=K$, having, for $z \to 0$, the semi-integer
limit 
\beq
k(z \to 0) \, = \, n \, + \, \frac{1}{2},
\eeq
and lying therefore exactly
at the middle of two consecutive 
resonant levels, $(n,l=0)$ and $(n+1,l=0)$.

Eq.(\ref{eq_mom_ord_Neq4}) is easily
transformed into the following monic
fourth-order algebraic equation in $\tan(w)$:
\beq
\tan^4(w) \, + \, \frac{4}{z w} \tan^3(w) \, - \, 10 \tan^2(w) 
\, - \, \frac{4}{z w} \tan(w) \, + \, 5 \, = \, 0.
\eeq
Note that the coupling $z$ appears in the coefficients
of the above polynomial always at the denominator,
making the study of the limit $z \to 0$ more
difficult than in the case $N=3$, where $z$ appears at the 
numerator (cfr. eq.(\ref{eq_mom_Neq3_pol_form})).

The usual decomposition of the particle
momenta $w=w_\nu$ reads:
\beq
\label{eq_Neq4_mom_solve}
w_\nu \, = \, \nu \, + \, \Delta w_\nu; 
\eeq
where:
\beq
\Delta w_\nu \, = \, f\left( z w_\nu \right).
\eeq
By exactly solving the above equation
in $\tan(w)$,
one obtains for the function 
$f = f_{\epsilon_1,\epsilon_2}$
the rather involved expression:
\beq
\label{eq_rather_involv}
f_{\epsilon_1,\epsilon_2}(\xi) 
\, \equiv \,
\arctan
\left\{
\frac{1}{\xi}
\left[
\epsilon_1 \, \Sigma^{1/2}(\xi) \, - \, 1
\, + \, \epsilon_2
\bigg(\,
3 \, + \, 5 \, \xi^2 \, - \, \Sigma(\xi)
\, - \, 2 \, \epsilon_1 \, 
\frac{ 1 \, + \, 2 \, \xi^2 }{
\Sigma^{1/2}(\xi) }
\bigg)^{1/2}
\right]
\right\}.
\eeq
The indices $\epsilon_1, \epsilon_2 = \pm 1$
are two independent sign determinations.
The branches of the square root, the cubic root 
and the arctangent function are the same as those
defined in sec.$\,$\ref{sec_triple_case}
(the case $N=3$).
The function $\Sigma(\xi) \equiv \xi^2 \, v_0/2$ is proportional to an (arbitrary) solution $v_0$ of the auxiliary 
third-order equation 
(see Appendix \ref{app_Card_ord4});
It is explicitly given by:
\beq
\Sigma(\xi)
\, \equiv \, 1 \, + \, \frac{5}{3} \, \xi^2
\, + \, \frac{\xi}{ \sqrt{3} }
\left[
\exp\left( + \, \frac{i \pi}{6} \right) 
F_+(\xi) \, + \, 
\exp\left( - \, \frac{i \pi}{6} \right) 
F_-(\xi)
\right].
\eeq
The functions $F_\pm(\xi)$ involve an overall cubic root
and differ from each other by a sign only:
\beq
F_\pm(\xi) \, \equiv \,
\left[  R(\xi) \, \mp \, \frac{i}{ 6 \sqrt{3} }
\, \xi\left( 25 \, \xi^2 \, + \, 18 \right)
\right]^{1/3}.
\eeq
The function $R(\xi)$ is the inner-most 
radical:
\beq
R(\xi) \, \equiv \,
\left(
 1 \, + \, 7 \, \xi^2 \, + \, 25 \, \xi^4
\, + \, \frac{125}{4} \, \xi^6    
\right)^{1/2}.
\eeq
Finally, the variable $\xi$ is the product of the
independent variable $z \equiv - g$ with the dependent
variable $w \equiv \pi \, k$:
\beq
\xi \, \equiv \, z \, w.
\eeq
By varying independently the signs $\epsilon_1$ 
and $\epsilon_2$ in the above formula,
one obtains four different levels for each value
of the principal index $n=1,2,3,\cdots$
--- namely one resonant level and three non resonant 
levels, as it should.
Note that the intermediate functions
(which we have introduced to have readable formulae),
are all normalized to one at $\xi=0$ (i.e. in the free limit):
\beq
\label{eq_Neq4_normal_interm}
R(0) \, = \, F_{\pm}(0) \, = \, \Sigma(0) \, = \, 1.
\eeq
Note also that:
\beq
\overline{ F_\pm(\xi) } \, = \, F_\mp(\xi)
\qquad (\xi \in \RR).
\eeq
The above relation implies that $\Sigma(\xi)$
is real for real $\xi$ (the physical case).

Because of eq.(\ref{eq_Neq4_normal_interm}),
we immediately find that, if $\epsilon_1=+1$,
the big round bracket
on the r.h.s. of eq.(\ref{eq_rather_involv})
exactly vanishes for $\xi \to 0$.
By using the small $\xi$ expansion for $\Sigma(\xi)$,
\beq
\Sigma(\xi) \, = \, 1 \, + \, \xi \, + \, 2 \, \xi^2 
\, + \, \mathcal{O}\left( \xi^3 \right),  
\eeq
as well as for $\sqrt{\Sigma(\xi)}$,
\beq
\sqrt{\Sigma(\xi)} \, = \, 1 \, + \, \frac{\xi}{2} 
\, + \, \frac{7}{8} \, \xi^2 
\, + \, \mathcal{O}\left( \xi^3 \right),  
\eeq
one finds that also the linear term in $\xi$
vanishes.
The round bracket is (approximately) equal to $\xi^2/4$ for small $\xi$:
\beq
\label{eq_BRB_correct}
\bigg(
3 \, + \, 5 \, \xi^2 \, + \, \cdots
\bigg)^{1/2}_{\epsilon_1=+1}
\, = \, 
\left( 
\frac{\xi^2}{4} \, + \, \mathcal{O}\left( \xi^3 \right) 
\right)^{1/2}
\, = \, \frac{\xi}{2}
\Big( 
1 \, + \, \mathcal{O}\left( \xi \right) 
\Big)^{1/2}.
\eeq
The last equality follows from the
fact that $\left( \xi^2 \right)^{1/2} = \xi$
for $\xi > 0$, so that, by analytic continuation,
$\left( \xi^2 \right)^{1/2} = \xi$ for any 
$\xi \in \CC$.
If $\epsilon_1=+1$,
because of a leading-order cancellation, 
also the function
in front of the big round bracket
vanishes linearly for $\xi \to 0$:
\beq
\Sigma^{1/2}(\xi) \, - \, 1
\, = \, 
\frac{\xi}{2}
\, + \, \mathcal{O}\left( \xi^2 \right).
\eeq
Therefore, if $\epsilon_1 = + \, 1$, 
the expansion in powers of $\xi$ of the argument of the arctangent
function, i.e. of the curly bracket
on the r.h.s. of eq.(\ref{eq_rather_involv}),
reads:
\beq
\label{eq_expansion_ep1eqp1}
\qquad
\Big\{ \cdots \Big\}_{\epsilon_1=+1}
\, = \, 
\frac{1}{\xi}
\Big[
\sqrt{\Sigma(\xi)} \, - \, 1 \, + \, \cdots
\Big] \, = \, 
\frac{1 \, + \, \epsilon_2}{2} 
\, + \, \frac{7 \, - \, 3 \, \epsilon_2}{8} \, \xi
\, + \, \mathcal{O}\left(\xi^2\right)
\qquad \left( \epsilon_1 = + 1 \right).
\eeq

\noindent
If $\epsilon_1=-1$, the big round bracket 
on the r.h.s. of eq.(\ref{eq_rather_involv})
is instead not vanishing in the free limit:
\beq
\lim_{\xi \to 0}
\Big( 3 + 5 \xi^2 + \cdots \Big)^{1/2}_{\epsilon_1=-1}
\, = \, + \, 2.
\eeq
Also the function in front of the big round bracket
is non-vanishing in the free limit:
\beq
\Big[ 
\epsilon_1 \, \Sigma^{1/2}(\xi) \, - \, 1 
\Big]_{\epsilon_1=-1}
\, = \,
 - \, \Sigma^{1/2}(\xi) \, - \, 1
\, \mapsto \, - \, 2
\qquad \mathrm{for} \,\, \xi \, \mapsto \, 0.
\eeq
Therefore, in the case $\epsilon_1=-1$,
the expansion for small $\xi$ of the curly bracket reads:
\beq
\label{eq_expansion_ep1eqm1}
\qquad
\Big\{ \cdots \Big\}_{\epsilon_1=-1}
\, = \,
\frac{2(\epsilon_2 \, - \, 1)}{\xi}
\, - \, \frac{\epsilon_2 \, + \, 1}{2}
\, + \, \frac{11 \epsilon_2 \, - \, 7 }{8} \, \xi
\, + \, \mathcal{O}\left(\xi^2\right)
\qquad\qquad \left( \epsilon_1 = - 1 \right).
\eeq
According to our general convention,
the non-resonant levels have the
following values of the sub-index $l$:
\beq
l \, = \, - \,  1, \,\, + \, 1, \,\, + \, 2.
\eeq
In the following sections we will
consider the above levels, by linking the
values of the index $l$ with the signs of 
$\epsilon_1$ and $\epsilon_2$. 

%
\begin{figure}[ht]
\begin{center}
\includegraphics[width=0.5\textwidth]{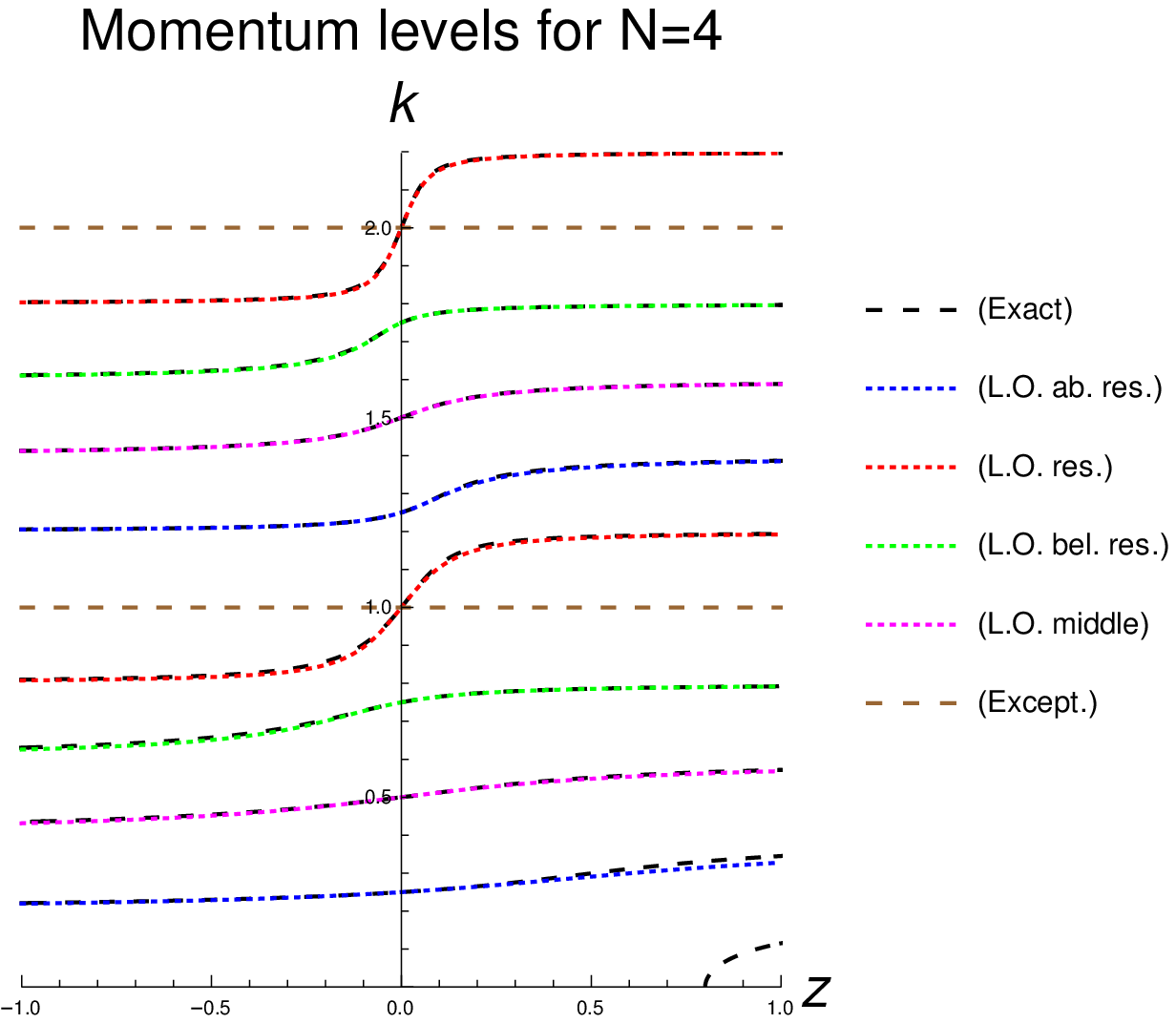}
\footnotesize
\caption{
\label{fig_plotNeq4}
\it
Lowest ten momentum levels for the case $N=4$.
The eight normal momentum levels 
$k_{1/4}(z)$, $k_{1/2}(z)$, $\cdots$, $k_2(z)$
are plotted as black dashed lines;
The two exceptional momenta $k_{exc} \equiv 1,2$ 
are plotted as brown dashed lines
The L.O. resonant levels, $k_1^{LO}(z)$ and $k_2^{LO}(z)$,  are given by red dotted lines;
The L.O. non-resonant levels above the resonant ones,
$k_{1/4}^{LO}(z)$ and $k_{5/4}^{LO}(z)$,
are given by blue dotted lines;
The L.O. non-resonant levels below the resonant ones,
$k_{3/4}^{LO}(z)$ and $k_{7/4}^{LO}(z)$,
are given by green dotted lines.
Finally, the L.O. non-resonant levels at the middle of two
resonant levels,
$k_{1/2}^{LO}(z)$ and $k_{3/2}^{LO}(z)$,
are given by magenta dotted lines.
The L.O. approximation works pretty well
for all the levels but the lowest-momentum one
$k_{1/4}(z)$ (the lower blue line), for which
there is a small discrepancy from the
exact level for $z \gsim 0.3$, as it was also 
the case with $N=3$.
}
\end{center}
\end{figure}
%


\subsection{Resonant levels}

By setting 
\beq
\epsilon_2 \, = \, - \, 1
\eeq
in eq.(\ref{eq_expansion_ep1eqp1}),
one obtains for the momentum shift:
\beq
\Delta w \, = \, 
\arctan\left[ \,
\frac{5}{4} \, \xi \, + \, \mathcal{O}\left(\xi^2\right)
\right]
\, = \,
\frac{5}{4} \, \xi \, + \, \mathcal{O}\left(\xi^2\right)
.
\eeq
Therefore, for
\beq
\epsilon_1 \, = \, + \, 1; 
\quad 
\epsilon_2 \, = \, - \, 1;
\eeq
the expansion for small coupling of the total momentum 
$w=w_\nu$
reads:
\beq
w_\nu \, = \, \nu \, + \, \frac{5}{4} \, \nu \, z 
\, + \, \mathcal{O}\left(z^2\right). 
\eeq
In the above equation we have taken into account that:
\beq
\xi \, = \, \xi_\nu \, = \, z \, w_\nu 
\, = \, z \, \nu \, + \, z \, \Delta w_\nu(z) 
\, = \, z \, \nu \, + \, \mathcal{O}\left(z^2\right).
\eeq
The above choice for the indices $\epsilon_1$ 
and $\epsilon_2$ then gives the resonant levels:
\beq
\epsilon_1 \, = \, + \, 1, 
\quad 
\epsilon_2 \, = \, - \, 1
\quad \Rightarrow \quad l \, = \, 0.
\eeq
Let us then define the function
\beq
\label{eq_Neq4_reson_case_ex}
H(\xi) \, \equiv \, f_{+1,-1}(\xi);
\eeq
with:
\beq
\label{eq_fp1m1}
f_{+1,-1}(\xi)
\, = \,
\arctan
\left\{
\frac{1}{\xi}
\left[
\Sigma^{1/2}(\xi) \, - \, 1
\, - \,
\bigg(
3 \, + \, 5 \, \xi^2 \, - \, \Sigma(\xi)
\, - \, 2 \, 
\frac{ 1 \, + \, 2 \, \xi^2 }{
\Sigma^{1/2}(\xi) }
\bigg)^{1/2}
\right]
\right\}.
\eeq
Note that:
\beq
H(0) \, = \, 0,
\eeq
as it should.
The (still exact) equation for the momentum shift $\Delta w = \Delta w_\nu$ is written:
\beq
\Delta w_\nu \, = \,
H( \eta \, + \, z \, \Delta w_\nu );
\eeq
where we have defined the variable:
\beq
\qquad\qquad\qquad
\eta \, \equiv \, \nu \, z \, = \, \pi \, n \, z;
\qquad\qquad\qquad
n = 1, 2, 3, \cdots.
\eeq
Note that:
\beq
\xi \, = \, \eta \, + \, z \, \Delta w_\nu,
\eeq
so that $\xi$ reduces to $\eta$
in the free limit $z \to 0$.

In terms of the complete momentum 
of the particle $w=w_\nu$,
the momentum equation above is rewritten:
\beq
w_\nu \, = \, 
\nu \, + \, \Delta w_\nu
\, = \, 
\nu \, + \, H( \eta \, + \, z \, \Delta w_\nu ),
\eeq
where the function $H$ has been defined
in eq.(\ref{eq_Neq4_reson_case_ex}).
The general theory of the solutions of the above
equation by means of a function series
or by means of a recursion equation
is treated in detail in section \ref{sect_gen_meth}, 
to which the reader is referred.

For concreteness sake, let us give the explicit expressions
of the three recursive solutions of lowest order
for the particle momentum:
\bea
w_\nu^{(1)} &=& \nu \, + \, H( \eta );
\\
w_\nu^{(2)} &=& \nu \, + \,  
H\left[ \eta \, + \, z \, H \left(\eta \right) \right];
\nonumber\\
w_\nu^{(3)} &=& \nu\, + \,
H\Big\{ 
\eta \, + \, z \, H\big[\eta \, + \, z \, H\left( \eta \right) \big] 
\Big\}.
\nonumber
\eea
The evaluation of higher-order recursions
is straightforward and is discussed, 
in a general form, in section$\,$\ref{eq_gen_meth_recursion}.
As in all previous cases,
one can also use the resummation scheme for the
particle momenta, which is based on a function series
expansion. 
In addition to a direct calculation,
the required functions can also be evaluated
by expanding the recursive solutions 
in powers of the coupling $z$, as described in
\ref{eq_gen_meth_func_series}.

Let us now discuss a problem occurring in the
numerical evaluation of the function $H(x)$,
which is needed for example for making 
a plot like fig.$\,$\ref{fig_plotNeq4}
(the red dotted curves).
The direct numerical evaluation of the
big round bracket on the r.h.s. of eq.(\ref{eq_rather_involv})
for $\epsilon_1=+1$ gives, for $|\xi|\ll 1$:
\beq
\Big( 3 \, + \, 5 \, \xi^2 \, + \, \cdots 
\Big)^{1/2}_{\epsilon_1=+1} 
\, = \, 
\left(
\frac{\xi^2}{4} \, + \, \mathcal{O}\left(\xi^3\right)
\right)^{1/2}
\, = \, 
\frac{|\xi|}{2} \, 
\Big( 1 \, + \, \mathcal{O}\left(\xi\right) \Big)^{1/2}.
\eeq
The result in the last member of the above
equation is incorrect for $\xi<0$ by an over-all sign, 
as seen by comparing with the last member of 
eq.(\ref{eq_BRB_correct}).
The error occurs because the program, for small $\xi$, first 
basically computes the square of $\xi$, always obtaining a 
positive number, and then computes the arithmetic (i.e. positive)
square root, always obtaining again a positive number,
even for $\xi<0$. 
To solve this problem,
one can rewrite the round bracket, 
for example, in the following form:
\beq
\Big( 3 \, + \, 5 \, \xi^2 \, + \, \cdots \Big)^{1/2} 
\, = \, \xi \,
\left( \frac{3 \, + \, 5 \, \xi^2 \, + \, \cdots}{\xi^2} 
\right)^{1/2}.
\eeq


\subsection{Non-resonant levels above the resonant ones $(l=+1)$}

By setting 
\beq
\epsilon_2 \, = \, + \, 1
\eeq
in eq.(\ref{eq_expansion_ep1eqp1}),
one obtains for the momentum shift:
\beq
\Delta w \, = \, 
\arctan\left[ \,
1 \, + \, \frac{\xi}{2} \, + \, \mathcal{O}\left(\xi^2\right)
\right] \, = \, \frac{\pi}{4} \, + \, \frac{\xi}{4}
\, + \, \mathcal{O}\left(\xi^2\right)
\, = \, \frac{\pi}{4} \, + \, \frac{1}{4} 
\left( \nu \, + \, \frac{\pi}{4} \right) z
\, + \, \mathcal{O}\left(z^2\right),
\eeq
since, at lowest order
\beq
w \, = \, \nu \, + \, \frac{\pi}{4} \, + \, \mathcal{O}\left(z\right),
\eeq
so that:
\beq
\xi \, \equiv \, w \, z \, = \, 
\left( \nu \, + \, \frac{\pi}{4} \right) z \, + \, 
\mathcal{O}\left(z^2\right).
\eeq
Then, by taking both signs positive
in the general formula (\ref{eq_rather_involv}), 
\beq
\epsilon_1 \, = \, + \, 1; 
\quad 
\epsilon_2 \, = \, + \, 1;
\eeq
the expansion of the corresponding momentum levels
reads:
\beq
w \, = \, w_{\hat{\nu}} \, = \,
\hat{\nu} \, + \, \frac{1}{4} \, \hat{\nu} \, z 
\, + \, \mathcal{O}\left(z^2\right);
\eeq
where we have absorbed the constant $\pi/4$ 
inside the index $\nu$ by introducing the new, 
shifted index:
\beq
\qquad\qquad\qquad
\hat{\nu} \, \equiv \, \nu \, + \, \frac{\pi}{4}
\, = \, 
\pi \left( n \, + \, \frac{1}{4} \right),
\qquad\qquad
n \, = \, 0, \, 1, \, 2, \, 3, \, \cdots.
\eeq
Therefore the positive choice for both indices gives 
the non-resonant levels right above the
resonant ones:
\beq
\epsilon_1 \, = \, + \, 1, 
\quad 
\epsilon_2 \, = \, + \, 1
\quad \Rightarrow \quad l \, = \, + \, 1.
\eeq
The function $\hat{H}=\hat{H}(\xi)$, giving the recursions,
is given by:
\beq
\hat{H}(\xi) \, = \, f_{+1,+1}(\xi) \, - \, \frac{\pi}{4};
\eeq
where:
\beq
f_{+1,+1}(\xi)
\, = \,
\arctan
\left\{
\frac{1}{\xi}
\left[
\Sigma^{1/2}(\xi) \, - \, 1 
\, + \,
\bigg(
3 \, + \, 5 \, \xi^2 \, - \, \Sigma(\xi)
\, - \, 2 \, 
\frac{ 1 \, + \, 2 \, \xi^2 }{
\Sigma^{1/2}(\xi) }
\bigg)^{1/2}
\right]
\right\}.
\eeq
Note that we have subtracted the constant $\pi/4$ 
from $f_{+1,+1}(\xi)$ to obtain the function $\hat{H}$,
since we have added the same quantity
to $\nu$ for going to $\hat{\nu}$.
With this subtraction, the function $\hat{H}(\xi)$
(linearly) goes to zero for $\xi \to 0$, as it should. 

For the non-resonant $l=+1$ levels, the (still exact) momentum equation
is conveniently written:
\beq
w_{\hat{\nu}} \, = \, 
\hat{\nu} \, + \, 
\hat{H}\left( \hat{\eta} \, + \, z \, \delta w_{\hat{\nu}} \right);
\eeq
where we have defined the new ad-hoc variable:
\beq
\hat{\eta} \, \equiv \, \hat{\nu} \, z
\, = \, \pi \left( n \, + \, \frac{1}{4} \right) z.
\eeq
For easy of reference, let us give the explicit
expressions of the three lowest recursions:
\bea
w_{\hat{\nu}}^{(1)} &=& \hat{\nu} \, + \, 
\hat{H}\left( \hat{\eta} \right);
\nonumber\\
w_{\hat{\nu}}^{(2)} &=& \hat{\nu} \, + \, 
\hat{H}\left[ \hat{\eta} \, + \, z \, \hat{H} \left(\hat{\eta} \right) \right];
\\
w_{\hat{\nu}}^{(3)} &=& \hat{\nu} \, + \,
\hat{H}\Big\{ 
\hat{\eta} \, + \, z \, \hat{H}\big[\hat{\eta} \, + \, z \, \hat{H}\left( \hat{\eta} \right) \big] 
\Big\}.
\nonumber
\eea
The evaluation of higher-order recursions
is discussed in general in section$\,$\ref{eq_gen_meth_recursion}.
In the resummation scheme given by a function
series, the required functions can be evaluated
by expanding the recursive solutions 
in powers of the coupling $z$, as described in
\ref{eq_gen_meth_func_series}.


\subsection{Non-resonant levels below the resonant ones $(l=-1)$}

By setting 
\beq
\epsilon_2 \, = \, + \, 1
\eeq
in eq.(\ref{eq_expansion_ep1eqm1}),
the pole term in the argument of the arctangent
vanishes, so that:
\beq
\Delta \, w = \, \arctan\left[
- \, 1 \, + \, \frac{\xi}{2} \, + \, \mathcal{O}\left(\xi^2\right)
\right]
\, = \, - \, \frac{\pi}{4} \, + \, \frac{\xi}{4} \, + \, \mathcal{O}(\xi^2).
\eeq
By absorbing the constant term $-\pi/4$
in the index $\nu$, the total momentum can be written:
\beq
w \, = \, \nu \, + \, \Delta w
\, = \, 
\check{\nu} \, + \, \frac{1}{4} \, \check{\nu} \, z 
\, + \, \mathcal{O}\left(z^2\right);
\eeq
where we have defined the new, shifted index:
\beq
\qquad\qquad\qquad
\check{\nu} \, \equiv \, \pi \left( n \, - \, \frac{1}{4} \right);
\qquad\qquad\qquad
n = 1, 2, 3, \cdots.
\eeq
Therefore the above sign choices for the indices
gives the non-resonant levels right below the
resonant ones:
\beq
\epsilon_1 \, = \, - \, 1, 
\quad 
\epsilon_2 \, = \, + \, 1
\quad \Rightarrow \quad l \, = \, - \, 1.
\eeq
The function $\check{H}=\check{H}(\xi)$ is given by:
\beq
\check{H}(\xi) \, = \, f_{-1,+1}(\xi) \, + \, \frac{\pi}{4};
\eeq
where:
\beq
f_{-1,+1}(\xi)
\, = \,
\arctan
\left\{
\frac{1}{\xi}
\left[
- \, 1
\, - \, \Sigma^{1/2}(\xi) 
\, + \,
\bigg(\,
3 \, + \, 5 \, \xi^2 \, - \, \Sigma(\xi)
\, + \, 2 \, 
\frac{ 1 \, + \, 2 \, \xi^2 }{
\Sigma^{1/2}(\xi) }
\bigg)^{1/2}
\right]
\right\}.
\eeq
It is immediate to check that:
\beq
\check{H}(0) \, = \, 0,
\eeq
as it should, to properly generate recursions
starting from the free-theory limit $z \to 0$.
For the constant $+\pi/4$ added to $f_{-1,+1}(\xi)$
and subtracted to $\nu \mapsto \check{\nu}$,
see the remark in the previous section.

For the recursions, the (still exact) momentum equation 
is conveniently written:
\beq
w_{\check{\nu}} \, = \, 
\check{\nu} \, + \, 
\check{H}\left( 
\check{\eta} \, + \, z \, \delta w_{\check{\nu}} 
\right);
\eeq
where:
\beq
\qquad\qquad\qquad
\check{\eta} \, \equiv \, \check{\nu} \, z
\, = \, \pi \left( n \, - \, \frac{1}{4} \right) z;
\qquad\qquad
n = 1, 2, 3,\cdots.
\eeq
According to the general theory described in 
section$\,$\ref{eq_gen_meth_recursion},
the lowest three recursions explicitly read:
\bea
w_{\check{\nu}}^{(1)} &=& \check{\nu} \, + \, 
\check{H}\left( \check{\eta} \right);
\nonumber\\
w_{\check{\nu}}^{(2)} &=& \check{\nu} \, + \, 
\check{H}\left[ \check{\eta} \, + \, z \, \check{H} \left(\check{\eta} \right) \right];
\\
w_{\check{\nu}}^{(3)} &=& \check{\nu} \, + \,
\check{H}\Big\{ 
\check{\eta} \, + \, z \, \check{H}\big[\check{\eta} \, + \, z \, \check{H}\left( \check{\eta} \right) \big] 
\Big\}.
\nonumber
\eea
The resummation scheme based on a function-series
expansion can be implemented as described in 
section$\,$\ref{eq_gen_meth_func_series}.


\subsection{Non-resonant levels at the middle of two resonant levels \\ $(l=+2)$}

Finally, by setting 
\beq
\epsilon_2 \, = \, - \, 1
\eeq
in eq.(\ref{eq_expansion_ep1eqm1}),
one obtains for the momentum shift:
\beq
\label{eq_mom_Neq4_nores_half}
\Delta w \, = \, \arctan\left[\, 
- \, \frac{4}{\xi} \, - \, \frac{9}{4} \, \xi 
\, + \, \mathcal{O}\left( \xi^2 \right)
\right]
\, = \, - \, \arctan\left[\, 
\frac{4}{\xi} \, + \, \frac{9}{4} \, \xi 
\, + \, \mathcal{O}\left( \xi^2 \right)
\right].  
\eeq
In the last equality we have taken into account
that our branch of the arctangent is an odd function.
By using eq.(\ref{eq_arctan_toinv}) for $\xi<0$
(the simpler repulsive case), one obtains:
\beq
\Delta w 
\, = \, \frac{\pi}{2} \, + \, 
\arctan\left( \frac{\xi}{4} \right)
\, + \, \mathcal{O}\left( \xi^2 \right)
\, = \, \frac{\pi}{2} \, + \, \frac{\xi}{4} 
\, + \,
\mathcal{O}\left( \xi^2 \right).
\eeq
In the above formula,
one can incorporate the constant $\pi/2$
inside the index $\nu$, by writing for the
total momentum: 
\beq
w_{\tilde{\nu}} \, = \, \tilde{\nu} \, + \, \frac{1}{4} \, \tilde{\nu} \, z 
\, + \, \mathcal{O}\left(z^2\right);
\eeq
where we have defined:
\beq
\qquad\qquad\qquad
\tilde{\nu} \, \equiv \, \pi \left( n \, + \, \frac{1}{2} \right),
\qquad\qquad\qquad
n \, = \, 0, 1, 2, 3, \cdots.
\eeq
Therefore the negative choice for both indices
gives the non-resonant levels at the middle of two
adjacent resonant levels:
\beq
\epsilon_1 \, = \, - \, 1, 
\quad 
\epsilon_2 \, = \, - \, 1
\quad \Rightarrow \quad l \, = \, + \, 2.
\eeq
The function $\tilde{H}=\tilde{H}(\xi)$, correctly vanishing at 
$\xi=0$, is given by:
\beq
\tilde{H}(\xi) \, = \, f_{-1,-1}(\xi) \, - \, \frac{\pi}{2};
\eeq
where:
\beq
f_{-1,-1}(\xi)
\, = \,
\arctan
\left\{
\frac{1}{\xi}
\left[
- \, 1
\, - \, \Sigma^{1/2}(\xi) 
\, - \,
\bigg(\,
3 \, + \, 5 \, \xi^2 \, - \, \Sigma(\xi)
\, + \, 2 \, 
\frac{ 1 \, + \, 2 \, \xi^2 }{
\Sigma^{1/2}(\xi) }
\bigg)^{1/2}
\right]
\right\}.
\eeq
To generate the recursions, the momentum equation
is conveniently written:
\beq
w_{\tilde{\nu}} \, = \, 
\tilde{\nu} \, + \, 
\tilde{H}\left( \tilde{\eta} \, + \, 
z \, \delta w_{\tilde{\nu}} \right);
\eeq
where:
\beq
\tilde{\eta} \, \equiv \, \tilde{\nu} \, z
\, = \, \pi \left( n \, + \, \frac{1}{2} \right) z.
\eeq
According to the general theory described in 
section$\,$\ref{eq_gen_meth_recursion},
the lowest three recursions explicitly read:
\bea
w_{\tilde{\nu}}^{(1)} &=& \tilde{\nu} \, + \, 
\tilde{H}\left( \tilde{\eta} \right);
\nonumber\\
w_{\tilde{\nu}}^{(2)} &=& \tilde{\nu} \, + \, 
\tilde{H}\left[ \tilde{\eta} \, + \, z \, \tilde{H} \left(\tilde{\eta} \right) \right];
\\
w_{\tilde{\nu}}^{(3)} &=& \tilde{\nu} \, + \,
\tilde{H}\Big\{ 
\tilde{\eta} \, + \, z \, \tilde{H}\big[\tilde{\eta} \, + \, z \, \tilde{H}\left( \tilde{\eta} \right) \big] 
\Big\}.
\nonumber
\eea
Higher-order recursions, as well as 
the functions to be computed in the function-series
resummation scheme,
can be evaluated as described in 
sections$\,$\ref{eq_gen_meth_recursion}
and \ref{eq_gen_meth_func_series} respectively.

Let us now discuss a problem occurring in
the numerical evaluation of the function $\tilde{H}(x)$.
According to the second member in 
eq.(\ref{eq_mom_Neq4_nores_half}), the argument of the arctangent function --- the curly bracket ---
has a simple pole at $\xi=0$, with a negative residue.
In order to (numerically) evaluate the arctangent of a small
argument for $\xi\approx 0$
(thus obtaining a continuous function at $\xi=0$),
it is convenient to transform the curly bracket to its inverse.
By assuming $\xi<0$ (repulsive interaction), one obtains: 
\bea
f_{-1,-1}(\xi)
&=& + \, \arctan\left\{ \frac{1}{\xi}
\Big[- \, 1 \, - \, \Sigma^{1/2}(\xi) \, + \, \cdots \Big] \right\} \, = \, 
\nonumber\\
&=& + \, \frac{\pi}{2} \, - \, 
\arctan
\left\{
\frac{\xi}{ \big[ - \, 1  \, - \, \Sigma^{1/2}(\xi)
\, + \, \cdots \big] }
\right\}.
\eea


\subsection{Discussion}

We think that the $N=4$ case can be reasonably
considered a large-$N$ case, 
with a weak-coupling dynamics ($|z|\ll 1$) 
resembling quasi-continuum resonance dynamics
($N \gg 1$).
As we have shown, a pair of contiguous resonant
levels, with indices $(n,\,l=0)$ and $(n+1,\,l=0)$
for $n=1,2,3,\cdots$,
is separated by three contiguous non-resonant levels,
with indices
$(n,\,l=+1)$, $(n,\,l=+2)$ and $(n+1,\,l=-1)$.
A non-resonant level right above a resonant one,
i.e. with $l=+1$, has most of its variation at $z>0$
and it is naturally associate to this resonant level.
Similarly, a non-resonant level right below a resonant one,
i.e. with $l=-1$, has most of its variation at $z<0$
and it is then naturally associated to such resonant level.
Therefore, as far as resonant behavior
is concerned, momentum levels group in triples,
containing a resonant level in the middle
and two adjacent non-resonant levels, 
as it occurs in the case $N=3$.
The non-resonant levels at the middle of
two resonant levels, i.e. with $l=+2$, 
are roughly symmetrical under a sign change
of the coupling, i.e. for $z \mapsto -z$.
They do not exhibit a resonant behavior
in any coupling region and therefore
cannot be naturally associated to any 
resonant level.  
We may say that the $l=+2$ levels,
which are the main novelty of the $N=4$ case,
have the role of "separators"
of contiguous resonance triplets.
They are qualitatively similar to
the non-resonant levels of the $N=2$ model.

If we consider an initial box eigenfunction
contained in the small resonating cavity,
analog to that in eq.(\ref{eq_wf_initial}), 
we expect the no-escape probability, for example, 
to exhibit an exponential decay in some
"intermediate" time region, as it occurs in
the continuous limit $(N=\infty)$.


\section{Conclusions}
\label{Sect_Concl}

The aim of this work was understanding how 
resonance dynamics emerges in finite-volume 
Winter (or $\delta$-shell) model,
from a simple oscillating behavior at $N=1$
(both resonant cavities having the same length),
by progressively increasing the ratio
$N$ of the length of the large
cavity over the small one.
By increasing this ratio from $N=1$ up to $N=2,3,4$,
we found new resonance phenomena emerging at each step.

%
%

Winter model in the symmetric case, $N=1$,
describes a system of two identical
resonant cavities, weakly-coupled to each other.
Therefore, in this smallest possible $N$ case,
the large cavity actually has the same
length as the small one.
The main property of the $N=1$ model
is that it has no non-resonant levels,
but only resonant and exceptional levels.

According to a simple statistical-mechanics
reasoning, the exponential decay of a resonant
state with time can be considered some
kind of thermalization process of this 
"hot" single state immersed in a "cold" background
(or reservoir), 
which is composed of the huge number of 
the available decay final states.
At infinite volume, $N=\infty$, 
where the exponential decay
of resonances occurs in a rather 
clean form and in a long time interval, 
the final decay states form indeed
a continuum (in energy space). 
At large volumes, $N\gg 1$, 
where the available decay final states 
form a quasi-continuum, exponential
decay of resonances is also expected
to occur in a sizable time region.
On the contrary, at $N=1$, the level density of
the large cavity is small, as it is equal
to the level density of the small one,
so that very few decay final states
are available to an initial box eigenfunction
contained in the small cavity.
As a consequence, if the above statistical picture
of resonance decay is correct,
an initial box eigenfunction is expected to produce
an oscillating behavior 
of the inside amplitude with time,
rather than an exponential decay.
The amplitude simply oscillates with time,
from one cavity to the other one, 
back and forth.

Winter model in the double case, $N=2$, 
in which the large cavity is two times larger 
than the small one, is the smallest-$N$ case 
expected to show some mild resonance behavior.
At $N=2$, the level density of the large 
cavity (in momentum space) is indeed two times larger 
than that of the small one, 
so that, according to the statistical
picture above, the number of available
final states to resonance decay is two times bigger 
with respect to the previous $N=1$ case.
By going from $N=1$ to $N=2$, the main novelty
is the appearance of non-resonant levels
in the spectrum of the model.
Resonant and non-resonant levels, as functions
of the coupling $z$, alternate, so that they 
separate each other (they never cross).
Since the non-resonant levels are rather symmetrical
under a sign change of the coupling, 
namely $z \mapsto -z$, their growth with $z$
in the positive-coupling region $(z>0)$ is similar 
to that in the negative-coupling one $(z<0)$. 
In general, non-resonant levels do not exhibit a
resonance behavior in any coupling region.
Therefore, as far as resonance dynamics is concerned,
it is not possible to associate, in a 
natural way, a non-resonant level to a resonant one.
We may say that, at $N=2$, non-resonant levels have the role
of "separators" of contiguous resonance levels.

%
%

The main novelty of Winter model in the triple case, $N=3$, with respect to the $N=2$ case,
is the "differentiation" of non-resonant levels.
At $N=3$, there are two non-resonant levels 
for each resonant level, with sub-index 
$l = \pm 1$.
The non-resonant level right above a resonant one,
i.e. with $l=+1$, exhibits a mild resonant behavior
(a roughly linear growth with the coupling $z$)
mostly in the positive coupling region $(z>0)$.
It is therefore natural to associate such 
non-resonant level to the resonant level
right below  it. 
As we know from general resonance
theory in the quasi-continuum $(N \gg 1)$,
a resonant behavior, exhibited by
a resonant level for small couplings is,
going to larger couplings, 
softened and transferred to the 
contiguous higher non-resonant $l=+1$ level.  
In a similar way, non-resonant levels
right below a resonant one, i.e. with $l=-1$,
present a mild resonant behavior for
negative couplings $(z<0)$ and
are naturally associated to such
resonant level. 
Therefore a resonance of 
standard Winter model $(N=\infty)$
corresponds, in the $N=3$ model,
to a triplet of momentum eigenstates, 
namely a resonant state and the two
adjacent non-resonant levels.
We expect an initial box eigenfunction,
contained in the small cavity,
to exhibit an appreciable resonant behavior,
i.e. an approximate exponential decay
in some limited time interval.
In general, we conjecture that the "transition" from
an oscillating behavior to a resonant one
is already occurred, to some extent, by going
from $N=1$ to $N=3$.

%
%

Finite-volume Winter model in the quadruple case, 
$N=4$, 
presents some new qualitative features 
with respect to the previous, $N=3$ case.
In the $N=4$ case, two contiguous
resonant levels are separated by
three non-resonant levels, 
with $l = \pm 1, + 2$.
The two non-resonant levels adjacent to
a resonant one, having $l=\pm 1$,
present a similar resonant behavior 
to the corresponding levels of the $N=3$ case.
Let us remark that the resonant behavior 
of the $l=\pm 1$ levels
is more marked {\it a fortiori} in the $N=4$ model 
than in the $N=3$ one.
The non-resonant levels at the middle of two
resonant levels, i.e. with $l=+2$, are the
main novelty of the $N=3\mapsto 4$ transition. 
These levels are quite symmetrical
under a change of sign of the coupling,
do not exhibit a resonant behavior
in any coupling region and are qualitatively
similar to the non-resonant levels of the
$N=2$ model.
The main role of the $l=+2$ levels
is that of separating different triplets 
of levels, having a related resonance behavior,
i.e. a triplet of levels with $(n, \, l=0,\pm 1)$
from triples of levels with $(n\pm 1,\, l=0,\pm 1)$
We expect an initial box eigenfunction
contained in the small cavity
to exhibit an exponential decay in some
intermediate time interval, i.e. after short-time
transients (the so-called Zeno effect) and before
power effects or Poincare' recurrence
take over.

%
%

To analytically understand the dynamics of the above models,
we have constructed a high-energy expansion
for the momenta $k$ of the particle 
as functions of the coupling $z$, 
namely $k=k(z)$.
This expansion realizes an approximate
resummation of the perturbative series
of the momenta, to all orders in the coupling 
$z$.
The high-energy expansion works much better than reasonably
expected {\it a priori}, as it also works for 
low-energy states.
Furthermore, it converges quickly with the order of the approximation and uniformly in the coupling 
$z \in\RR$.
The fact that our expansion also converges 
for intermediate and large couplings $(|z|\gsim 1)$ is really a 
second, unexpected "bonus", 
because it is initially derived by assuming to be 
in the small-coupling domain $(|z|\ll 1)$.
Technically, the good behavior of our expansion
at large couplings may be due to the appearance
of arctangent functions containing the coupling $z$
inside their argument.

To go in some details, we have constructed two different resummation schemes. 
The first scheme is based on a  
function-series expansion 
(necessarily truncated at some finite order)
and has a very clear physical picture. 
The second scheme is based on a
recursion equation, which is necessarily 
solved in an approximate way, 
by computing a finite number of recursions
from the initial condition. 
At the first, leading order, the two 
resummation schemes give identical results, 
while they slightly differ from next-to-leading 
order on.
Beyond leading order, it turns out that 
the recursive scheme realizes an approximate resummation,
to all orders, of the function series generated within the first scheme.
With a suggestive language, we may say that
the recursive scheme "resums the resummation"
defining the function-series scheme.
Because of that, we believe that the recursive scheme 
generally offers a more accurate approximation
to the particle momenta than the function-series scheme. 

%
%

Let us finally discuss some possible developments
of our work which, quite generally, can be continued 
along different lines of investigation.
Finite-volume Winter model at $N=4$ is the largest-$N$
case which we have analyzed with our expansion
and which can be treated by means
of elementary functions.
By means of special functions, our
high-energy expansion for the particle momenta,
can be extended to analytically investigate finite-volume
Winter models with $N>4$.
Technically the analysis is expected to be
considerably more complicated than the ones 
in this work but is, in principle, feasible.

A similar analysis to the one presented
in this work could also be made for the
model considered in ref.\cite{Bender-Cooper},
where two harmonic oscillators are coupled
to each other by cutting the tails of their potentials
and gluing them together. 
One could study, in particular, whether
resonant states are described by clusters 
of energy levels centered around specific levels,
as it happens in finite-volume Winter model.
If finite-volume resonance dynamics turned 
out to be the same, one would get some hints 
about its universality.
In the oscillator case, the unperturbed spectrum
(free oscillators) 
is evenly spaced in energy space, rather
than in momentum space.

A different development line involves
the explicit investigation of various, intriguing
time-dependent phenomena occurring in finite-volume Winter
model.  
By considering, for example, initial box eigenfunctions 
localized in the small cavity, one can study
the depletion and the filling of this cavity with time.
Typical phenomena of resonances in a box,
such as the Poincare' recurrence or the
limited decay, can be analyzed.
For the same initial wavefunction,
one can study the temporal evolution
for different values of the coupling $z$
and of the cavities length ratio $N$.



\appendix


\section{Reduction of the tangent of a multiple angle}
\label{app_tang_mult_ang}

In this appendix we derive a general formula
for the reduction of the tangent of a multiple
angle --- namely the reduction of $\tan(Nw)$
to a rational function of $\tan(w)$ of degree $N$,
where $N$ is any positive integer.
This formula is used in the main body of the article
for the crucial simplification of the momentum
spectrum equation of the particle in all the $N>1$ cases.

According to De-Moivre formula:
\bea
\cos (N w) \, + \, i \sin (N w) &=& \exp( i N w ) 
\, = \, \big[ \exp(iw) \big]^N
\, = \, \big[ \cos(w) \, + \, i \sin(w) \big]^N 
\, =
\nonumber\\
&=& \cos(w)^N \big[ 1 \, + \, i \tan(w) \big]^N.
\eea
By equating the first and the last member of the
above chain of equality's
and taking the real part on both sides
of the resulting equation, 
one immediately obtains:
\beq
\cos(N w) \, = \, \cos(w)^N \,
\mathrm{Re}\left\{ \big[ 1 \, + \, i \tan(w) \big]^N \right\}
\qquad (w \in \RR).
\eeq
In a similar way, by equating the imaginary parts:
\beq
\sin(N w) \, = \, \cos(w)^N \,
\mathrm{Im}\left\{ \big[ 1 \, + \, i \tan(w) \big]^N \right\}
\qquad (w \in \RR).
\eeq
By dividing at each member the last equation
by the previous one, one obtains:
\beq
\label{eq_tanNw_interm}
\tan(N w) \, = \, 
\frac{\mathrm{Im}\left\{ 
\big[ 1 \, + \, i \tan(w) \big]^N \right\}}{\mathrm{Re}\left\{ 
\big[ 1 \, + \, i \tan(w) \big]^N \right\}}.
\eeq	
Since $N$ is integer by assumption,
we can apply the binomial formula 
to the expression in curly brackets above:
\bea
\left( 1 \, + \, i \, t \right)^N \, = \, 
\sum_{k=0}^N \binom{N}{k} \left( i t \right)^k 
&=& 
1 \, + \, i \, \binom{N}{1} t \, - \, \binom{N}{2} t^2 
\, - \, i \, \binom{N}{3} t^3 \, + \, \binom{N}{4} t^4
\, + \, \cdots 
\nonumber\\
&& \,\,\,\,\,\, + \, \cdots \, + \, \binom{N}{N-1}
\left( i t \right)^{N-1}
\, + \, \left( it \right)^N;
\eea
where, to simplify the notation, we have defined
\beq
t \, \equiv \, \tan(w).
\eeq
Since:
\beq
i^k \, = \,  
\begin{cases}
(-1)^{k/2} \qquad \quad \mathrm{for}\,\, k \,\, \mathrm{even},  
\\
(-1)^{(k-1) /2} \, i\quad \mathrm{for}\,\, k \,\, \mathrm{odd},
\end{cases}
\eeq
the terms with even index $k$ above are real,
with alternating sign, while the odd-index terms 
are purely imaginary, again with alternating sign.
We may therefore write:
\bea
\mathrm{Re}\left[ \big( 1 \, + \, i \, t \big)^N \right] 
&=& 
+ \, 1 \, - \, \binom{N}{2} \, t^2 \, + \, \binom{N}{4} \, t^4 
\, + \, \cdots \, + \, 
\binom{N}{2 \lfloor N/2 \rfloor } \, 
\left( - t^2 \right)^{ \lfloor N/2 \rfloor };
\nonumber\\
\mathrm{Im}\left[ \big( 1 \, + \, i \, t \big)^N \right] 
&=& + \, \binom{N}{1} \, t \, - \, \binom{N}{3} \, t^3 \, + \, \binom{N}{5} \, t^5 \, + \, \cdots 
\, + \, 
\\
&& \qquad\qquad\qquad\qquad\,\,\,
+ \, \binom{N}{2 \lfloor (N-1)/2 \rfloor + 1} 
\, t \, (-t^2)^{ \lfloor (N-1)/2 \rfloor };
\nonumber
\eea
where by $\lfloor \alpha  \rfloor$ we denote the integer part
of $\alpha$, i.e. the largest integer that is
smaller than, or equal to, $\alpha$.
By replacing the above expansions on the r.h.s. of eq.(\ref{eq_tanNw_interm}),
the final reduction formula for the tangent
of a multiple angle is obtained:
\beq
\label{eq_TanNwExpansion}
\tan \left( N w\right) 
\, = \, R_N\left[ \tan(w) \right],
\qquad N \in \NN,
\eeq
where:
\beq
\label{eq_R_fromPandQ}
R_N(t) \, \equiv \, t \, 
\frac{ P_{2\lfloor (N-1)/2 \rfloor}(t) }{ 
Q_{2\lfloor N/2 \rfloor }(t) }.
\eeq
The polynomial at the numerator is given by:
\beq
\label{eq_P_polin}
P_{2\lfloor (N-1)/2 \rfloor}(t) 
\, \equiv \,
\sum_{r=0}^{\lfloor (N-1)/2 \rfloor}
(-1)^r \, \binom{N}{2r+1} \, t^{2r};
\eeq
while the polynomial at the denominator 
reads:
\beq
\label{eq_Q_polin}
Q_{2 \lfloor N/2 \rfloor}(t)
\, \equiv \,
\sum_{r=0}^{\lfloor N/2 \rfloor}
(-1)^r \, \binom{N}{2r}
\, t^{2r}.
\eeq
For even $N$, the polynomial at the numerator
of the rational function, 
$t P_{2\lfloor(N-1)/2\rfloor}(t)$,
has degree $N-1$, while the polynomial
at the denominator, $Q_{2\lfloor N/2 \rfloor}(t)$,  
has degree $N$.
For odd $N$, degrees are interchanged:
$t P_{2\lfloor (N-1)/2 \rfloor}(t)$
has degree $N$, while $Q_{2\lfloor N/2 \rfloor}(t)$ has degree $N-1$.
Therefore, for any $N$, the rational function
$R_N(t)$ has degree $N$.

By specifying eq.(\ref{eq_TanNwExpansion}) 
to the cases $N=2,\,3,\,4$, we immediately obtain:
\bea
\label{eq_tan_red_partic}
\tan \left( 2w \right) 
&=& \frac{2 \tan(w)}{1 \, - \, \tan^2(w)} ;
\nonumber\\
\tan \left( 3w \right) \, 
&=& \tan(w) \, \frac{3 \, - \, \tan^2(w)}{
1 \, - \, 3 \tan^2(w)} ;
\\
\tan \left( 4w \right) \, 
&=& 4 \tan(w) \, 
\frac{1 \, - \, \tan^2(w)}{
1 \, - \, 6\tan^2(w) \, + \, \tan^4(w)}.
\nonumber
\eea


\section{General third-order algebraic equation}
\label{app_Card_ord3}

In this appendix we present a derivation
of the formula for the zeroes of a general third-order
algebraic equation. 
This formula involves nested square and cubic
roots and rational operations and has been used
to determine the momentum spectrum
of Winter model in the case $N=3$.
As we are going to show, the solution
involves a second-order auxiliary equation.

Let us then consider a general, monic third-order 
algebraic equation:
\beq
\label{eq_Neq3_general}
x^3 \, + \, a \, x^2 \, + \, b \, x \, + \, c 
\, = \, 0,
\eeq
where $a,b,c$ are arbitrary complex numbers.
If $a \ne 0$, the term proportional to $x^2$,
next to the highest power, can always
be sent to zero by means of the shift
\beq
x \, = \, y \, - \, \frac{a}{3},
\eeq
bringing the equation to the form
(called reduced form):
\beq
\label{eq_Neq3_reduce}
y^3 \, + \, p \, y \, + \, q \, = \, 0,
\eeq
where:
\bea
\label{FromabcTopq}
p &=& p(a,b ) \, = \, b \, - \, \frac{a^2}{3};
\nonumber\\
q &=& q(a,b,c) \, = \, c \, - \, \frac{a \, b}{3} \, + \, \frac{2}{27} a^3. 
\eea
Next one uses the identity (cube of the binomial):
\beq
(u+v)^3 \, - \, 3 u v (u+v) \, - \, u^3 \, - \, v^3 \, = \, 0;
\eeq
where $u$ and $v$ are two arbitrary complex
numbers.
By setting:
\beq
y \, = \, u \, + \, v,
\eeq
the above identity is re-written:
\beq
y^3 \, + \, p \, y \, + \, q \, = \, 0;
\eeq
where we have defined:
\bea
p &\equiv& - \, 3 uv;
\nonumber\\
q &\equiv& - \, u^3 \, - \, v^3.
\eea
The two equations above are easily
solved for $u$ and $v$.
For example, one takes the third power of the
first equation,
\beq
p^3 \, = \, - \, 27 u^3 v^3,
\eeq
and solves it with respect to $v^3$:
\beq
v^3 \, = \, - \, \frac{p^3}{27 \, u^3} \qquad (u \ne 0).
\eeq
By substituting the above solution in the second equation
of the system, namely $u^3+v^3+q=0$,
one obtains the equation
\beq
u^3 \, - \, \frac{p^3}{27 \, u^3} \, + \, q \, = \, 0;
\eeq
which is basically the bi-cubic equation
\beq
\label{eq_auxiliary_Neq3}
t^2 \, + \, q \, t \, - \, \frac{p^3}{27} \, = \, 0,
\eeq
where we have defined:
\beq
t \, \equiv \, u^3.
\eeq
Eq.(\ref{eq_auxiliary_Neq3}) is an auxiliary second-order
algebraic equation in $t$, which is easily solved to give:
\beq
t \, = \, - \, \frac{q}{2} \, \pm \, \sqrt{\frac{q^2}{4} \, + \, \frac{p^3}{27}}.
\eeq
The solution for the original variable $u$
is simply obtained by taking a cubic root
on both sides of the above equation:
\beq
u \, = \, \sqrt[3]{t} \, = \,
\sqrt[3]{- \, \frac{q}{2} \, \pm \, \sqrt{\frac{q^2}{4} \, + \, \frac{p^3}{27}}}.
\eeq
Since $uv=-p/3$, the quantity $v$ has
the opposite sign determination of the
square root with respect to $u$:
\beq
v \, = \, 
\sqrt[3]{- \, \frac{q}{2} \, \mp \, \sqrt{\frac{q^2}{4} \, + \, \frac{p^3}{27}}}.
\eeq
The solution of the reduced equation (\ref{eq_Neq3_reduce})
therefore reads:
\beq
y \, = \, u \, + \, v \, = \,
\sqrt[3]{- \, \frac{q}{2} \, + \, \sqrt{\frac{q^2}{4} \, + \, \frac{p^3}{27}}}
\, + \,
\sqrt[3]{- \, \frac{q}{2} \, - \, \sqrt{\frac{q^2}{4} \, + \, \frac{p^3}{27}}}.
\eeq
By introducing the primitive cubic root
of unity,
\beq
f \, = \, \exp\left( \frac{2 \pi i}{3} \right),
\eeq
$(f^3 \, = \, 1; \,\, \bar{f} \, = \, f^2; \,\, \bar{f}^2 \, = \, f)$,
all the three solutions of the reduced equation can be 
explicitly written as:
\beq
y_n \, = \,
f^n \, \sqrt[3]{- \, \frac{q}{2} \, + \, 
\sqrt{\frac{q^2}{4} \, + \,  \frac{p^3}{27}}}
\, + \,
{\bar{f}}^n \, \sqrt[3]{- \, \frac{q}{2} \, - \, 
\sqrt{\frac{q^2}{4} \, + \, \frac{p^3}{27}}};
\qquad\qquad n = 0, 1, 2.
\eeq
The solution of the original equation (\ref{eq_Neq3_general})
is simply obtained by subtracting $a/3$ to $y_n$:
\beq
x \, = \, x_n \, = \, y_n \, - \, \frac{a}{3} \, = \,
f^n \,
\sqrt[3]{- \, \frac{q}{2} \, + \, \sqrt{\frac{q^2}{4} \, + \, \frac{p^3}{27}}}
\, + \,
\bar{f}^n \,
\sqrt[3]{- \, \frac{q}{2} \, - \, \sqrt{\frac{q^2}{4} \, + \, \frac{p^3}{27}}}
\, - \, \frac{a}{3};
\qquad
\eeq
where $n=0,1,2$, and the coefficients $p$ and $q$ are
given in terms of the original coefficients
$a,b,c$ in eqs.(\ref{FromabcTopq}).


\section{General fourth-order algebraic equation}
\label{app_Card_ord4}

In this appendix we sketch the derivation
of the formula for the zeroes of a general fourth-order
algebraic equation. 
Similarly to the third-order case treated
in the previous appendix,
this formula involves nested square and cubic
roots and rational operations.
It is however much more complicated than the 
third-order one, because it involves an auxiliary 
equation of third-order (rather than of second order).
The general solution of the fourth-order
equation has been used 
to determine the momentum spectrum
of Winter model in the case $N=4$.

A general monic fourth-order algebraic equation
reads:
\beq
\label{eq_gen_IVorder}
x^4 \, + \, a \, x^3 \, + \, 
b \, x^2 \, + \, c \, x \, + \, d \, = \, 0,
\eeq
where $a,b,c,d$ are given complex numbers.
If $a \ne 0$, by means of the shift
\beq
x \, = \, y \, - \, \frac{a}{4},
\eeq
one obtains the reduced fourth-order equation:
\beq
y^4 \, + \, p \, y^2 \, + \, q \, y \, + \, r \, = \, 0;
\eeq
where:
\bea
\label{eq_fromabcdtopqr}
p&=&p(a,b) \, = \, b \, - \, \frac{3}{8} a^2 ;
\nonumber\\
q&=&q(a,b,c) \, = \, c \, - \, \frac{a \, b}{2} + \frac{a^3}{8};
\\
r&=&r(a,b,c,d) \, = \, d \, + \, \frac{a^2 \, b}{16} \, - \, \frac{a \, c}{4}
\, - \, \frac{3}{256} a^4.
\nonumber
\eea
The general idea is to rewrite the reduced equation
in such a way that both its left-hand-side and
its right-hand-side become squares
of polynomials of lower degree.
Let us begin with the two higher powers of $y$,
which we keep on the l.h.s.:
\beq
y^4 \, + \, p \, y^2 \, = \, - \, q \, y \, - \, r .
\eeq
By adding $p^2/4$ to both sides of the
above equation, the latter is  written:
\beq
\label{eq_deg4_interm}
\left( y^2 \, + \, \frac{p}{2} \right)^2
\, = \, \frac{p^2}{4} \, - \, q \, y \, - \, r .
\eeq
So, we succeeded in writing the l.h.s.
as the square of a quadratic (i.e. second-degree) polynomial
in $y$ (without the linear term).
Now we want to write the r.h.s. as
the square of a linear (i.e. first degree)
polynomial in $y$.
It is clear that we have to add a term 
quadratic in $y$ to both sides of the above
equation; in doing that, we must not spoil
the perfect square on the l.h.s..
Let us add therefore to both sides
of eq.(\ref{eq_deg4_interm}) the quantity
\beq
v^2 \, + \, 2 \, v \left( y^2 \, + \, \frac{p}{2} \right),
\eeq
with $v$ a still undetermined quantity,
to obtain:
\beq
\left( y^2 \, + \, \frac{p}{2} \, + \, v \right)^2
\, = \, 2 \, v \, y^2 \, - \, q \, y \, + \,
\frac{p^2}{4} \, + \, p \, v \, + \, v^2 \, - \, r .
\eeq
According to the above idea,
we now simply choose $v$
in such a way that the r.h.s.
is the square of a linear polynomial:
\bea
&& 2v \left( 
y^2 \, - \, \frac{q}{2 \, v} \, y \, + \, 
\frac{v^2 \, + \, p \, v \, + \, p^2/4 \, - \, r}{2 \, v}
\right)
\, = \, 2 \, v \left( y \, - \, \frac{q}{4v} \right)^2 \, =
\nonumber\\
&=& 
2 \, v \left( 
y^2 \, - \, \frac{q}{2 \, v} \, y \, + \, \frac{q^2}{16 \, v^2} 
\right).
\eea
By equating the first member in the
above chain of equalities with the last one, 
one obtains the following auxiliary
(general) third-order equation in $v$:
\beq
\label{eq_auxiliary_ordIII}
v^3 \, + \, \tilde{a} \, v^2 
\, + \, \tilde{b} \, v
\, + \, \tilde{c} \, = \, 0;
\eeq
with the coefficients given by:
\bea
\label{eq_aux_ordIII_coef}
\tilde{a} &=& + \, p;
\nonumber\\
\tilde{b} &=& + \, \frac{p^2}{4} \, - \, r;
\nonumber\\
\tilde{c} &=& - \, \frac{q^2}{8}.
\eea
The above equation can be solved by means of the general 
method described in the previous appendix$\,$\ref{app_Card_ord3}.
Let us call
\beq
v_0 \, = \, \tilde{v}_0\left(\tilde{a},\tilde{b},\tilde{c} \right) 
\, = \, v_0\left(p,q,r\right)
\eeq
anyone of its solutions
(note that we are free to choose any solution
we like, as any solution would work).
The fourth-order equation is then written:
\beq
\left( y^2 \, + \, \frac{p}{2} \, + \, v_0 \right)^2
\, = \, 2 \, v_0 \left( y \, - \, \frac{q}{4 \, v_0} \right)^2.
\eeq
By taking the square root on both sides,
one obtains two second-order equations in $y$:
\beq
y^2 \, + \, \frac{p}{2} \, + \, v_0
\, = \, \pm \, \sqrt{2 \, v_0} \left( y \, - \, \frac{q}{4 \, v_0} \right).
\eeq
The solutions of the above equations explicitly read:
\beq
y \, = \, \pm \, \frac{\sqrt{2 \, v_0}}{2}
\, (\pm) \, 
\sqrt{- \, \frac{p \, + \, v_0}{2} \, \mp \, \frac{q}{2 \, \sqrt{2 \, v_0}}}.
\eeq 
The sign determination in front of the big square
root has been put in parenthesis to indicate
that it is an independent determination with respect
to the other one. 
In general, therefore, four different roots
are obtained, as it should.

The zeroes of the original fourth-order equation 
(\ref{eq_gen_IVorder}) are given by:
\beq
x \, = \, y \, - \, \frac{a}{4}
\, = \, 
\pm \, \frac{\sqrt{2 \, v_0}}{2}
\, (\pm) \, 
\sqrt{- \, \frac{p \, + \, v_0}{2} \, \mp \, \frac{q}{2 \, \sqrt{2 \, v_0}}}
\, - \, \frac{a}{4},
\eeq
where $v_0$ is a solution of the auxiliary third-order
equation (\ref{eq_auxiliary_ordIII}).
The coefficients $\tilde{a},\tilde{b},\tilde{c}$ of this equation are given in terms of the coefficients $p,q,r$
of the reduced fourth-order equation in 
eqs.(\ref{eq_aux_ordIII_coef}).
The latter, in turn, are given 
in terms of the coefficients $a,b,c,d$
of the original fourth-order equation in eqs.(\ref{eq_fromabcdtopqr}).

\end{document}